\pdfoutput=1
\documentclass[11pt,twoside,a4paper,cmspaper,final,collab]{cms-tdr}
\def\svnVersion{67dd6f1}\def\svnDate{2026/03/25}\def\cmsCernNoTag{CERN-EP-2025-241}\def\cmsCernDate{\today}\def\cmsMessage{Published in the Journal of High Energy Physics as \href{http://dx.doi.org/10.1007/JHEP03(2026)189}{\doi{10.1007/JHEP03(2026)189}.}}
\begin{document}\cmsNoteHeader{EXO-24-008}

\newcommand{\Pomega}{\ensuremath{\HepSusyParticle{\PGo}{}{}}\xspace}
\newcommand{\Peta}{\ensuremath{\HepSusyParticle{\PGh}{}{}}\xspace}
\newcommand{\TLambda}{\ensuremath{\tilde{\Lambda}}\xspace}
\newcommand{\Ppi}{\ensuremath{\HepSusyParticle{\PGp}{}{}}\xspace}
\newcommand{\dpart}{\ensuremath{\psi}\xspace}
\newcommand{\dApart}{\ensuremath{\bar{\dpart}}\xspace}
\newcommand{\dphoton}{\ensuremath{\mathrm{A}'}\xspace}
\newlength\cmsTabSkip\setlength{\cmsTabSkip}{1ex}
\providecommand{\cmsTable}[1]{\resizebox{\textwidth}{!}{#1}}

\cmsNoteHeader{EXO-24-008}
\title{Search for low-mass hidden-valley dark showers with non-prompt muon pairs in proton-proton collisions at \texorpdfstring{$\sqrt{s} = 13\TeV$}{sqrt(s) = 13 TeV}}

\date{\today}

\abstract{ A search for signatures of a dark analog to quantum chromodynamics is performed. The analysis targets long-lived dark mesons that decay into standard-model particles, with a high branching fraction of the dark mesons decaying into muons. The dark mesons are formed by the hadronisation of dark partons, which are produced by a decay of the Higgs boson. The search is performed using a data set corresponding to an integrated luminosity of 41.6\fbinv, which was collected in proton-proton collisions at $\sqrt{s}=13\TeV$ by the CMS experiment at the CERN LHC in 2018 using non-prompt muon triggers. The search is based on resonant muon pair signatures. Machine-learning techniques are employed in the analysis, utilising boosted decision trees to discriminate between signal and background. No significant excess is observed above the standard model expectation. Upper limits on the branching fraction of the Higgs boson decaying to dark partons are determined to be as low as $10^{-4}$ at 95\% confidence level, surpassing and extending the existing limits on models with dark \Pomega mesons for mean proper decay lengths of less than 500\mm and for \Pomega masses down to 0.3\GeV. First limits are set for extended dark-shower models with two dark flavours that contain dark photons, probing their masses down to 0.33\GeV.}

\hypersetup{%
pdfauthor={CMS Collaboration},%
pdftitle={Search for low-mass hidden-valley dark showers with non-prompt muon pairs in proton-proton collisions at sqrt(s) = 13 TeV},%
pdfsubject={CMS},%
pdfkeywords={CMS, exotica, long-lived, LLP}}

\maketitle

\section{Introduction}
The standard model (SM) of particle physics has made many successful predictions; however, there remain a number of open questions to be answered. A major unresolved question concerns the nature of dark matter. From various astrophysical observations \cite{Rubin:1980zd, Markevitch:2003at, Planck:2018vyg}, the existence of a ``dark'' form of matter is inferred, which does not interact with the electromagnetic force but has mass. About 27\% of the energy content of our universe is believed to be made up of dark matter, compared to only 5\% being visible matter \cite{Rubin:1980zd,Planck:2018vyg,Arbey:2021gdg,Persic:1995ru}.

Many theories beyond the SM try to account for dark matter. For example, hidden-valley models, which extend the SM to incorporate a new, non-Abelian gauge group \cite{Strassler:2006im, Hur:2007uz, Kribs:2009fy, Beauchesne:2018myj, Francis:2018xjd, Bernreuther:2019pfb, Cheng:2021kjg, Cheng:2024hvq}. Dark matter particles are charged under this new, dark gauge group. Hidden-valley models can produce events with high-multiplicity sprays of dark particles, known as dark showers. In hidden-valley models with confined dark charges, the dark particles assemble themselves to form dark hadrons. Dark hadrons can decay into SM particles after a measurable lifetime, creating displaced signatures in detectors. Dark showers containing sprays of dark hadrons are expected to create high-multiplicity displaced signatures \cite{Knapen:2021eip, Strassler:2006im}. This analysis targets long-lived dark mesons that decay with a high branching fraction into muons, by searching for displaced vertices formed from resonant muon pairs.

 Three broad classes of dark-sector models are explored: the vector portal model with a long-lived dark vector meson \cite{Knapen:2021eip}; a model with a long-lived dark photon, Scenario A \cite{Born:2023vll}; and a model with a long-lived dark pion, Scenario B1 \cite{Born:2023vll}. The nomenclature of the latter two models follows that in Ref.~\cite{Born:2023vll}. The production of dark showers from decays of the SM Higgs boson is considered because it has a narrow width and is a sensitive probe for low-mass SM singlet states \cite{Curtin:2013fra}. 
 
\begin{figure}[ht] 
\centering
\includegraphics[width=0.5\textwidth]{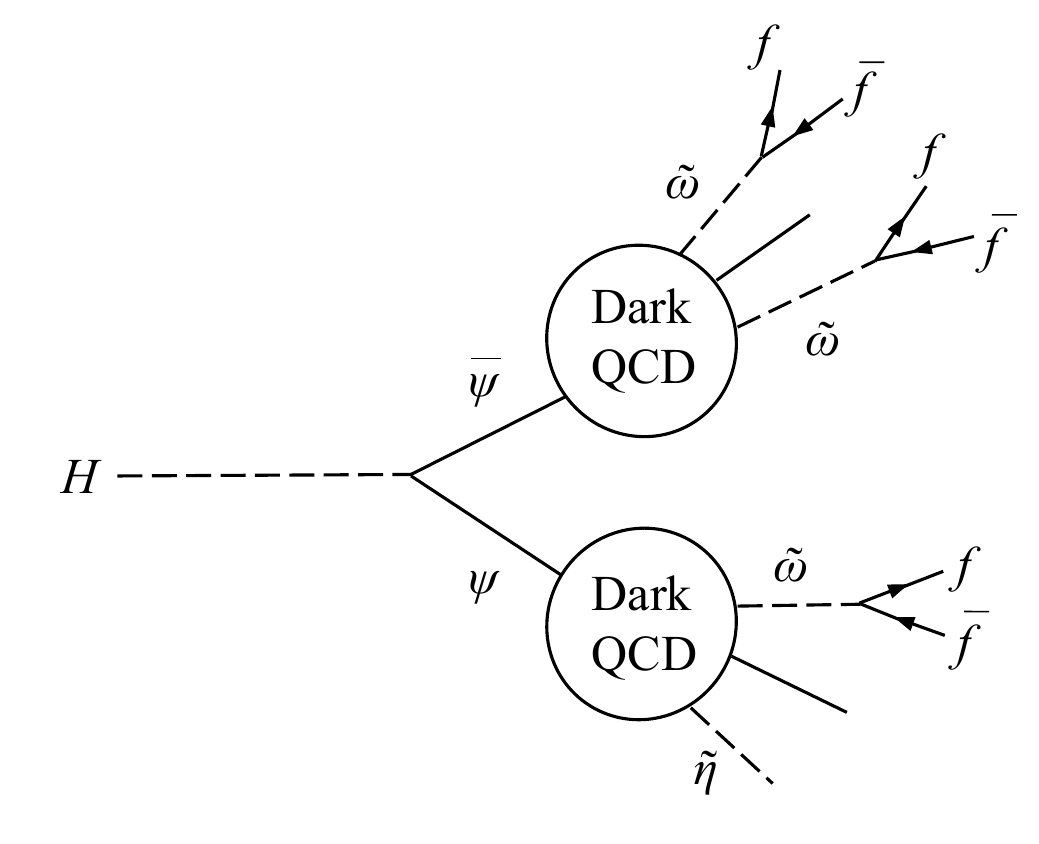}
\caption{Diagram for the vector portal model. An SM Higgs boson decays to dark partons $\dpart\dApart$, which then hadronise to form dark hadrons including dark vector mesons \Pomega and dark pseudoscalar mesons \Peta. The \Pomega then undergoes displaced decay into SM fermions $f\bar{f}$.}
\label{Feynman_diagram_vector_portal}
\end{figure}

 In the vector portal model, there are two types of dark mesons (\Peta and \Pomega), and it is assumed that only one of them has a measurable lifetime. This dark meson is referred to as the visibly-decaying dark particle. The other dark meson is assumed to either decay promptly to the visibly-decaying dark particle or to escape the detector and thus create an apparent momentum imbalance. Even though this choice does not account for the full range of possible dark-shower topologies, it still encompasses the phenomenology of a broad class of models and describes a wide range of distinct signatures with a minimal number of arbitrary parameters \cite{Knapen:2021eip}. A few theory assumptions are made when constructing the model. It is assumed that there are no new sources of SM flavour violation. Operators up to dimension-5 in the infrared effective theory are allowed \cite{Knapen:2021eip}. A diagram for the vector portal model is shown in Fig.~\ref{Feynman_diagram_vector_portal}. The Higgs boson first decays into a pair of dark partons $\dpart\dApart$, which then hadronise through a dark analog to quantum chromodynamics (QCD) into dark hadrons, among them dark vector mesons \Pomega and dark pseudoscalar mesons \Peta. The \Pomega has a non-prompt decay into SM particles, with a significant branching fraction of 10--30\% for the decay into muons.

\begin{figure}[ht] 
\centering
\includegraphics[width=0.46\textwidth]{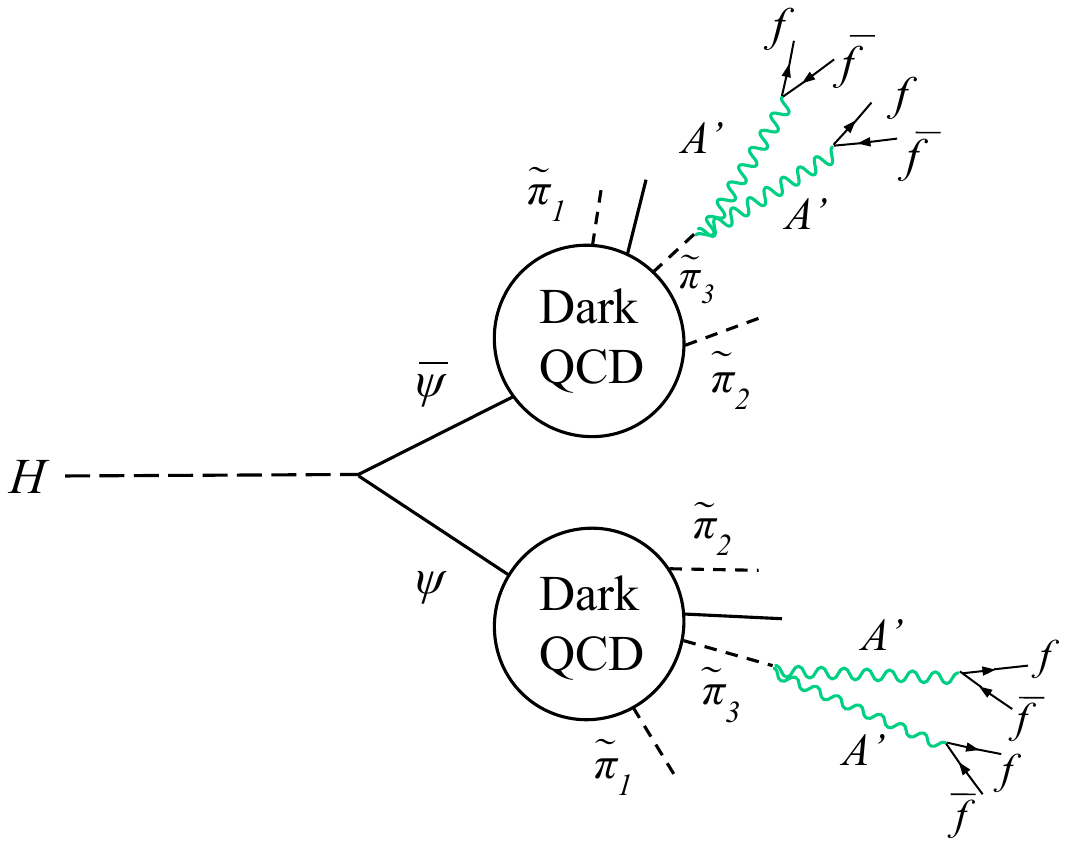}
\hfill
\includegraphics[width=0.46\textwidth]{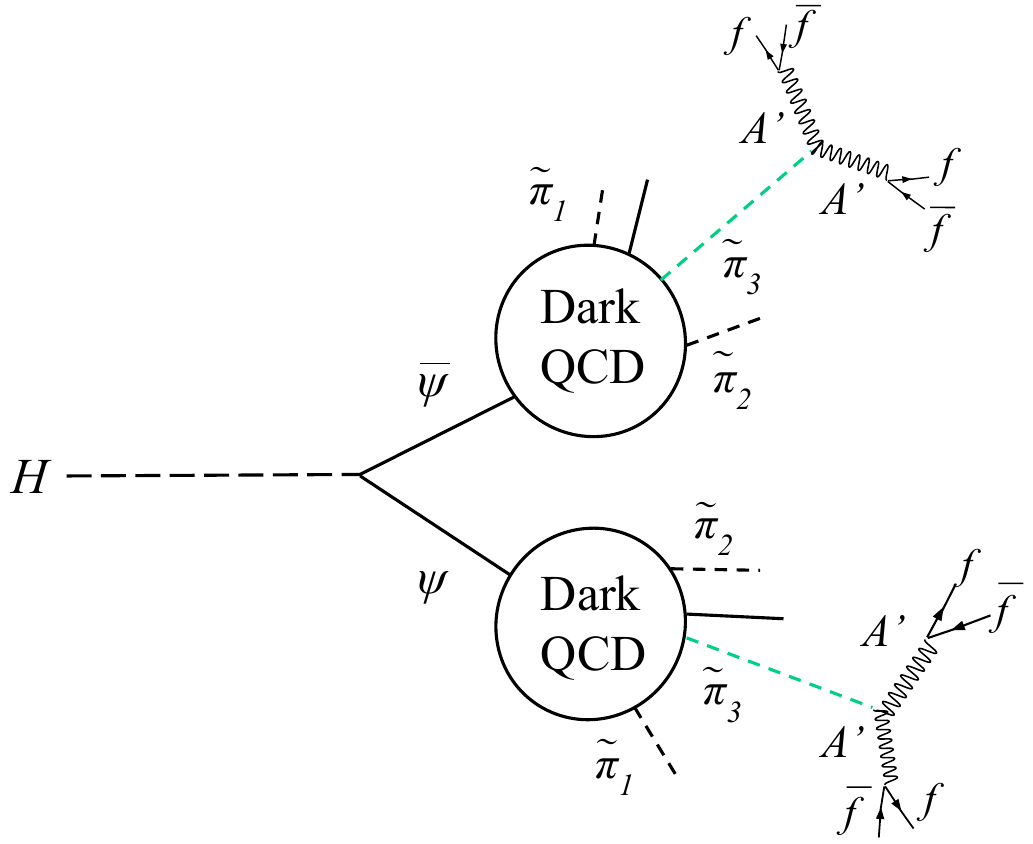}
\caption{Diagrams for the Scenario A model (left) and the Scenario B1 model (right). In these extended models the dark hadronisation produces a spectrum of dark mesons, including the dark pions $\Ppi_{1}$, $\Ppi_{2}$ and $\Ppi_{3}$. The $\Ppi_{3}$ then decays into SM fermions $f\bar{f}$ through the dark photon \dphoton. Green is used to indicate a long-lived particle. The \dphoton is a long-lived particle in Scenario A, while $\Ppi_{3}$ is a long-lived particle in Scenario B1.}
\label{Feynman_diagram_scenarioX}
\end{figure}

Scenarios A and B1 consist of extended hidden valley models with more complex decay topologies \cite{Born:2023vll}. These models have two dark flavours instead of one, which leads to a wider spectrum of dark hadrons, including three dark pions ($\Ppi_{1}$, $\Ppi_{2}$, and $\Ppi_{3}$) and a heavier pseudoscalar meson \Peta. The benchmark models include both pointing and non-pointing decay topologies, which refer to whether the dimuon vertices from the decays point back to the beam axis or not. The analysis is designed to be sensitive to both topologies. Diagrams for the decays in these scenarios are shown in Fig.~\ref{Feynman_diagram_scenarioX}. In both scenarios, the pseudoscalar \Peta is assumed to decay promptly into three dark pions, and the dark pions $\Ppi_{1}$ and $\Ppi_{2}$ are assumed to be stable. In Scenario A, the dark pion $\Ppi_{3}$ decays promptly into two dark photons \dphoton, which can then each undergo a non-prompt decay into muons. The vector sum of the momenta of the two dimuon vertices points back to the beam axis, so it is referred to as the pointing scenario. In Scenario B1, the dark pion $\Ppi_{3}$ undergoes a displaced decay into two dark photons \dphoton, which can then decay promptly into muons. The $\Ppi_{3}$ is a long-lived particle in Scenario B1, while \dphoton is a long-lived particle in Scenario A. The dark photons are produced in the non-prompt decay of a long-lived parent particle in Scenario B1, and the momentum vectors of the resulting dimuon vertices do not point back to the beam axis, so this is referred to as the non-pointing scenario. As the dark photon decay is taken to be prompt, the dimuon vertices overlap with each other in this scenario. The pointing angle, defined as the angle between the momentum vector of the dimuon system and the displacement vector from the primary vertex to the secondary vertex, is used to characterise the decay topology. The primary vertex is taken to be the vertex corresponding to the hardest scattering in the event, evaluated using tracking information alone, as described in Section 9.4.1 of Ref.~\cite{CMS-TDR-15-02}.

The search presented in this paper is performed using a data set collected in 2018 by the CMS experiment at the CERN LHC using a ``data parking'' strategy at a proton-proton (pp) center-of-mass energy of
13\TeV, corresponding to an integrated luminosity of 41.6\fbinv \cite{CMS:2024zhe,Bainbridge:2730111}. For the \PQB-parking data set, the thresholds used by the trigger algorithms are lowered with respect to standard data sets, facilitating the selection of low transverse momentum (\pt) muons that are displaced from the primary vertex. This data set enables searches for low-mass, long-lived dark-sector particle decays into displaced muons. The present analysis probes a dimuon invariant mass range of 0.3 to 20\GeV. Machine-learning techniques are applied to perform background rejection. The \PQB-parking data set includes full event reconstruction, which provides input for machine learning in this analysis. 

Compared to previous dimuon searches performed at the CERN LHC \cite{CMS:2021sch,CMS:2024qxz,ATLAS:2018rjc,LHCb:2020ysn}, the study presented in this paper is the first to target hidden-valley dark-shower models with dark-sector particles decaying into pairs of muons. The sensitivities of previous searches in the low-mass regime were limited by dimuon mass restrictions. Ref.~\cite{CMS:2021sch} requires the four-muon mass of muon pairs from two displaced vertices to be compatible with the Higgs boson mass. In Ref.~\cite{CMS:2024qxz} and Ref.~\cite{ATLAS:2018rjc} there are requirements on the dimuon mass to be larger than 10\GeV and 15\GeV respectively. Compared to the long-lived particle search in Ref.~\cite{CMS:2024bvl} performed by the CMS experiment in the muon detectors, this analysis makes use of the tracker, which complements the sensitivity in the lower displacement region. This analysis exploits a different signature from those in Ref.~\cite{CMS:2024gxp,CMS:2018bvr}, which search for dark showers through the emerging jet signature \cite{Schwaller:2015gea,Renner:2018fhh}.  

Tabulated results for this analysis are provided in the HEPData record \cite{HEPData}.


\section{The CMS detector}
The CMS apparatus~\cite{CMS:2008xjf,CMS:2023gfb} is a multipurpose, nearly hermetic detector, designed to trigger on~\cite{CMS:2020cmk,CMS:2016ngn} and identify electrons, muons, photons, and (charged and neutral) hadrons~\cite{CMS:2020uim,CMS:2018rym,CMS:2014pgm}. A global ``particle-flow'' algorithm~\cite{CMS:2017yfk} aims to reconstruct all individual particles in an event, combining information provided by the all-silicon inner tracker and by the crystal electromagnetic and brass-scintillator hadron calorimeters, operating inside a 3.8\unit{T} superconducting solenoid, with data from the gas-ionisation muon detectors embedded in the flux-return yoke outside the solenoid. The reconstructed particles are used to identify jets and leptons and to measure missing transverse momentum~\cite{CMS:2018jrd,CMS:2016lmd,CMS:2019ctu}.  

Events of interest are selected using a two-tiered trigger system. The first level (L1), composed of custom hardware processors, uses information from the calorimeters and muon detectors to select events at a rate of around 100\unit{kHz} within a fixed latency of 4\mus~\cite{CMS:2020cmk}. The second level, known as the high-level trigger (HLT), consists of a farm of processors running a version of the full event reconstruction software optimised for fast processing, and reduces the event rate to a few\unit{kHz} before data storage~\cite{CMS:2016ngn}. 

In standard data sets, full offline reconstruction proceeds soon after the data are recorded. The data-parking strategy was designed to overcome the limitation in the ability to perform prompt reconstruction \cite{Bainbridge:2730111}. The trigger thresholds are lowered in the parking strategy, leading to a higher trigger rate, which exceeds the computing capacity available for prompt event reconstruction. The data stream is therefore transferred, unprocessed, to tape storage (the data stream is "parked"). It is kept in a raw format until sufficient computing resources become available for the events to be reconstructed, for example during shutdowns between data-taking periods. 

Muons are identified as tracks in the inner tracker that are consistent with either a track or a segment made up of several hits in the muon system. The energy of muons is obtained from the corresponding track momentum, which is measured from the curvature of the track. Muons are measured in the range of $\abs{\eta}<2.4$. The single-muon trigger efficiency exceeds 90\% over the full $\eta$ range. The efficiency to reconstruct and identify muons with $\pt>20\GeV$ is greater than 96\%. Matching muons to tracks measured in the silicon tracker results in a relative \pt resolution of 1\% in the barrel and 3\% in the endcaps for muons with \pt up to 100\GeV~\cite{CMS:2018rym}.

\section{Data sets, triggers and Monte Carlo simulation}
\label{datasets}
For this analysis, the \PQB-parking data set collected by CMS during the 2018 data-taking period \cite{CMS:2024zhe,Bainbridge:2730111} is used. In order to enrich the data set in \PQB hadron decays, the events collected contain at least one muon that satisfies kinematic requirements on the transverse momentum and the impact parameter significance. The latter is defined as ${\mathrm{IP}}_{\text{sig}}=\mathrm{IP}/\sigma_{\mathrm{IP}}$, where $\mathrm{IP}$ is the impact parameter of the muon with respect to the beam axis in the transverse plane, and $\sigma_{\mathrm{IP}}$ is its uncertainty. The impact parameter is defined as the distance of closest approach of the muon track to the beam axis. The minimum requirement on $\mathrm{IP}_{\text{sig}}$ serves the dual purpose of reducing the prompt muon background while enhancing the potential contribution from new long-lived particles. The \pt thresholds in the triggers range from 7 to 12\GeV, and the thresholds on ${\mathrm{IP}}_{\text{sig}}$ vary between 3 and 6. The sample contains ${\sim}1.3\times10^{10}$ events, with ${\sim}70\%$ being B hadron decays.

Simulated signal samples are produced with \PYTHIA 8.240 \cite{Sjostrand:2014zea} for the vector portal model, and with \PYTHIA 8.309 \cite{Bierlich:2022pfr} for Scenario A and Scenario B1 models. The NNPDF3.1 parton distribution functions (PDFs) at next-to-next-to-leading order are used \cite{NNPDF:2017mvq}. The search explores a dimuon mass range between 0.3 and 20\GeV. A variety of mean proper decay lengths ($c\tau$) of long-lived particles in these three models are considered, ranging from 0.1 to 500\mm. The search targets low-mass dimuon vertices, which are highly Lorentz boosted. The range of $c\tau$ considered is constrained by the drop in sensitivity at higher displacement. The gluon-gluon fusion production mode of the Higgs boson is considered, which has a cross section of 43.9\unit{pb}. A branching fraction of 0.01 is assumed for the Higgs boson decaying into dark partons to provide an illustration of a possible signal. The parameters of the models are introduced in Table \ref{model_parameters}.

The vector portal model utilises three dark colours, and for simplicity, one dark flavour. The resulting dark hadron spectrum consists of two dark flavour singlet states: a spin-zero meson \Peta, and a spin-one
 meson \Pomega. The dark quark mass is set to be approximately half the mass of \Peta \cite{Carloni:2011kk}. In the benchmark models it is assumed that $m_{\Pomega}=\TLambda=m_{\Peta}$, where \TLambda is the dark sector confinement scale, so that the decay of \Pomega into \Peta is kinematically forbidden, preventing the \Pomega from decaying into other dark-sector states \cite{Knapen:2021eip}. The branching fractions of the \Pomega meson decaying into SM fermions are set according to Ref.~\cite{Knapen:2021eip}, and they depend on the \Pomega mass.

For Scenario A and Scenario B1, the number of dark flavours is set to two and the number of dark colours is set to three. It is assumed that $m_{\Peta}=\TLambda$. There is a mixing angle $\theta$ that parametrises the isospin violation in the coupling of the dark photon to dark-sector quarks. The mixing angle is taken to be $\sin\theta=0.1$ so that the isospin breaking is relatively small, which results in the dark pions having similar masses, \ie $m_{\Ppi_{1}}\approx m_{\Ppi_{2}}\approx m_{\Ppi_{3}}$. This gives a simpler mass spectrum, and also ensures the simulation from \PYTHIA provides a reasonable approximation to the hadronisation process. The branching fraction $\mathcal{B}(\Ppi_{3}\to \dphoton\dphoton)$ is assumed to be 1 \cite{Born:2023vll}. The dark photon \dphoton decays into SM fermions by kinematically mixing with the SM photon, and the branching fractions are given by Ref.~\cite{Curtin:2014cca}. The assumptions made for all the models are theoretically well-motivated and offer simplified benchmark models for interpretation.

For the background Monte Carlo (MC) simulations, a set of muon-enriched QCD background samples are produced, which contain events with at least one muon with $\pt>5\GeV$. The samples are generated with \PYTHIA 8.240, and they are binned in $\widehat{p}_{\mathrm{T}}$, which is the scale of momentum transfer of the QCD interactions.

Simulated minimum bias events are combined with simulated signal and background events to
 describe the effect of additional proton-proton interactions within the same or neighboring bunch crossings, which are referred to as pileup interactions.

\begin{table}[ht]
\centering
\topcaption{Model parameters of the different classes of signal models interpreted by the analysis.}
\label{model_parameters}
\cmsTable{
\begin{tabular}{l c c} 
{Signal model} & {Model parameter} & {Range} \\  
 \hline
 Vector portal & $m_{\Pomega}$: mass of the dark vector meson & 0.3--20\GeV \\
 & $c\tau$: mean proper decay length of the dark vector meson & 0.1--500\mm \\
 [\cmsTabSkip]
 Scenario A/B1 & $m_{\Ppi_{3}}$: mass of the dark pion $\Ppi_{3}$ & 1--12\GeV \\
 & $m_{\dphoton}$: mass of the dark photon & 0.33--2.5\GeV \\
 & $c\tau$: mean proper decay length of the dark photon (Scenario A) & 0.1--100\mm\\
 & or the dark pion $\Ppi_{3}$ (Scenario B1) & \\
\end{tabular}}
\end{table}

\section{Search strategy}
The search targets the displaced decays of dark mesons and dark photons into pairs of muons. The dark mesons are produced through hadronisation within dark parton showers that are initiated by the decays of SM Higgs bosons. Final states with multiple non-prompt muon pairs are expected to result from the signal models. 

\begin{figure}[ht] 
\centering
\includegraphics[width=0.66\textwidth]{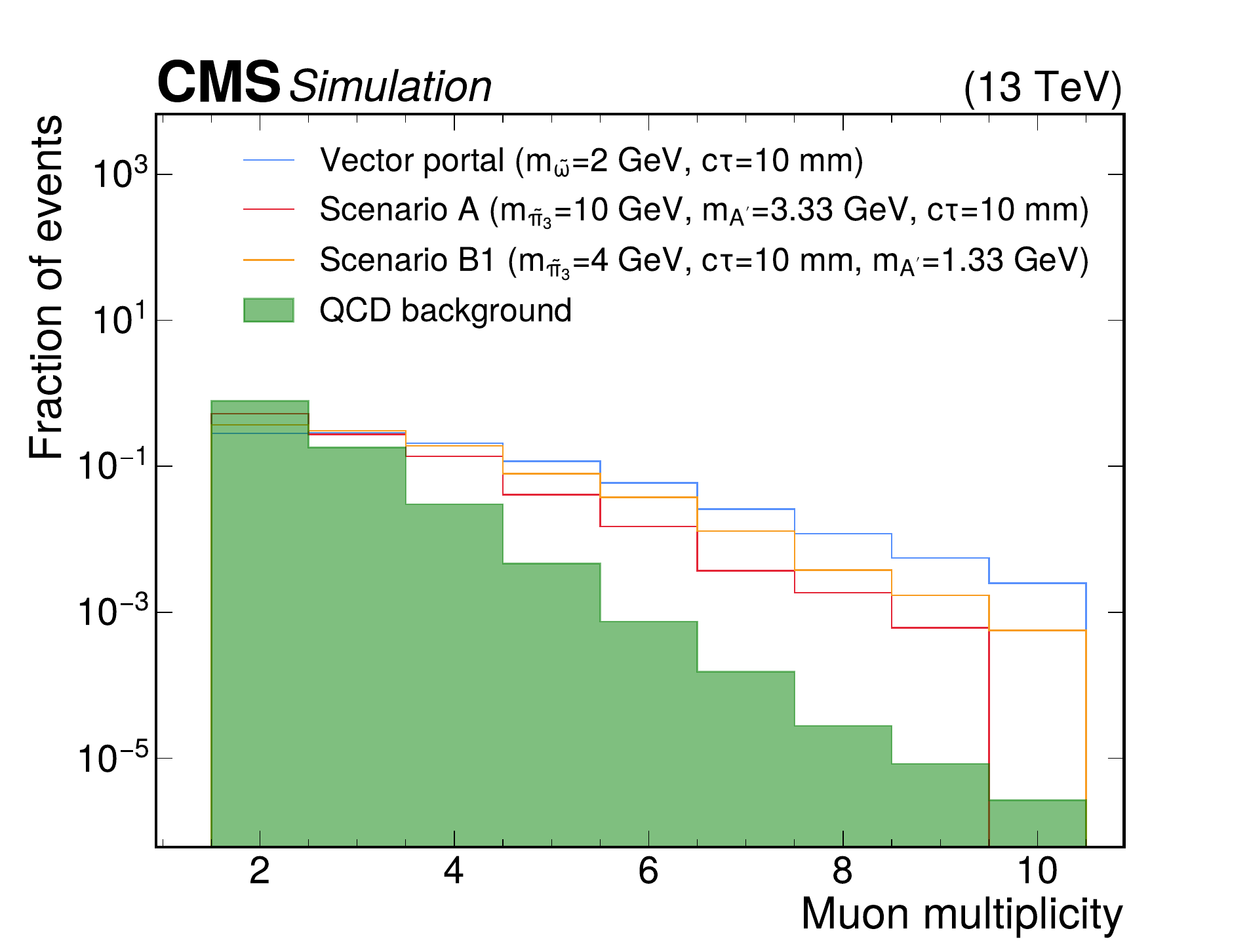}
\includegraphics[width=0.74\textwidth]{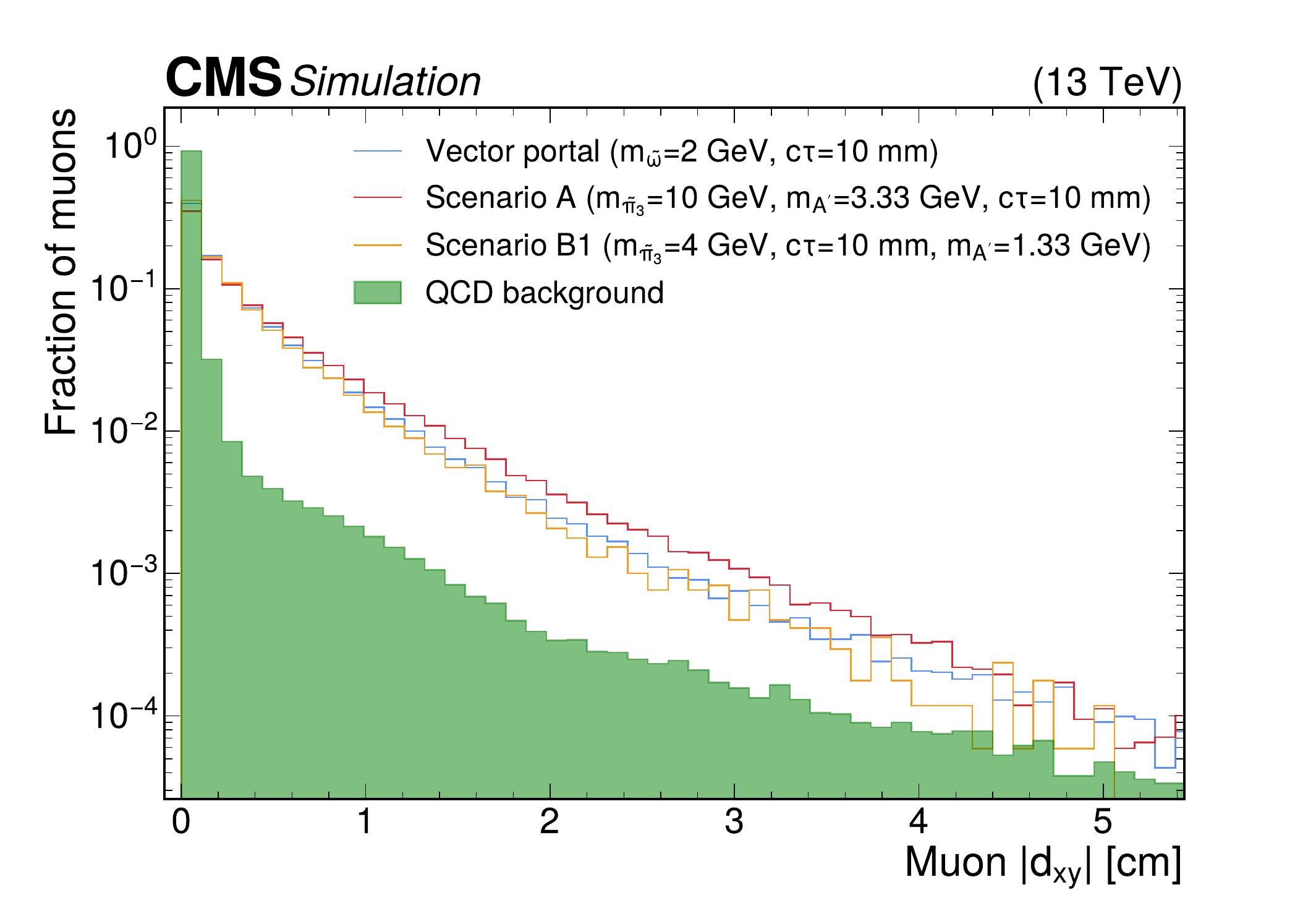}
\caption{Distributions of examples of variables that are used in the BDT training for the QCD background and benchmark signal models: muon multiplicity (upper) and muon transverse impact parameter (lower).}
\label{BDT_features}
\end{figure}

An event-level boosted decision tree (BDT) is used to separate the signal from the heavy-flavour QCD background. Minimal selections are first applied to the events. For simplicity, in the pre-selections for the BDT training, a single non-prompt muon trigger with a muon \pt threshold of 9\GeV and an impact parameter significance threshold of 6 is applied. In addition, at least one dimuon secondary vertex (with two oppositely charged muons originating from a common vertex) that contains a muon with $\pt>5\GeV$ is required. Features that discriminate between signal and background are then studied. Examples of these features are the muon multiplicity shown in Fig.~\ref{BDT_features} (upper), and the muon transverse impact parameter $d_{xy}$ shown in Fig.~\ref{BDT_features} (lower). It can be seen that there is a much larger number of muons in the signal events as compared with the background events. It is also found that muon variables provide excellent discrimination between signal and background, for example kinematic variables and variables related to the impact parameter. The multivariate input includes variables of muons, secondary vertices (produced from tracks of all types of charged particles), and muon secondary vertices (produced from muon tracks). Vertex variables include the $\chi^{2}$, pointing angle and displacement. Altogether there are a maximum of 390 input variables to the BDT. Where the value for a given feature does not exist for a particular event, it is set to a default value. As a result, the average number of input variables to the BDT with physical values is about 100. Jet variables are not included to make the BDT more model-independent. The variables providing the greatest discrimination are those associated with muons and muon secondary vertices. Good agreement between the data and the background Monte Carlo is observed for the BDT input variables. 

\begin{figure}[ht] 
\centering
\includegraphics[width=0.7\textwidth]{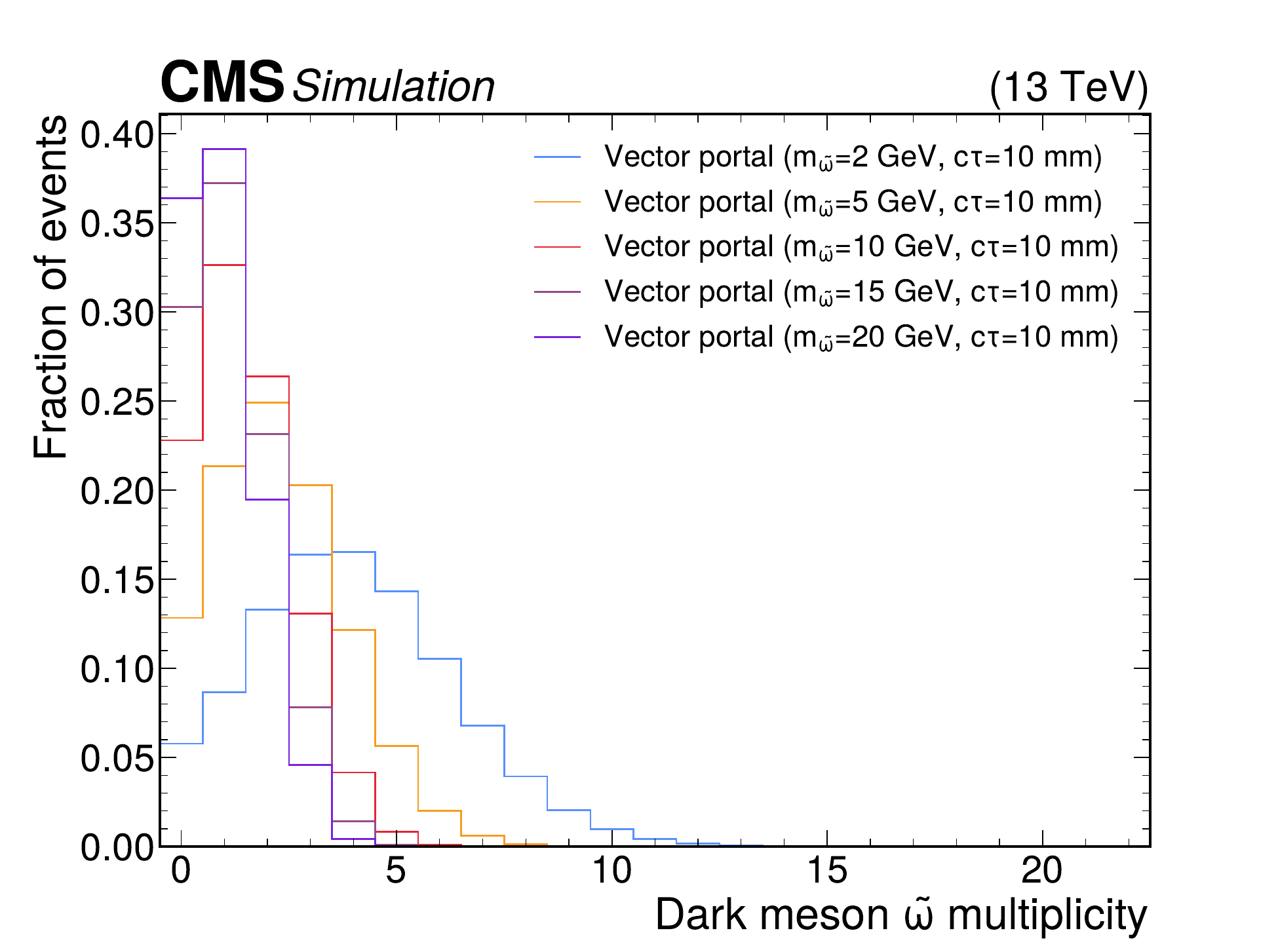}
\includegraphics[width=0.7\textwidth]{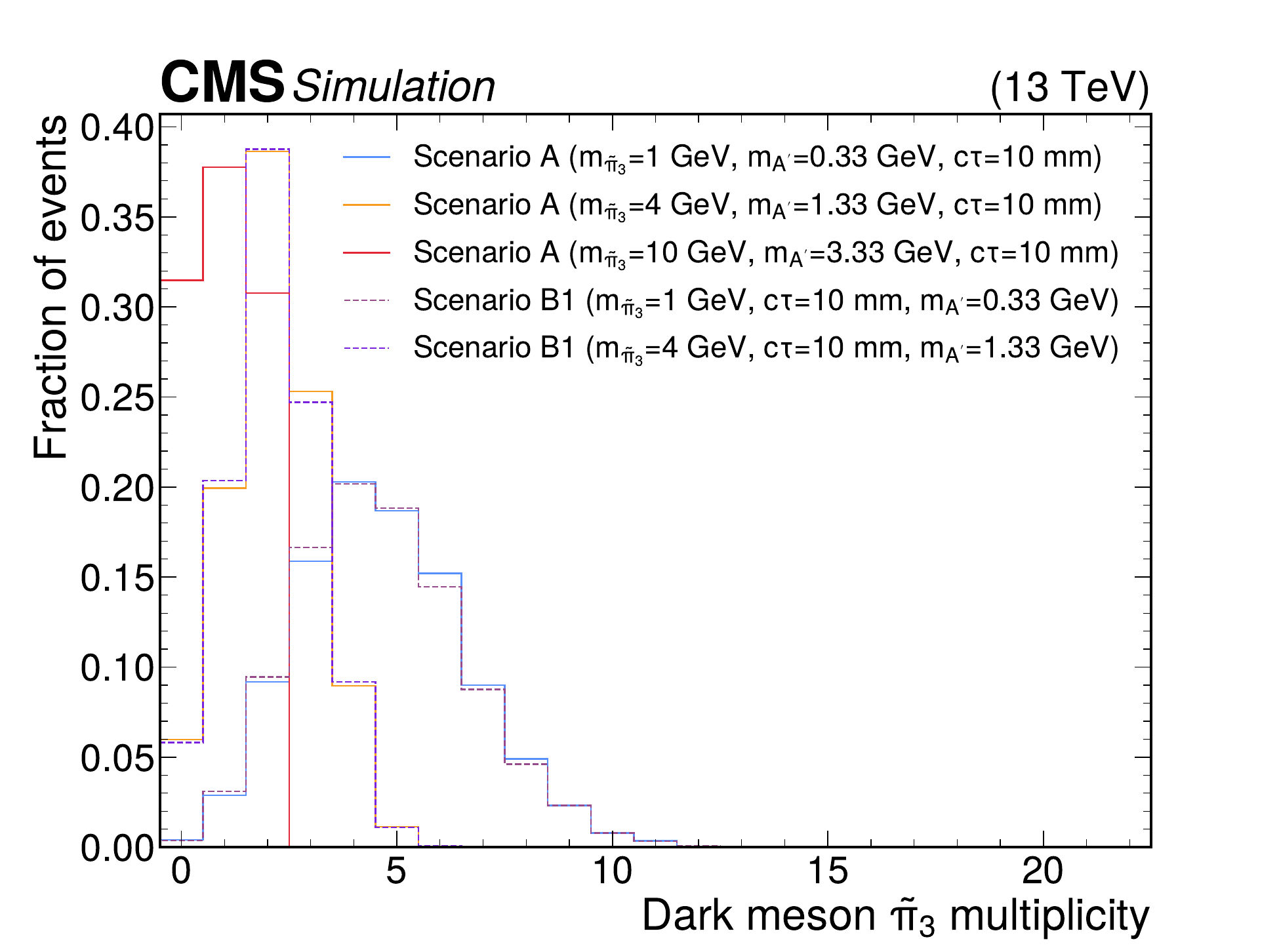}
\caption{Distributions of the multiplicity of dark vector mesons \Pomega for representative vector portal models (upper), and the multiplicity of dark mesons $\Ppi_{3}$ for representative Scenario A and B1 models (lower).}
\label{dark_meson_multiplicities}
\end{figure}

The high muon multiplicity in the signal comes from the high multiplicity of dark mesons in the events, with up to an average of five dark mesons per event in Scenario A for example. Figure~\ref{dark_meson_multiplicities} shows the multiplicity of dark mesons for representative model parameters for the vector portal model (upper), as well as for Scenario A and B1 (lower). The multiplicity increases with decreasing mass of the dark meson, as expected, because less energy is required to produce a lighter meson. The multiplicity is found to be very similar between Scenario A and Scenario B1 for the same $\Ppi_{3}$ meson mass, as the two models have similar decay processes. Moreover, there is a high branching fraction for the dark mesons to decay into muons \cite{Knapen:2021eip}. The branching fraction to muons drops with increasing mass of the dark meson because of competing decays to $\PQc\PAQc$, $\PQb\PAQb$, and $\PGtp\PGtm$. For example, in the vector portal model, the branching fraction to muons is 25\% for $m_{\Pomega}=2\GeV$, while it is 15\% for $m_{\Pomega}=20\GeV$. The high muon multiplicity in the dark-shower signal models allows a dedicated BDT to be trained for each of the three classes of models using a variety of input features to suppress the background.

The \textsc{XGBoost} package \cite{Chen_2016} is used for the BDT training. Most of the hyperparameters are fixed as their default values \cite{XGBoostmanual}, while some are fine-tuned to optimise the performance for each of the three signal models. For example, the number of boost rounds is changed to adapt to the size of the training samples. Other hyperparameters that are tuned are the learning rate, gamma, the maximum tree depth, the maximum delta step, subsampling ratios, and regularisation terms. Signal and QCD background samples are used for the training. The signal and background samples are weighted to give equal contributions to the training by using a binary cross-entropy loss function which is weighted event-by-event. Each of the $N$ signal model samples (where, for example, $N = 25$ for the vector portal) is reweighted to give a contribution of $1/N$ to the total signal class after the preselections have been applied. The BDT is applied after the basic analysis selection criteria, which will be described in Section \ref{event-selection}.

\section{Event selection and categorisation}
The event sample is selected by requiring the presence of two oppositely charged muons that form a secondary vertex (SV) and by applying a minimum BDT score threshold. Events are categorised according to the number of vertices and in bins of transverse displacement ($l_{xy}$) and pointing angle to increase sensitivity to a signal.

\subsection{Event selection}
\label{event-selection}
Selected events are required to pass at least one of the \PQB-parking triggers and contain at least one reconstructed muon candidate satisfying $\pt > 9\GeV$, $\mathrm{IP}_{\text{sig}}> 6$, and $\abs{\eta}<1.5$ that is matched to the muon that passes the trigger requirement by requesting $\Delta R<0.1$, where $\Delta R=\sqrt{\smash[b]{(\Delta\eta)^2+(\Delta\phi)^2}}$. Furthermore, the muon must pass the loose muon identification (ID) requirements~\cite{CMS:2018rym}.

At least one SV is required to be formed by a pair of oppositely charged muons. The vertex is reconstructed with a standard Kalman filter \cite{Fruhwirth:1987fm} algorithm, with the range in the distance from the beam axis extended \cite{Speer:927395}. This modification improves the reconstruction efficiency for the present search, which probes a transverse displacement of up to about 15\cm, by more than 40\% relative to using the vertexing algorithm with standard parameters for different displacements. Quality criteria are then applied to the muon SVs, which are $\Delta R(\mu, \mu)<1.2$, $\chi^{2} < 10$, one muon matching the one that triggered the event, and both muons passing the loose ID \cite{CMS:2018rym}. If more than one muon SV passes the selections, only the muon SV with the lowest $\chi^{2}$ is used for the final fit. 

A selection is applied on the BDT scores of the events, with a threshold of 0.997 (0.999 for signal mass hypotheses less than 5\GeV) for the vector portal model. The BDT thresholds for Scenario A and Scenario B1 models are 0.98 and 0.991 respectively. The BDT cut alone gives a signal efficiency of ${\approx}$30--60\% and a background efficiency of ${\approx}10^{-4}$ (or ${\approx}10^{-5}$ for signal mass hypotheses less than 5\GeV in the vector portal model). The BDT thresholds are optimised using control regions in data (data sidebands) for representative signal mass models. The data sidebands for each representative signal mass hypothesis are regions in the dimuon mass distribution that are outside the signal region, which is located at the centre of the mass window. The BDT selections are varied to find the optimal expected upper limits on the branching fraction of the Higgs decaying to dark partons. 

The $\chi^{2}$ of the muon SVs is found to be one of the most important variables of the BDT. The signal muon SVs are better fitted and have a lower $\chi^{2}$ compared to the background, which is dominated by muon vertices from random muon combinations. The muon SV's transverse displacement and its significance are also found to be important variables. This is expected as the signal models contain non-prompt decays, while the QCD background is dominated by prompt processes. The pointing angle is also a good discriminating variable.

\subsection{Event categorisation}
Events are categorised based on whether there are at least two vertices with similar dimuon mass. An event is classified as a multi-vertex event if there is at least one pair of muon SVs with dimuon masses within 3\% of each other, which is about three times the width of the signal. Otherwise the event is classified as a single-vertex event.

Both the single-vertex and multi-vertex events are further categorised by the $l_{xy}$ and pointing angle of the muon SV with the lowest $\chi^{2}$. The transverse displacement is defined as the distance in the transverse plane between the primary vertex and the secondary vertex. At increasing displacement from the primary vertex, the QCD background processes are expected to diminish. Events are categorised in the transverse displacement in order to increase the sensitivity to different proper lifetimes of the signal models. Categorising in transverse displacement is found to be more beneficial to the sensitivity than categorising in transverse displacement significance. The $l_{xy}$ categories are defined to be $[0.0, 1.0]\cm$, $(1.0, 10.0]\cm$ and beyond 10\cm. The signal efficiency drops considerably for $l_{xy}\gtrsim20\cm$.

\begin{figure}[ht!] 
\includegraphics[width=0.5\textwidth]{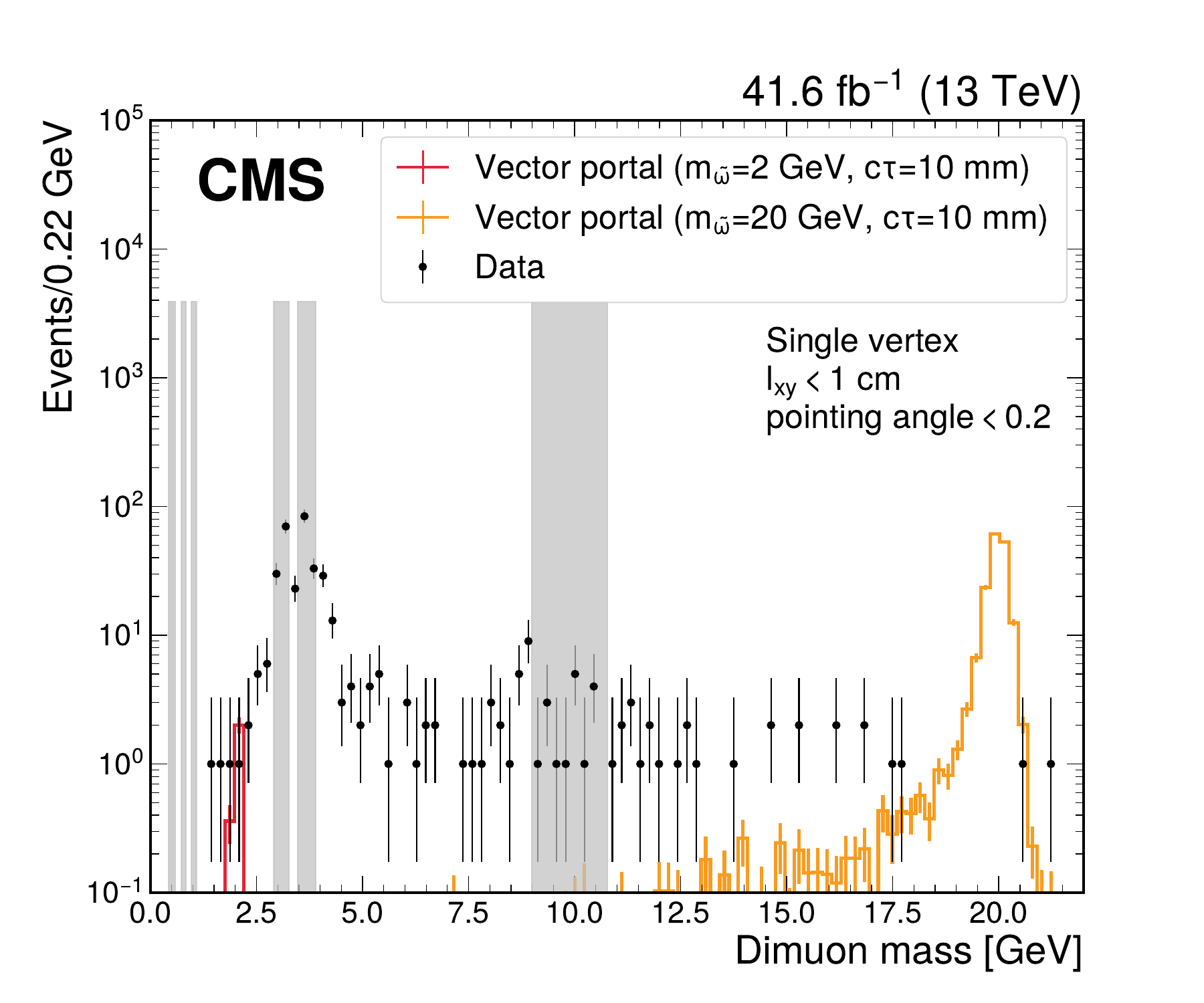}
\includegraphics[width=0.5\textwidth]{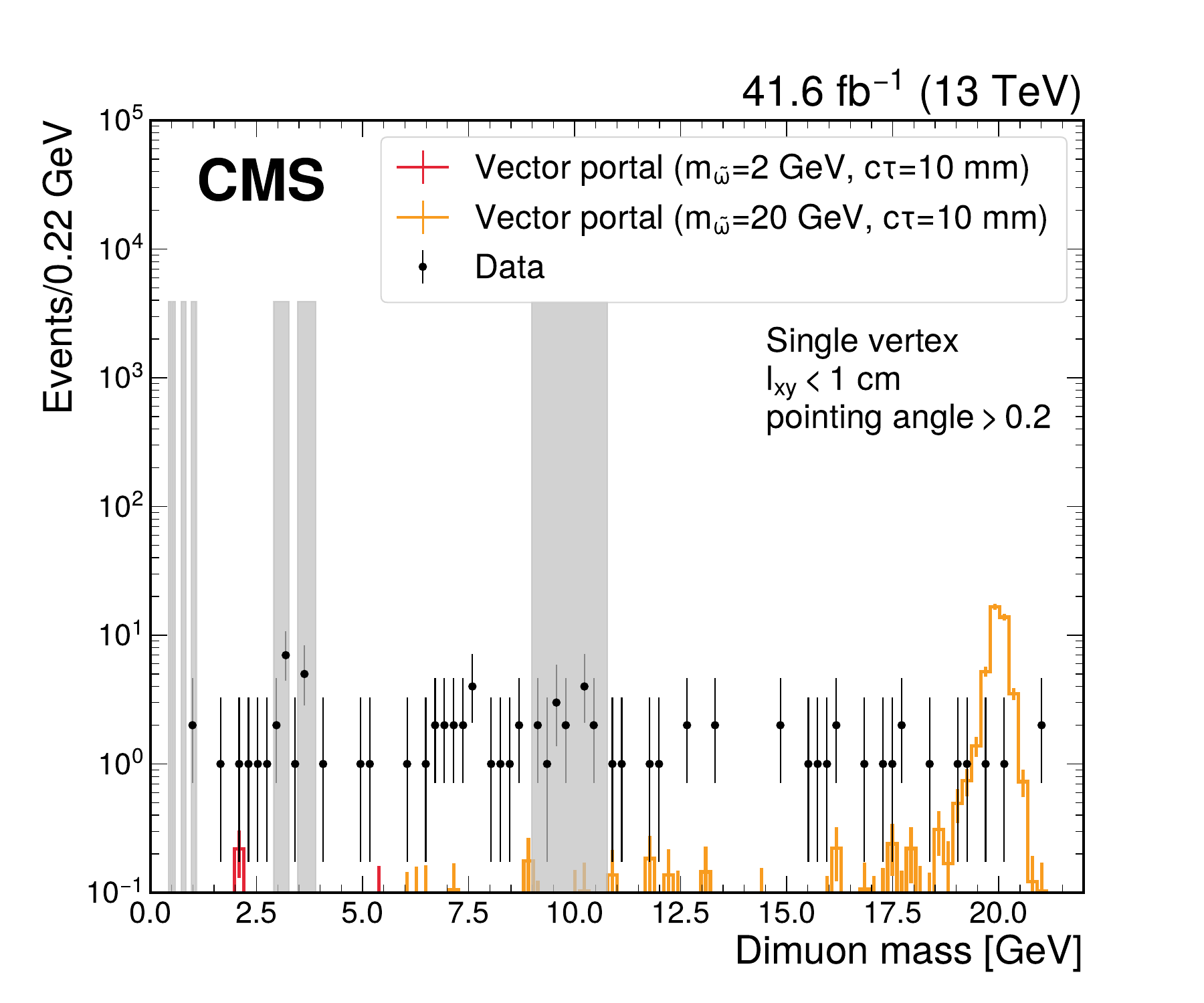}
\includegraphics[width=0.5\textwidth]{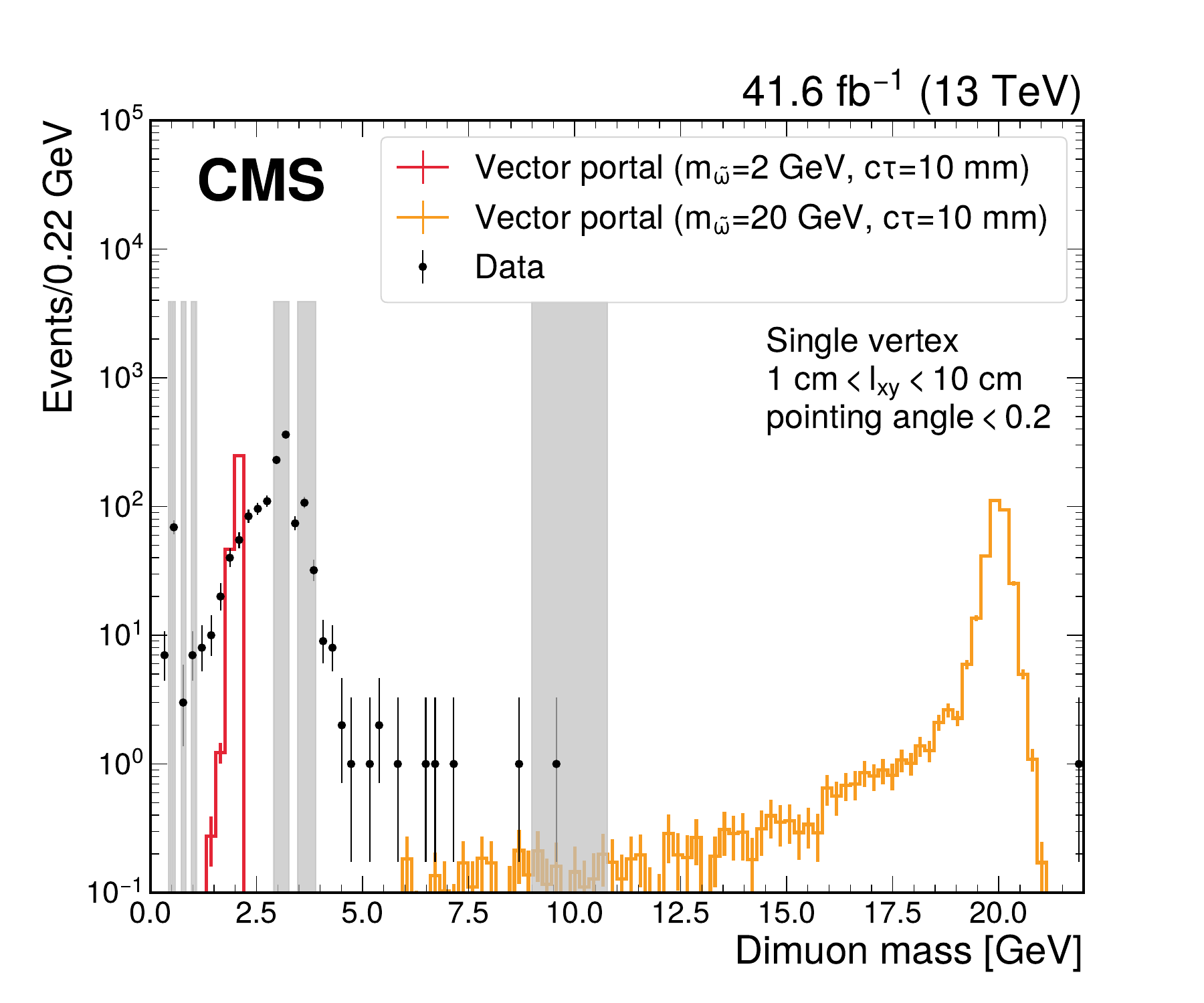}
\includegraphics[width=0.5\textwidth]{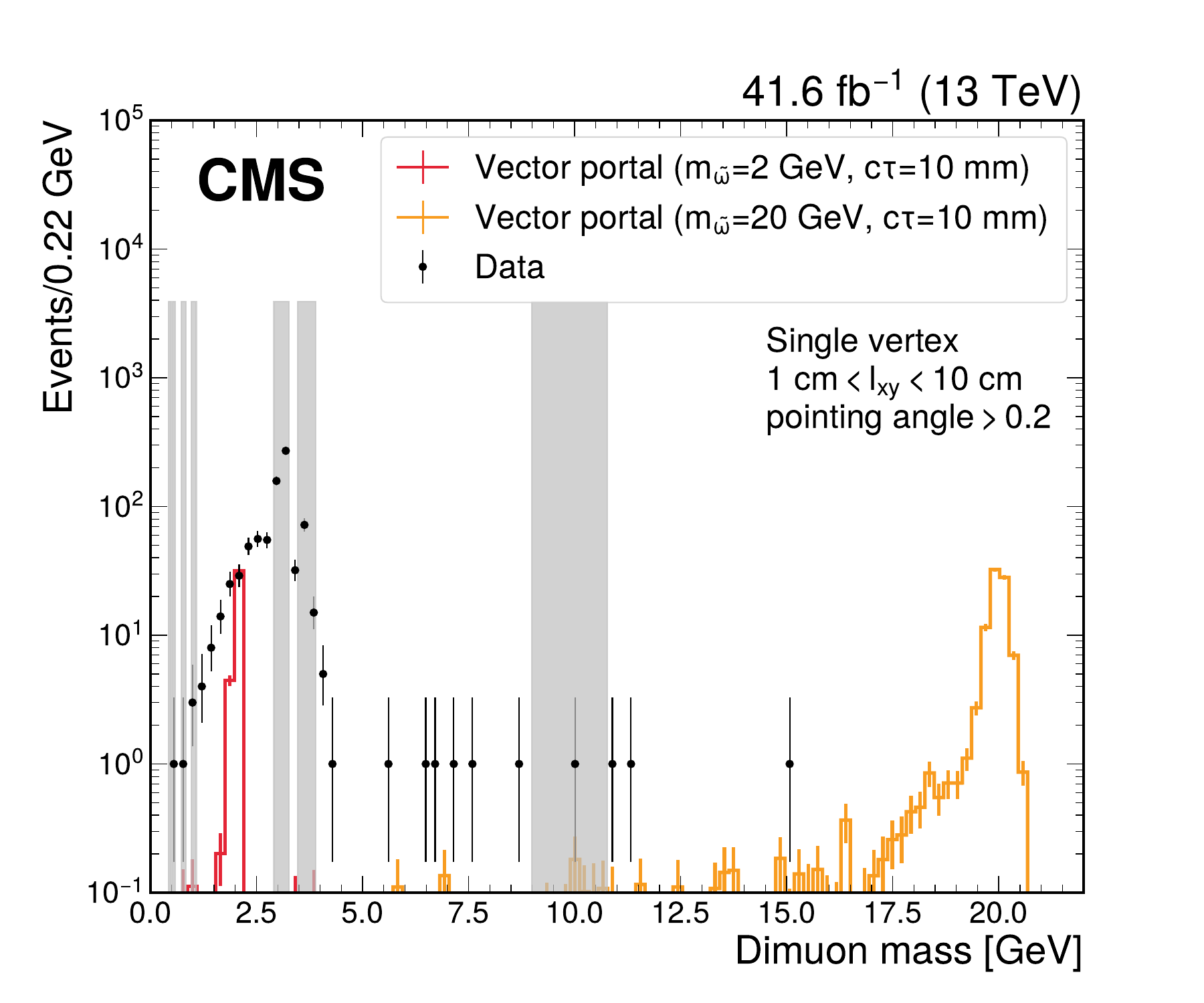}
\includegraphics[width=0.5\textwidth]{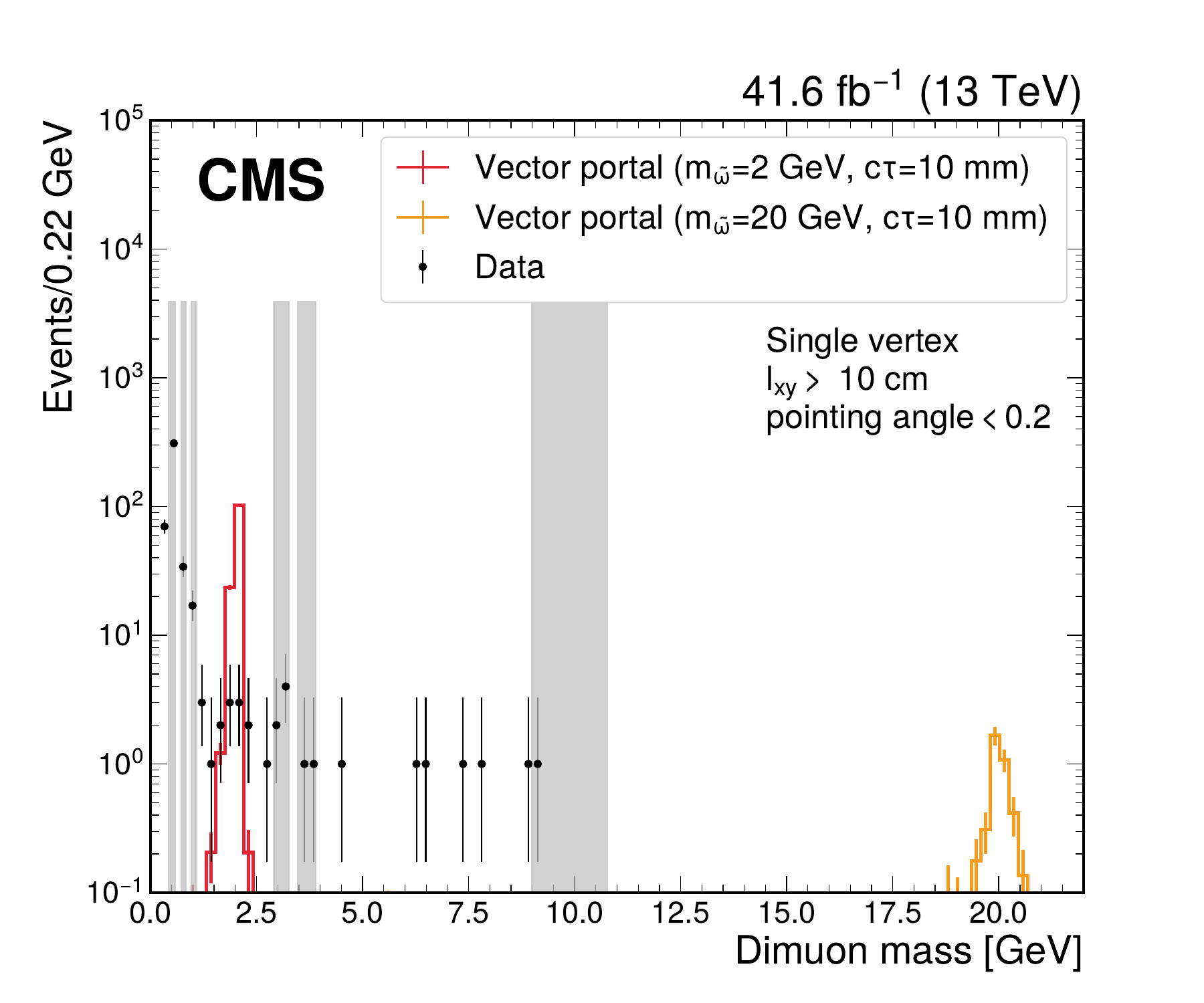}
\includegraphics[width=0.5\textwidth]{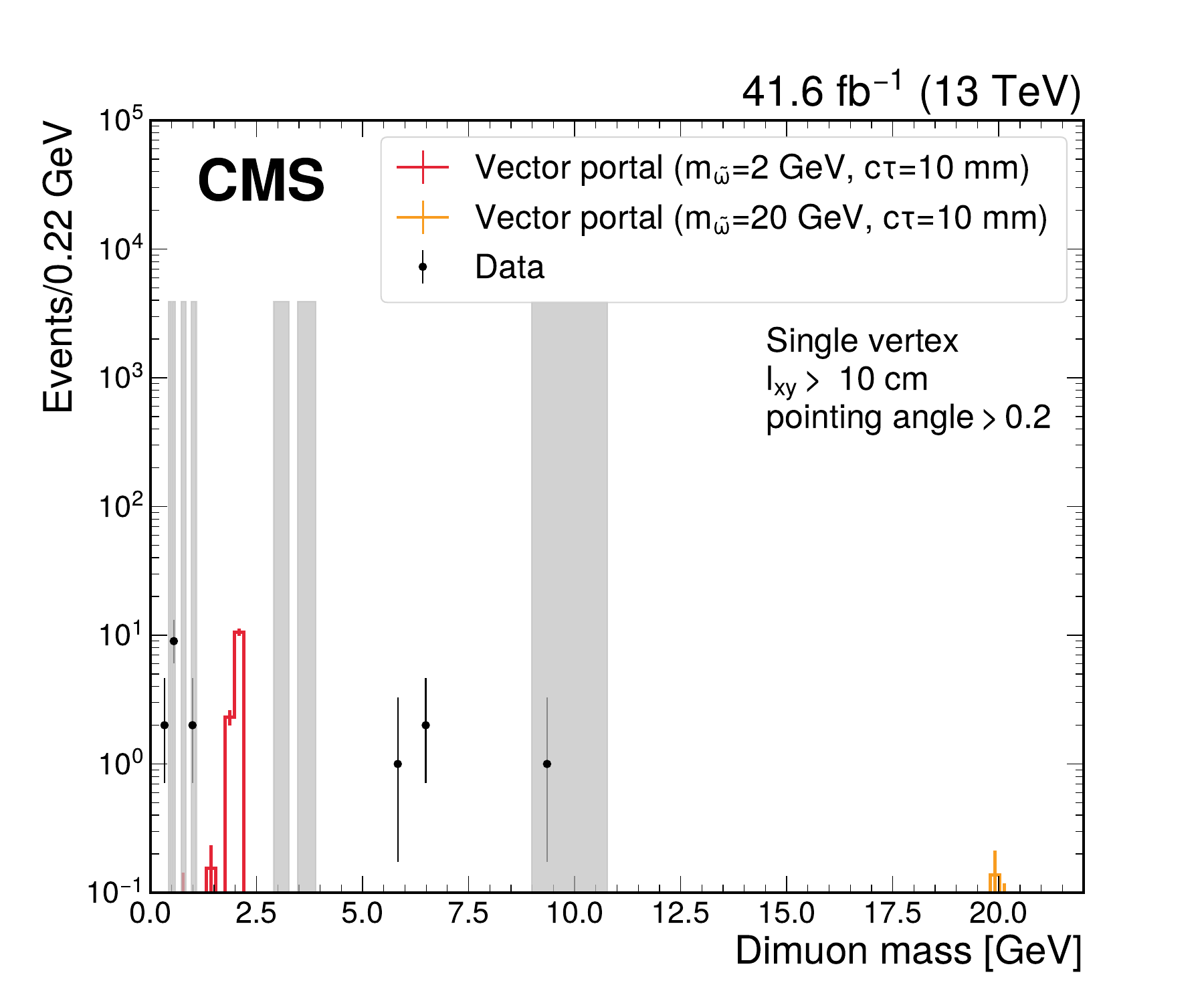}
\caption{Dimuon invariant mass distributions for each single-vertex category for data and two vector portal model signal benchmarks. The shaded regions indicate mass regions of known SM resonances, which are masked in the search. A branching fraction of 0.01 is assumed for the Higgs boson decaying into dark partons for illustrating the signals.}
\label{fig:single_cats}
\end{figure}

\begin{figure}[ht!] 
\includegraphics[width=0.5\textwidth]{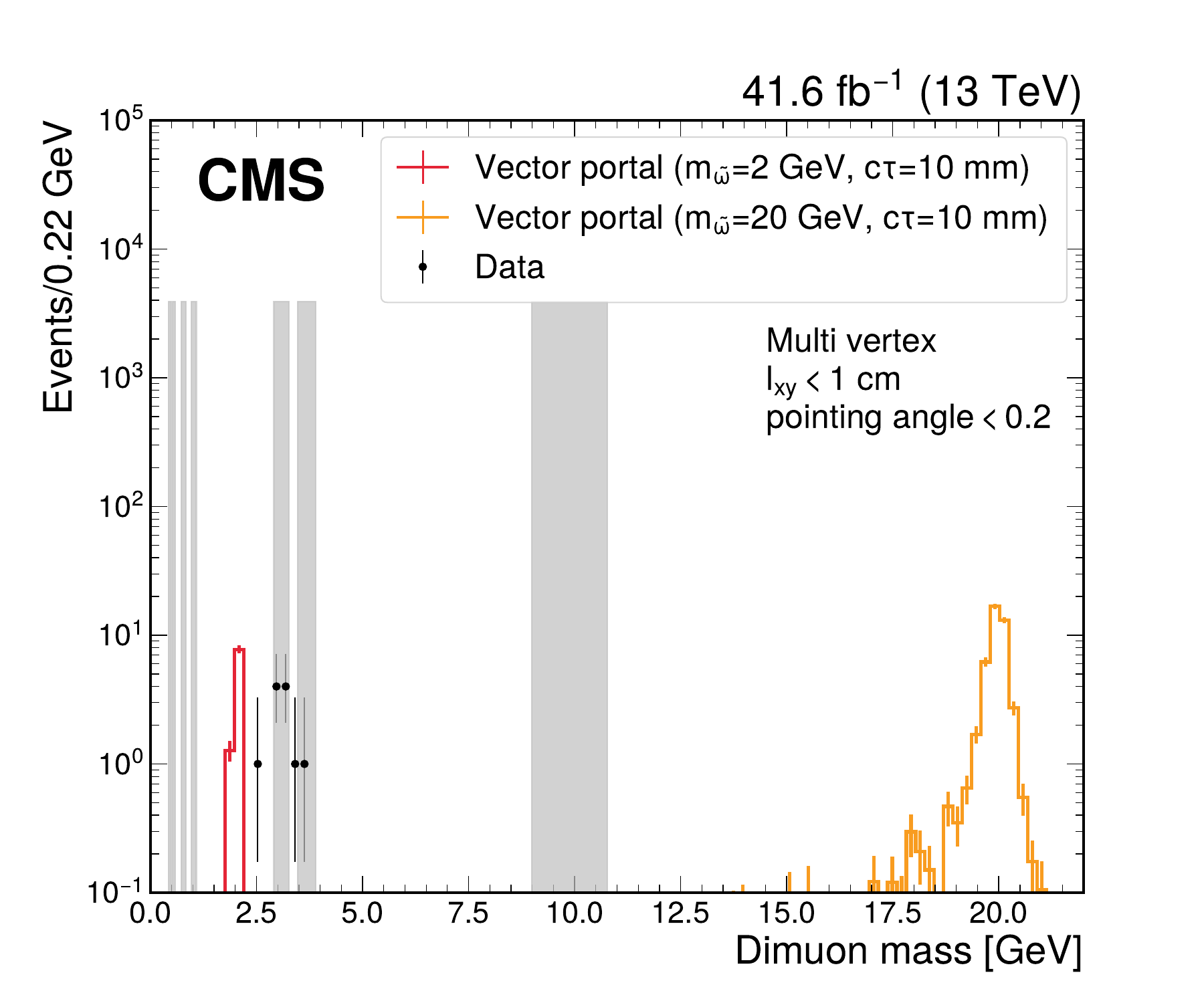}
\includegraphics[width=0.5\textwidth]{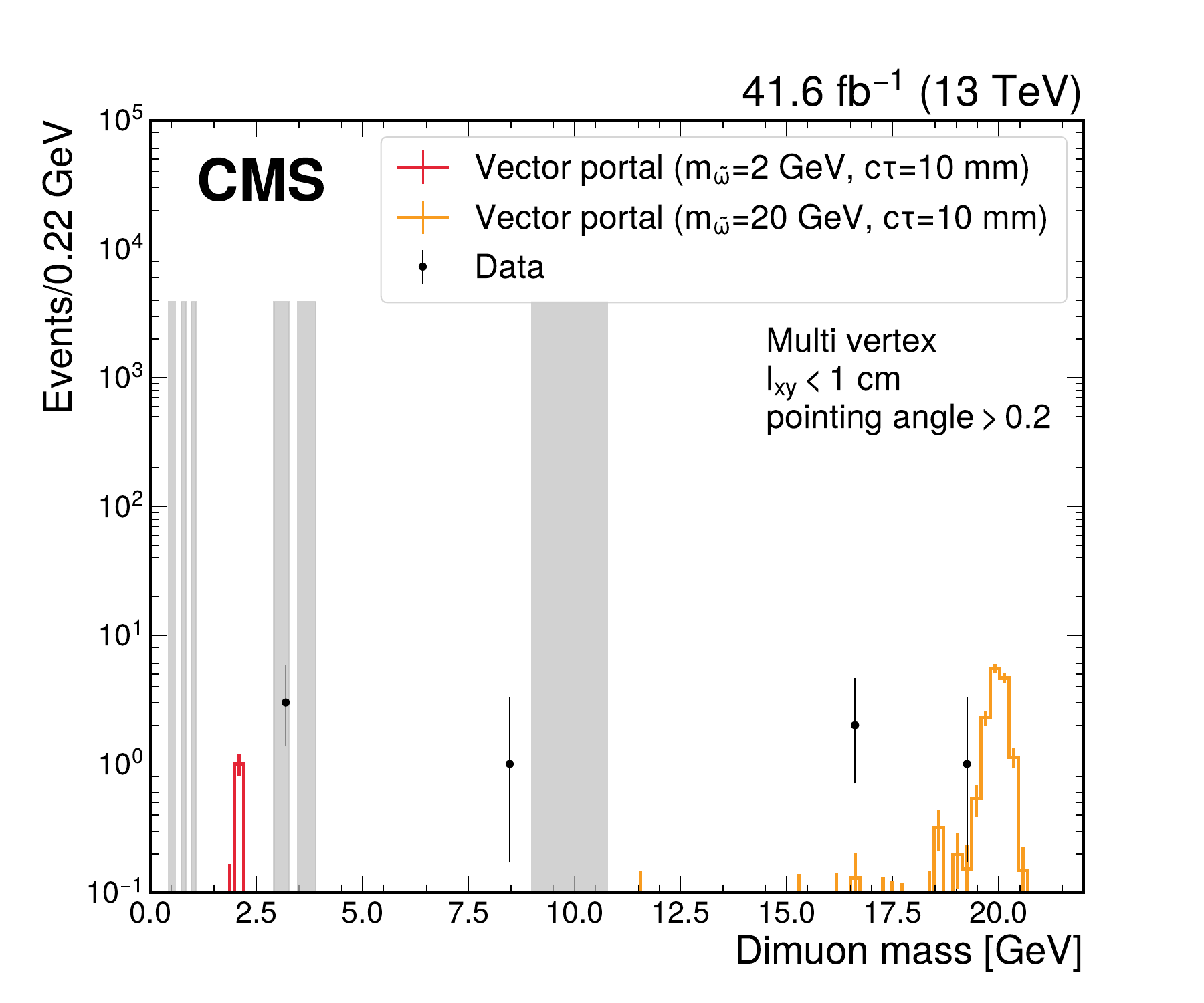}
\includegraphics[width=0.5\textwidth]{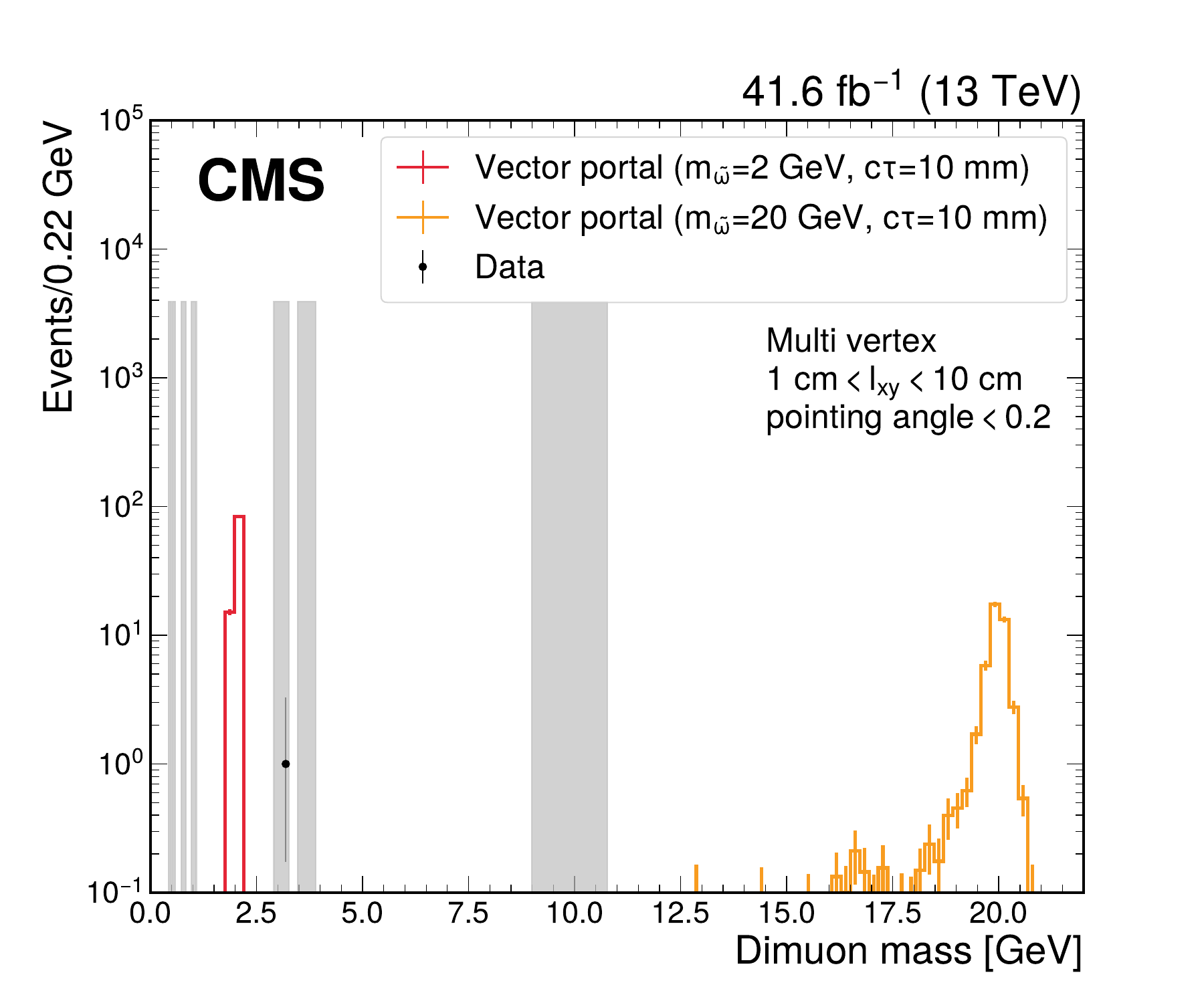}
\includegraphics[width=0.5\textwidth]{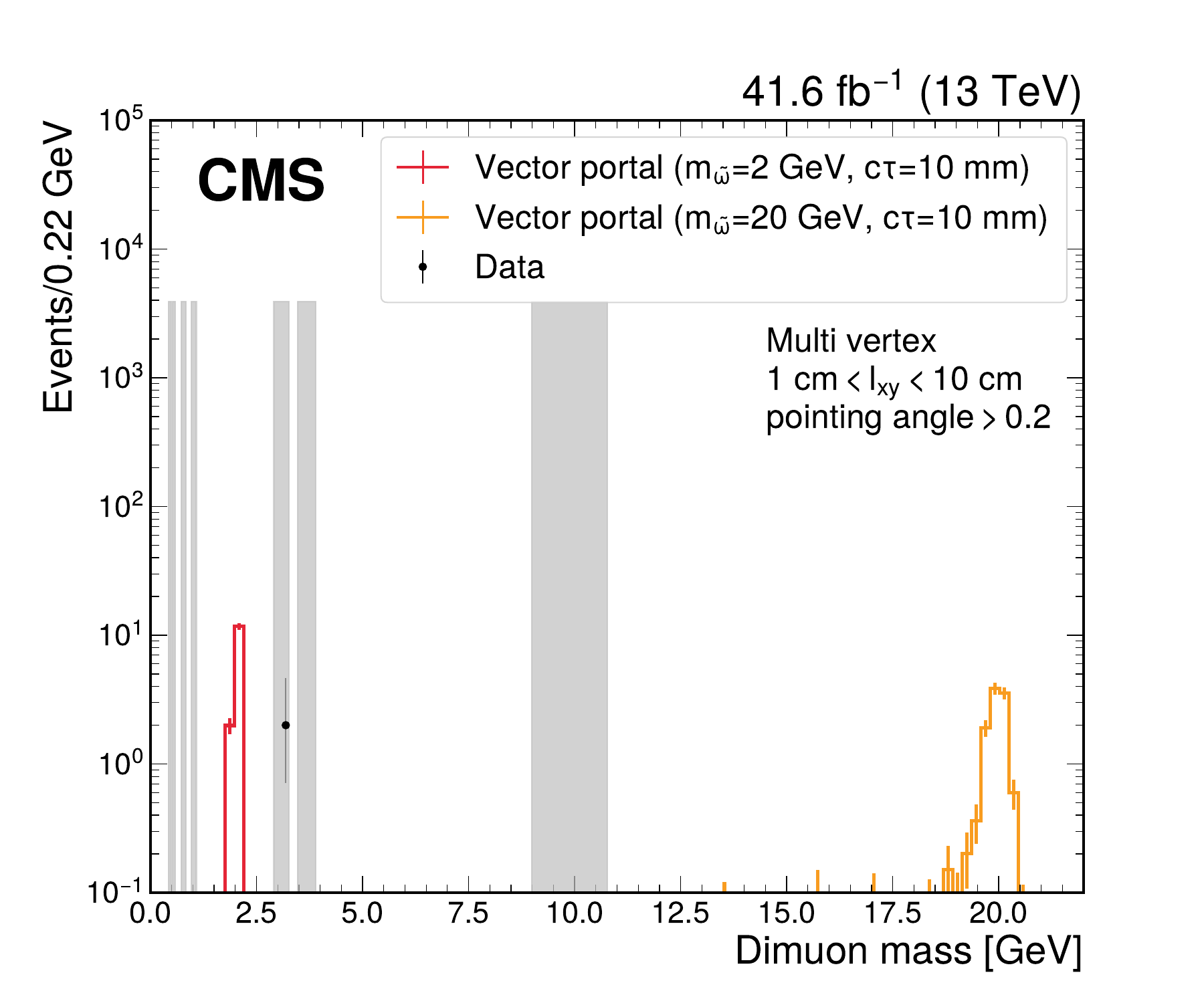}
\includegraphics[width=0.5\textwidth]{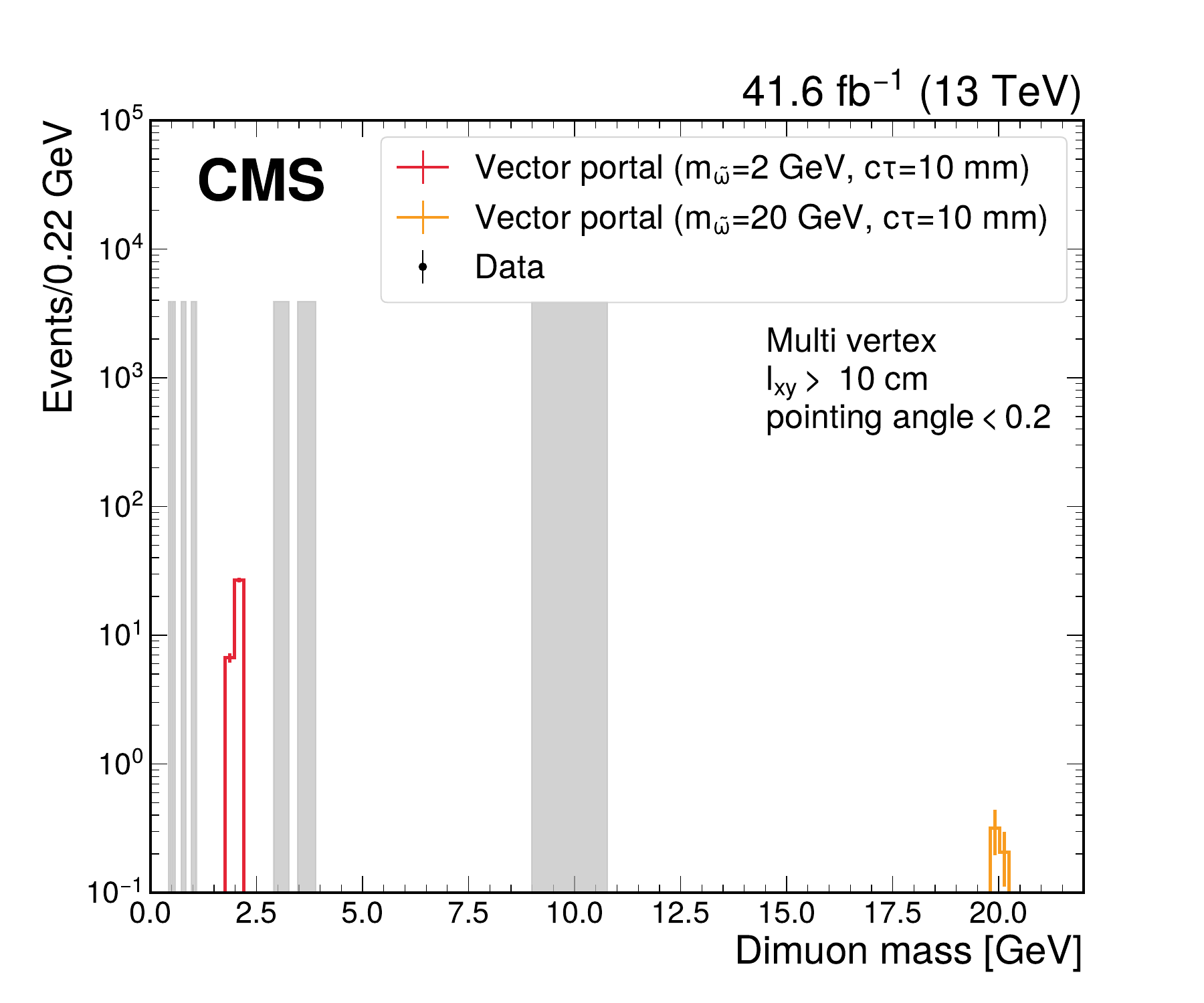}
\includegraphics[width=0.5\textwidth]{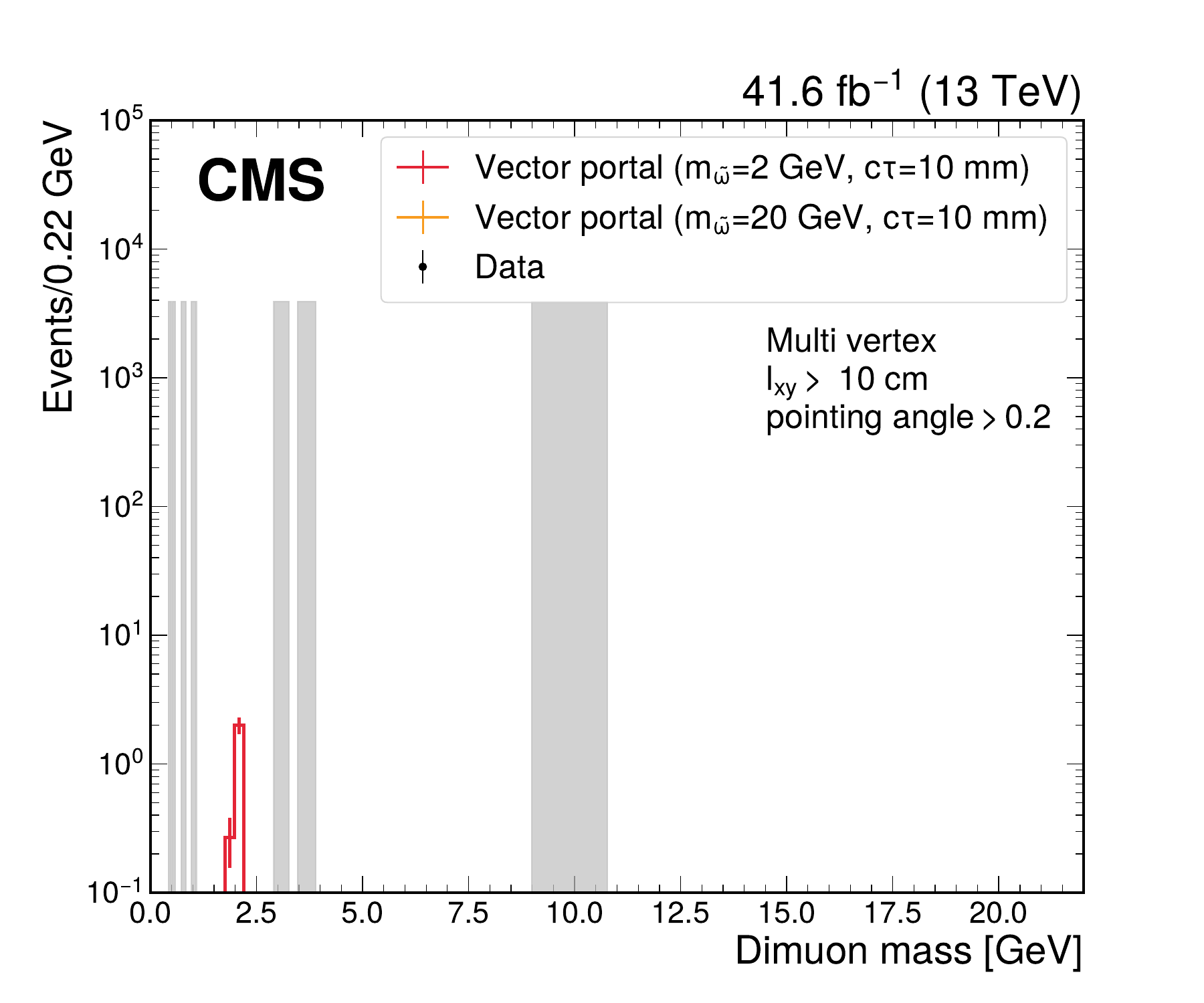}
\caption{Dimuon invariant mass distributions for each multi-vertex category for data and two vector portal model signal benchmarks. The shaded regions indicate mass regions of known SM resonances, which are masked in the search. A branching fraction of 0.01 is assumed for the Higgs boson decaying into dark partons for illustrating the signals.}
\label{fig:multi_cats}
\end{figure}

In each $l_{xy}$ bin, events are further divided into bins of the pointing angle. This is to increase sensitivity to signal models that are more pointing than the background. The pointing angle bins are chosen to be $[0.0, 0.2]$ and $(0.2, \pi]$ to maintain sensitivity to non-pointing signal scenarios. Twelve categories are defined in total. The dimuon mass distributions for all categories are shown in Figs.~\ref{fig:single_cats} and~\ref{fig:multi_cats} for data and two vector portal model signal benchmarks. The grey bands in the plots correspond to mass regions of known SM resonances, which are masked in the search, as listed in Table \ref{masked_regions}. This is to avoid misidentifying the SM resonances as the signal. It can be seen that some of the categories are background-free, which shows the capability of the BDT in rejecting the background.

\begin{table}[ht!]
\centering
\topcaption{The SM resonances and the corresponding mass windows that are masked in the analysis.}
 \label{masked_regions}
\begin{tabular}{l c c} 
 
 Resonance & Lower bound [\GeVns{}] & Upper bound [\GeVns{}]  \\
 \hline
 \PKS & 0.43 & 0.49  \\
 \PGh & 0.52 & 0.58 \\
  $\PGp / \PGo$ & 0.73 & 0.84\\
 \PGf & 0.96 & 1.08\\ 
  \PJGy &  2.91 & 3.27 \\
  \PGyP{2S} &  3.47 & 3.89 \\
  \PGUP{1S} & 8.99 & 9.87 \\
  \PGUP{1S} & 9.61 & 10.39 \\
  \PGUP{3S} & 9.87 & 10.77 \\
  \end{tabular}
\end{table}

\section{Parametrisation of the dimuon mass distribution}
Using the dimuon mass distribution in each category, a parametric fit is performed for different signal mass and lifetime hypotheses for each of the three models. A Voigtian function (which is the convolution of a Gaussian function and a Breit-Wigner function) is used to fit the signal dimuon mass distribution. From the fits, it is found that the Gaussian width of the signal, $\sigma$, is about 0.75\% of the mass for the vector portal, and about 0.85\% of the mass for Scenario A and Scenario B1, each with an uncertainty of about $15\%$. 

To perform the search, a sliding fit window of ${\pm}5\times$HWHM of the signal around the signal mass hypothesis is chosen, where HWHM is the half width at half maximum of the signal resonance and is estimated to be 1\% of the mass. To model the background, data sidebands (which are defined as mass regions in the fit window, outside of the region within ${\pm}2\times$HWHM of the signal) are fitted with an envelope of functions, which are the first order polynomial, exponential, and power law functions. At each point in the likelihood scan, the particular fitting function that maximises the likelihood is chosen. Discrete profiling is used to account for the background systematic uncertainty from the choice of the function \cite{Dauncey:2014xga}. The parameters of each function are fixed in the likelihood fit, while the normalisation is allowed to vary. Goodness of fit tests have been performed, which show the functions can model the background appropriately. Bias tests have also been performed, and the fit is found to be sufficiently unbiased in the measurement of a potential signal. 

For mass windows with very low background, a flat function is used instead of the envelope of functions to ensure stability of fit, following the fitting strategy used in previous dimuon searches \cite{CMS:2021sch, CMS:2023byi}. It was found that fitting with this approach gives upper limits that agree with those obtained from a counting experiment.

\begin{figure}[ht!] 
\centering
\includegraphics[width=0.77\textwidth]{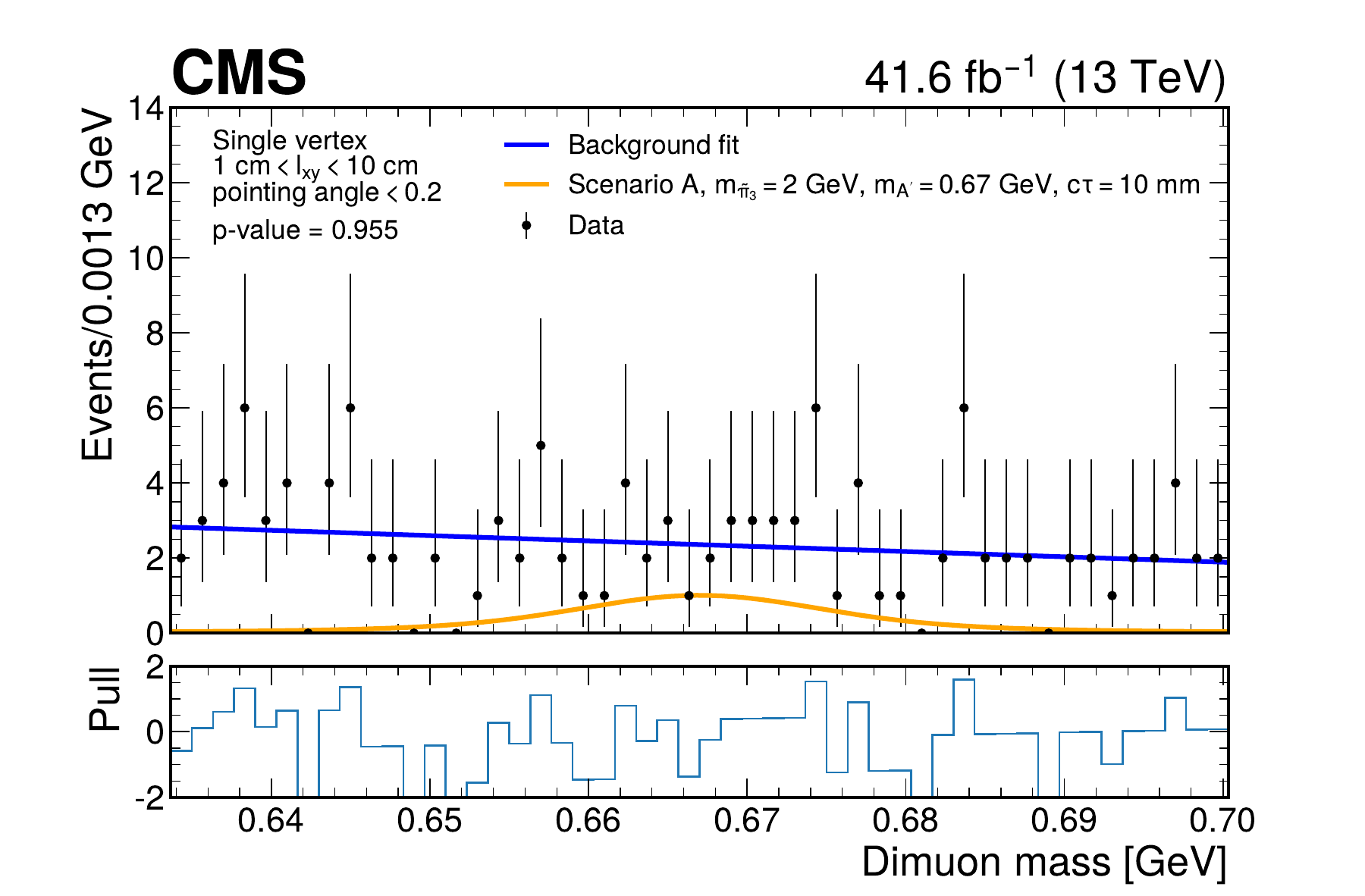}
\includegraphics[width=0.77\textwidth]{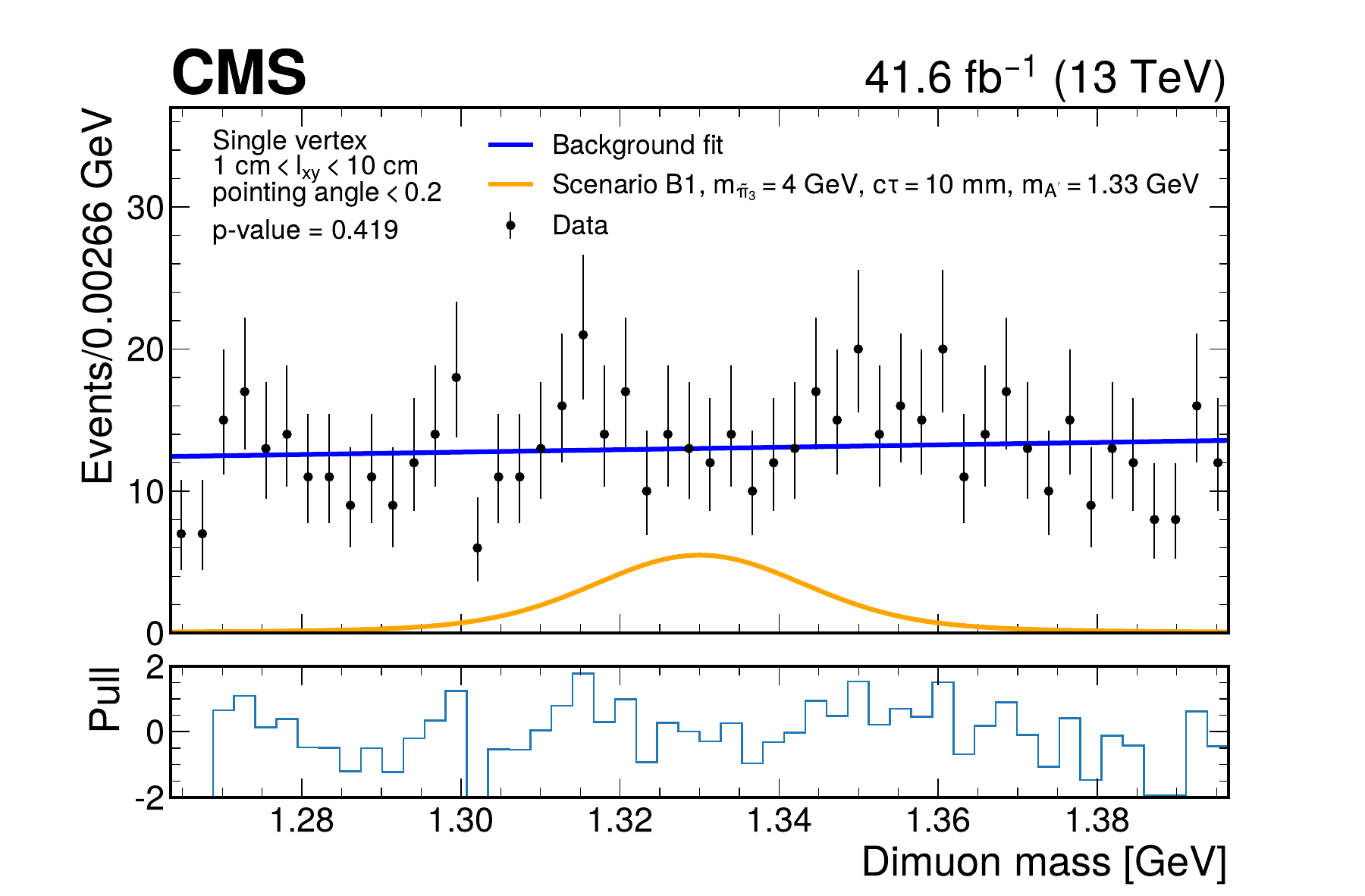}
\caption{Dimuon invariant mass distributions in mass windows around 0.67\GeV (upper) and 1.33\GeV (lower), in the single-vertex category with $1\cm<l_{xy}<10\cm$ and $\text{pointing angle}<0.2$. The background fit is shown together with the signal expected for a representative Scenario A model (upper) and a representative Scenario B1 model (lower). A branching fraction of 0.01 is assumed for the Higgs boson decaying into dark partons for illustrating the signals. The lower panel in each plot shows the pull distribution, defined as the difference between the data and the background fit in each bin divided by the statistical uncertainty.}
\label{sig-bkg-fits}
\end{figure}

The background-only fit is shown in Fig.~\ref{sig-bkg-fits} for two mass windows in an example category, together with the signal expected from representative signal models.

\section{Results and interpretation}
Maximum likelihood fits to the data are performed under the background-only or background-plus-signal hypotheses, using a sliding mass window for different signal mass hypotheses. The fits are carried out simultaneously for all the categories which contain non-zero expected signal.

\subsection{Systematic uncertainties}
Various sources of systematic uncertainties in the signal modelling are considered. The systematic uncertainty in non-prompt muon ID efficiency is measured using the ``tag and probe'' technique \cite{CMS-PAS-EWK-10-005} in the \PJGy mass region to study the differences between data and simulation. \PJGy from heavy flavour decay is used as a ``proxy'' for the signal as it is a known displaced resonance that lies in the mass range of the search. Scale factors are derived to account for the differences, and the systematic uncertainty in the ID efficiency is found by varying the scale factors with respect to the nominal values. The uncertainty is found to be between 1\% and 17\%, depending on the \pt of the muon and the transverse displacement of the muon SV. For the trigger selection, the same technique is used to compare the trigger efficiencies in data and simulation. The systematic uncertainty in the trigger efficiency typically lies between 1\% and 2\%, depending on the \pt and the impact parameter significance of the muon. The BDT selection efficiency is measured in the \PJGy mass region for data and simulation. The systematic uncertainty in the BDT selection efficiency is found to be 10\%. The systematic uncertainty from pileup is obtained by considering the pileup distribution during the time when the \PQB-parking triggers were active. Pileup weights are applied to Monte Carlo events such that the number of interactions per bunch crossing agrees with that observed in data. The pileup modelling uncertainty is found, by varying the weights, to be about 5\%. The systematic uncertainty in the luminosity is 2.5\% \cite{CMS:2019jhq}. Various sources of systematic uncertainties in the gluon-gluon fusion Higgs boson production cross section are considered. These include a theory uncertainty of 3.9\% due to missing higher order terms in the QCD calculation, and uncertainties in the factorisation and renormalisation scales. There is also an uncertainty of 1.9\% from the choice of PDFs, and an uncertainty of 2.6\% from the strong coupling constant value \cite{LHCHiggsCrossSectionWorkingGroup:2016ypw}. The various sources of systematic uncertainties for the signal are summarised in Table~\ref{Systematics_table}. 

The background systematic uncertainty is obtained from discrete profiling of the envelope of fitting functions used to model the background, which predominantly originates from QCD processes. The background uncertainty is found to be negligible compared to the other systematic uncertainties.

Uncertainties in the trigger efficiency, non-prompt muon ID efficiency and pileup modelling are treated as uncorrelated across categories. Other uncertainties are treated as correlated. The total systematic uncertainty is in the range 13--21\%, which is less than the statistical uncertainty in the background normalisation.

\begin{table}[ht!]
\centering
\topcaption{Summary of the systematic uncertainties in the signal yield expectation.}
 \label{Systematics_table}
\begin{tabular}{lc} 
 Source & Systematic uncertainty [\%] \\
 \hline
 Non-prompt muon identification & $\sim$1--17 \\
 Trigger & $\sim$1--2 \\
 BDT & 10 \\
 Pileup & ${\sim}5$ \\
 Theory uncertainty in ggF Higgs production & 3.9 \\
 Strong coupling constant & 2.6 \\
 PDF & 1.9 \\
 Luminosity & 2.5 \\
 \end{tabular}
 \end{table}

\subsection{Constraints on new physics models}
\begin{figure}[ht!]
\centering
\includegraphics[width=0.48\textwidth]{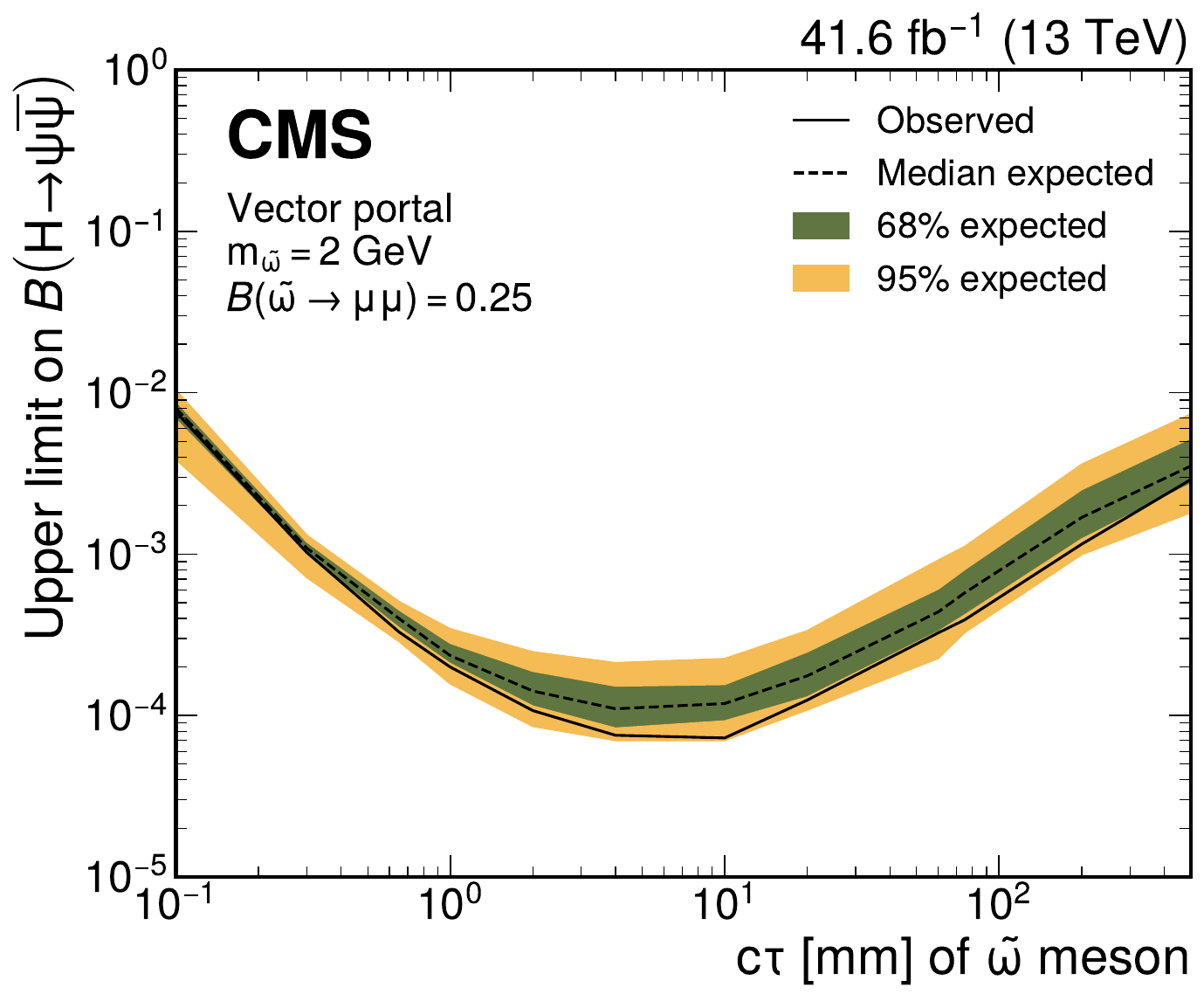}~~
\includegraphics[width=0.48\textwidth]{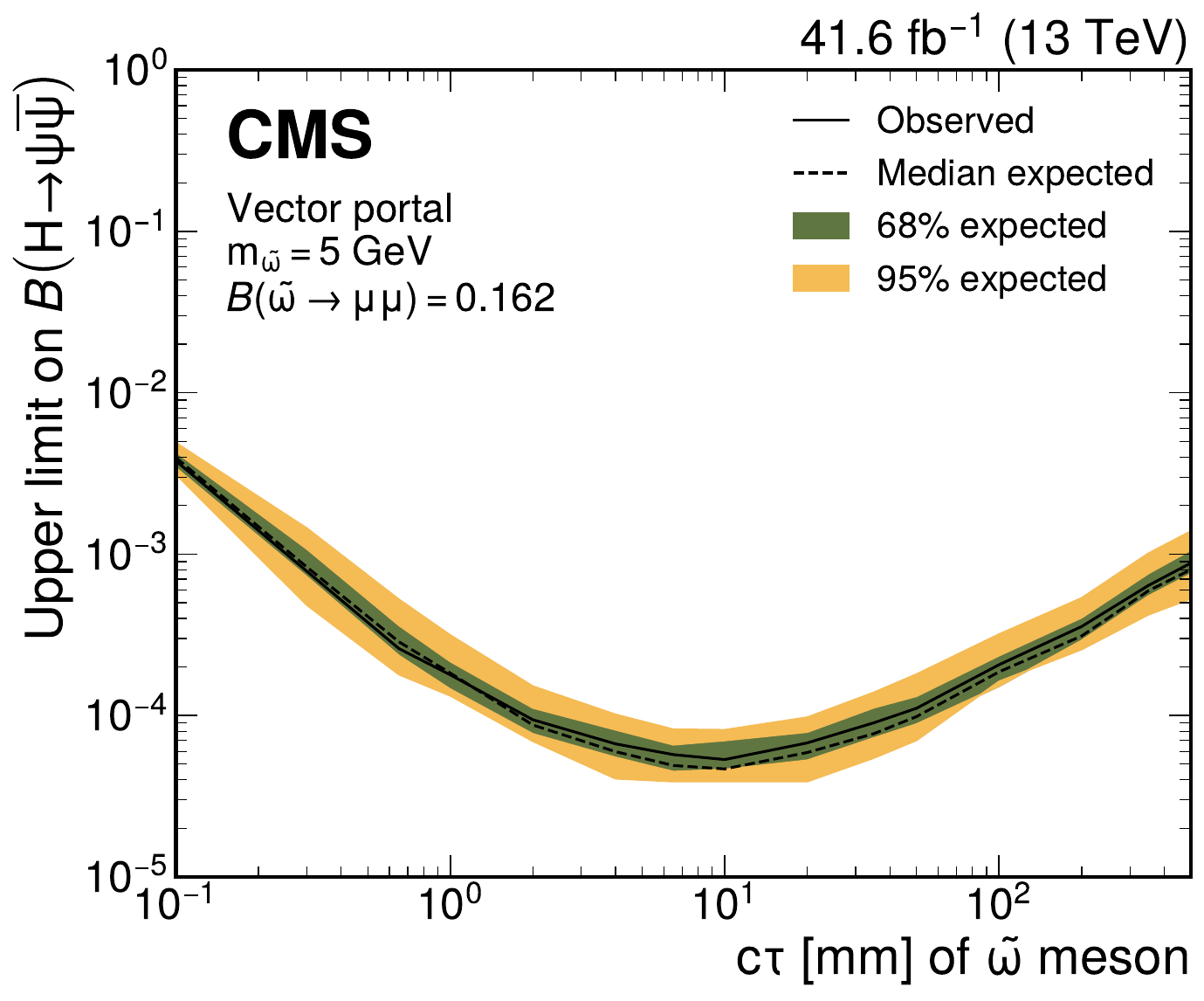}
\includegraphics[width=0.48\textwidth]{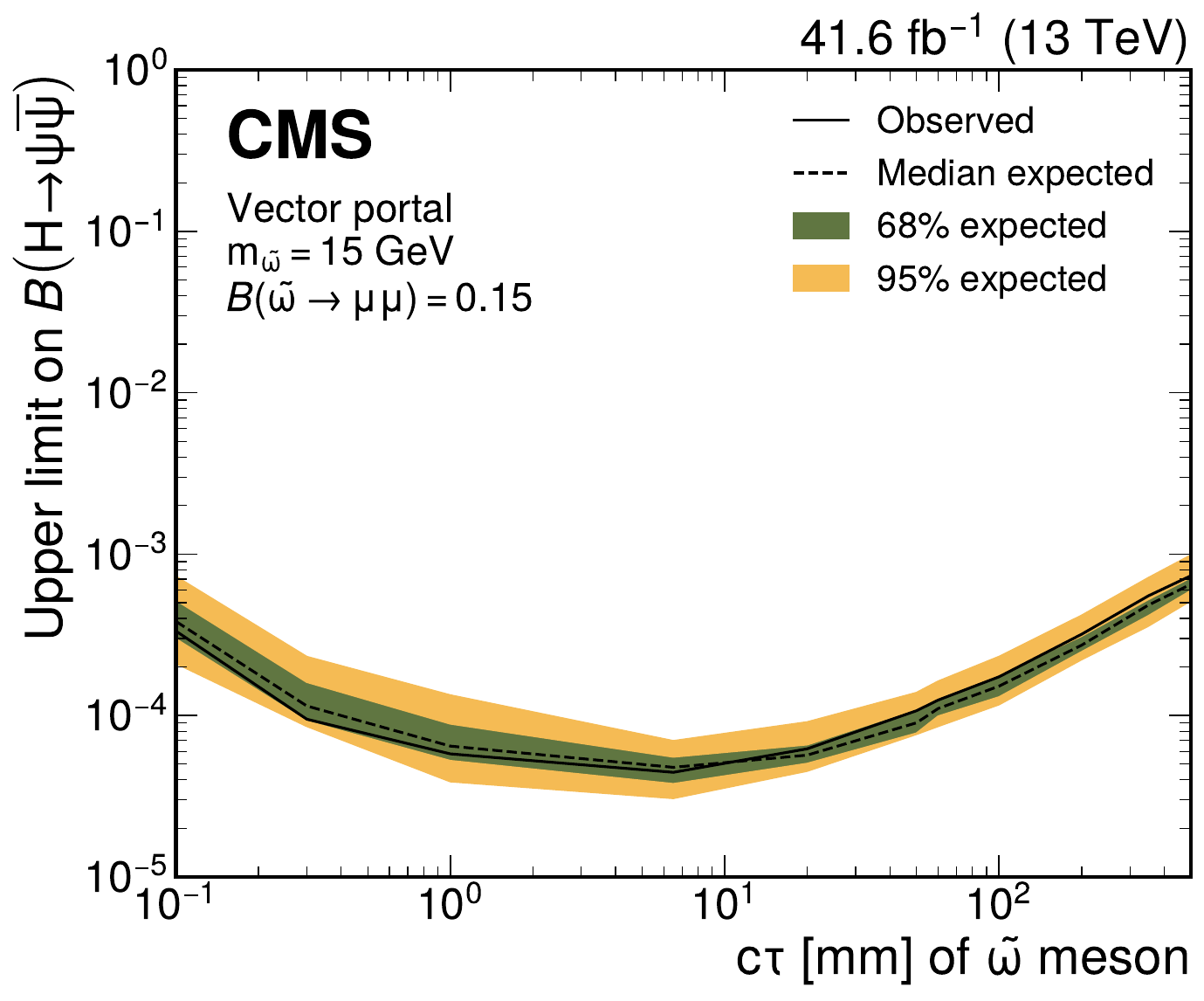}~~
\includegraphics[width=0.48\textwidth]{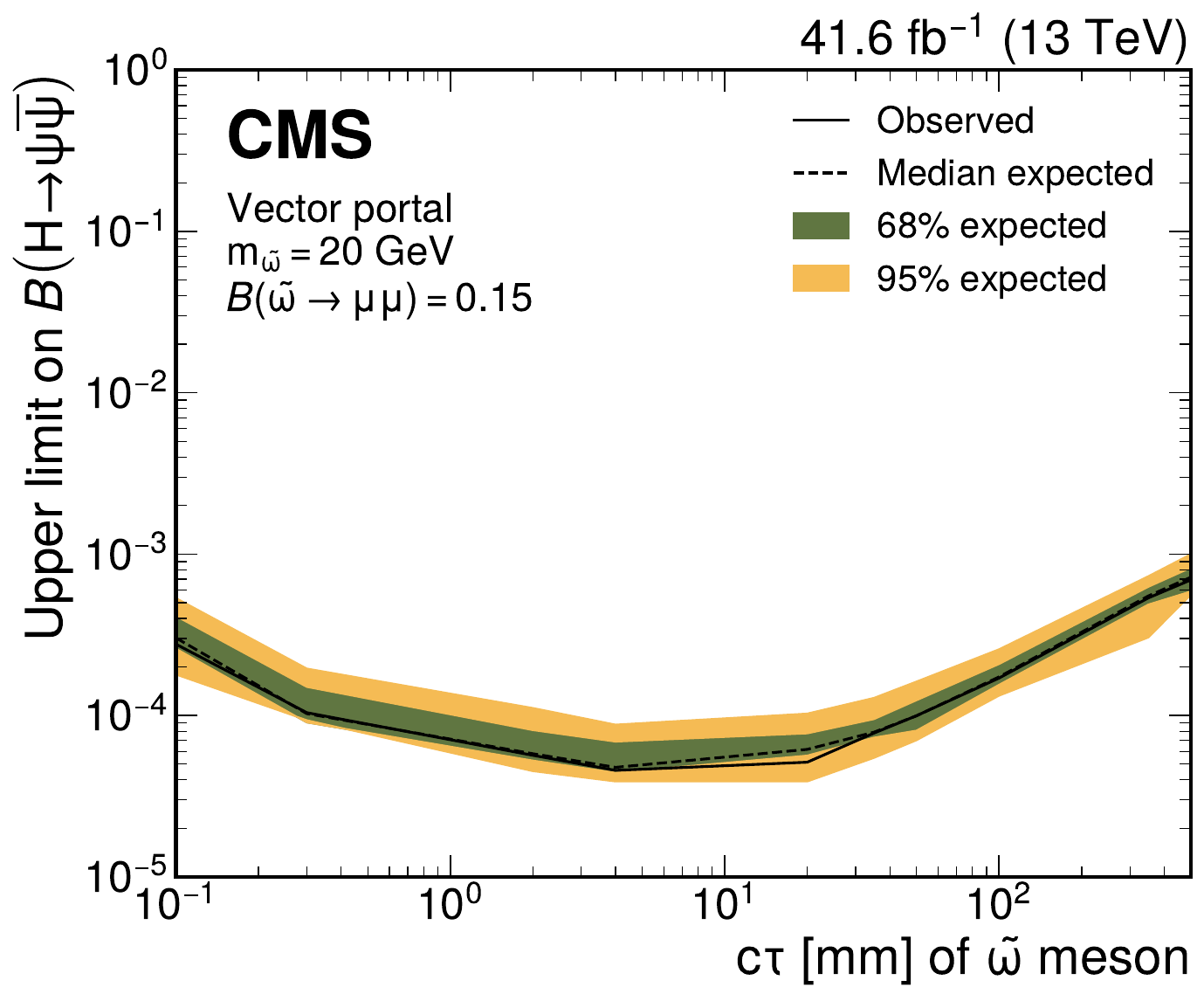}
\caption{Upper limits at 95\% \CL on the branching fraction $\mathcal{B}(\PH\to\dpart\overline{\dpart})$ as functions of the \Pomega meson $c\tau$ for representative \Pomega mass hypotheses and branching fractions for decaying into a muon pair in the vector portal model. It is assumed that $m_{\Pomega}=\TLambda=m_{\Peta}$, where $m_{\Pomega}$, \TLambda, and $m_{\Peta}$ are parameters of the dark sector: the mass of the spin-one meson, confinement scale, and the mass of the spin-zero meson, respectively.}
\label{vector_limits_per_mass}
\end{figure}

No significant excess beyond the standard model expectation is observed, and 95\% confidence level (\CL) upper limits are set on the branching fraction of the Higgs boson decaying to dark partons. Limits are estimated using a modified frequentist approach with the \CLs criterion \cite{Cowan:2010js, Junk:1999kv, Read:2002hq}. The limits are obtained for different mass and lifetime hypotheses of the dark shower signal models. For the vector portal, the upper limits on the branching fraction $\mathcal{B}(\PH\to\dpart\overline{\dpart})$ as functions of the dark vector meson proper decay length are shown in Fig.~\ref{vector_limits_per_mass} for four representative mass hypotheses. The limits on the branching fraction are in the range $10^{-3}$ to $10^{-5}$ throughout most of the parameter space.

\begin{figure}[ht!]
\centering
\includegraphics[width=0.48\textwidth]{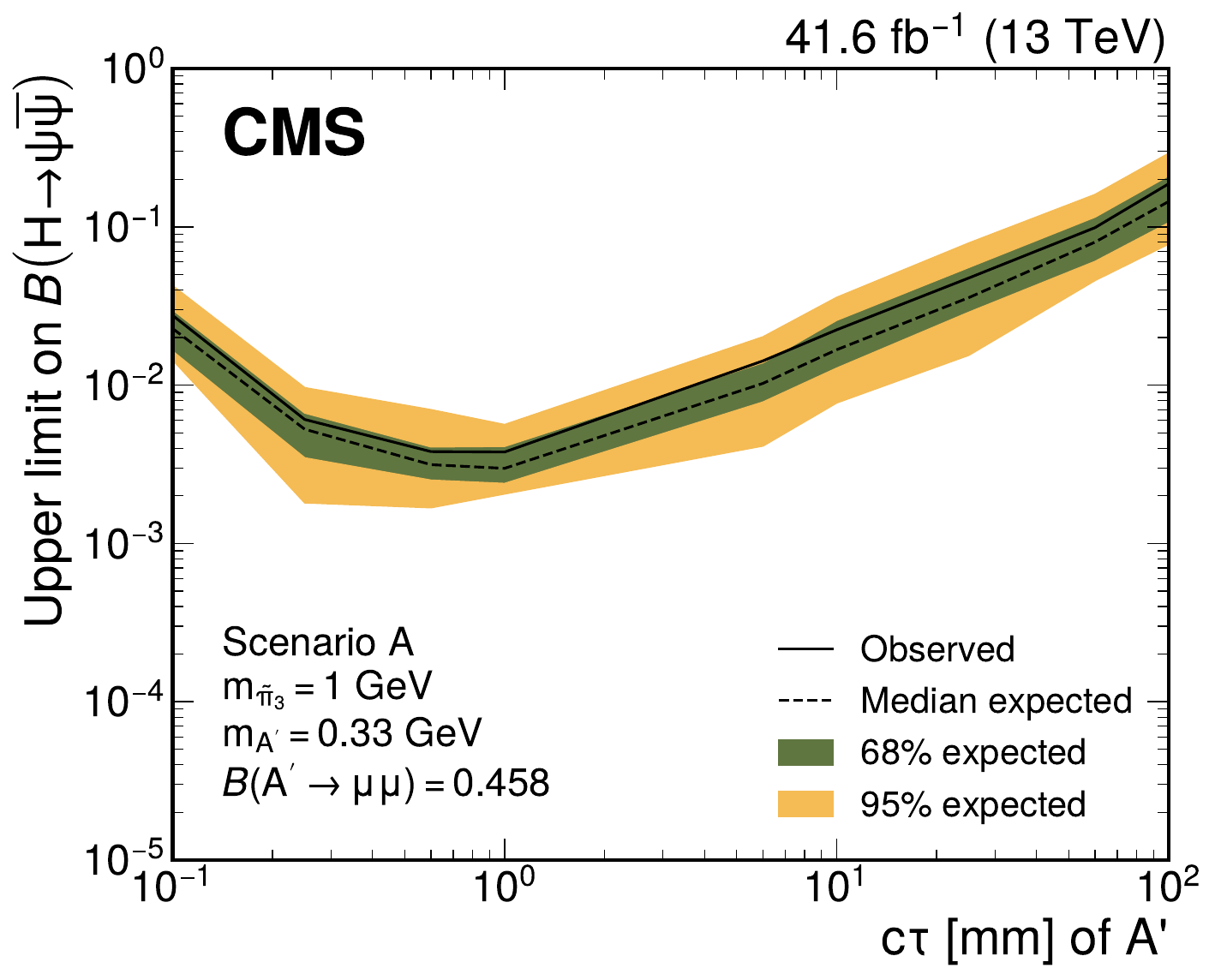}~~
\includegraphics[width=0.48\textwidth]{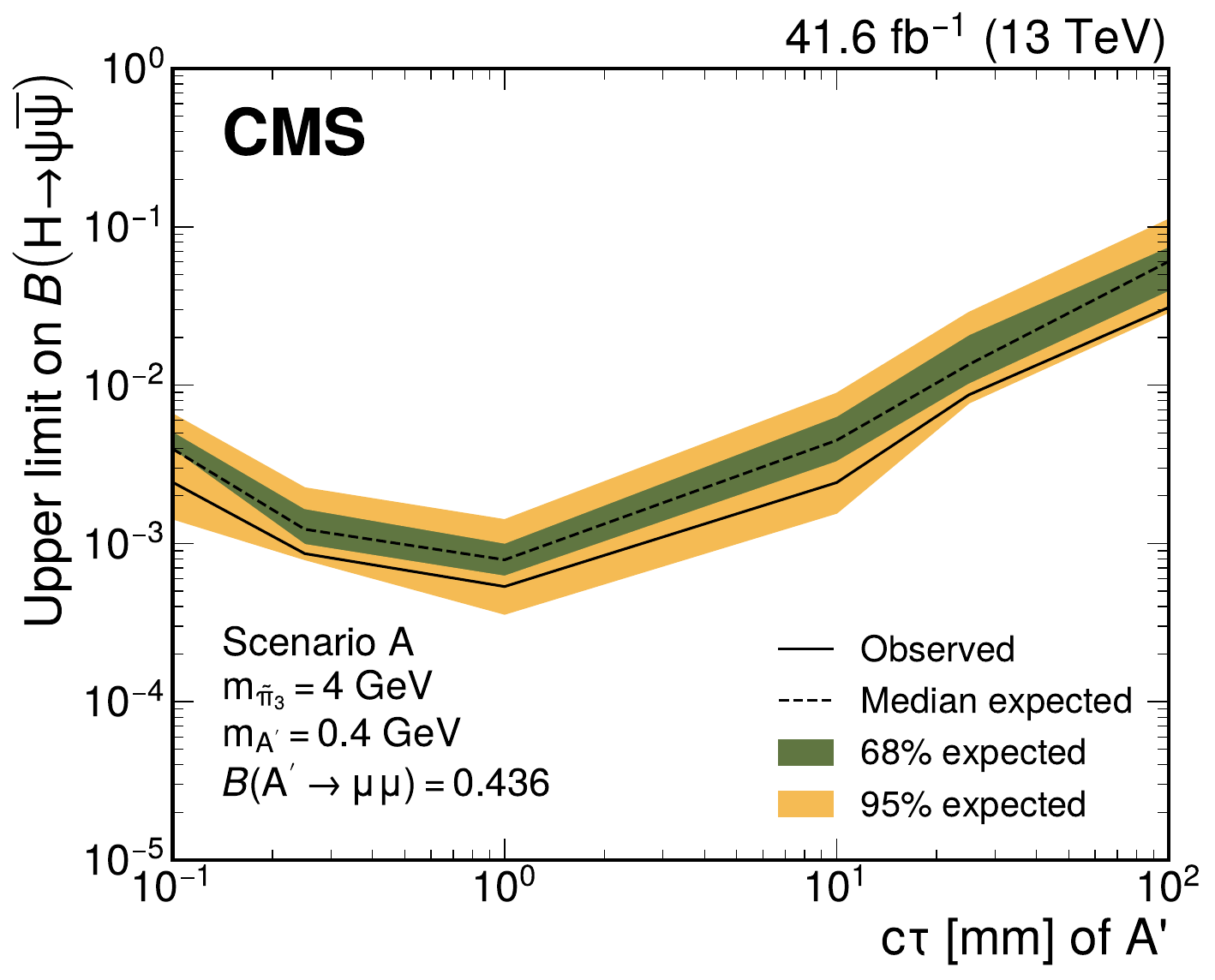}\\
\includegraphics[width=0.48\textwidth]{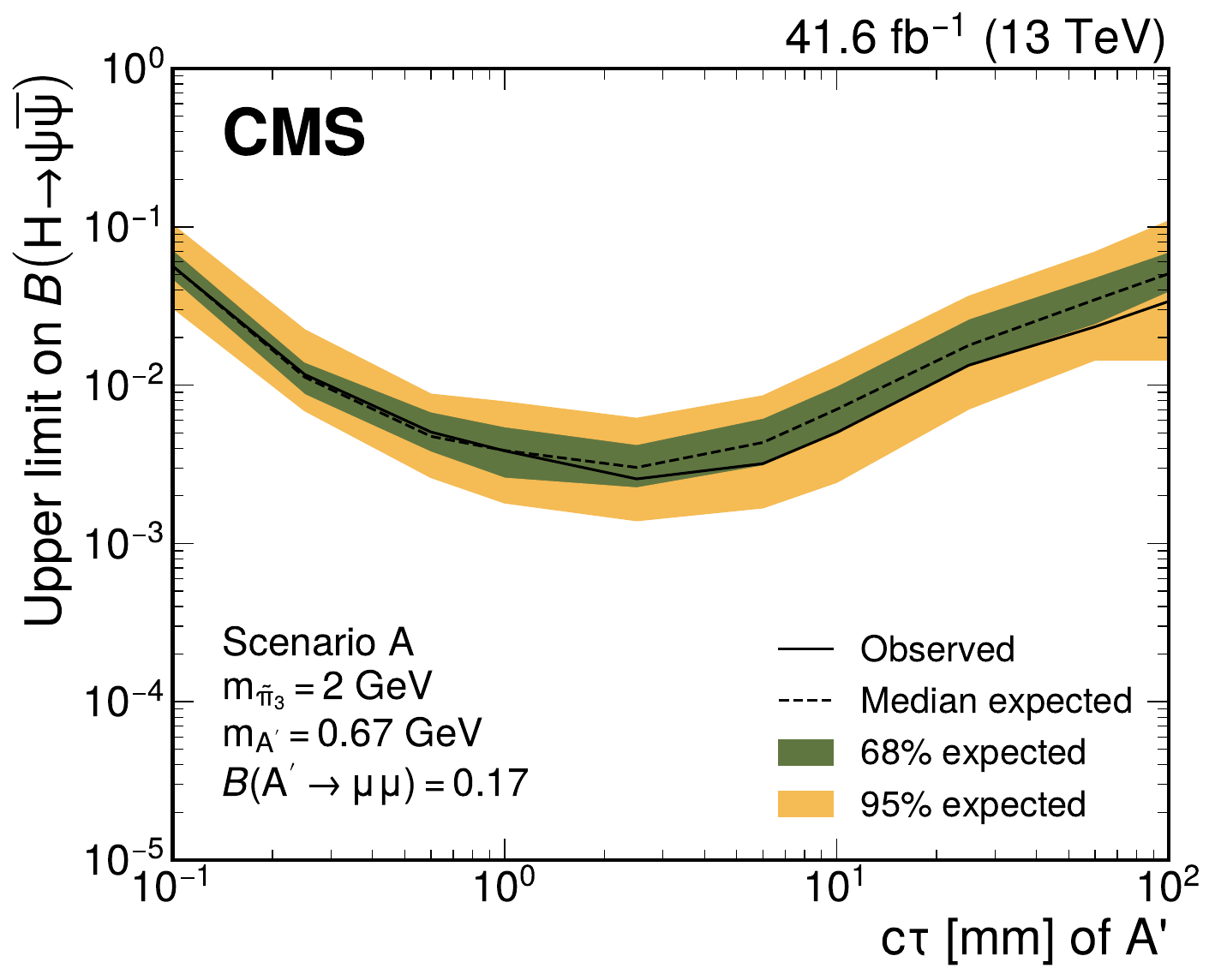}~~
\includegraphics[width=0.48\textwidth]{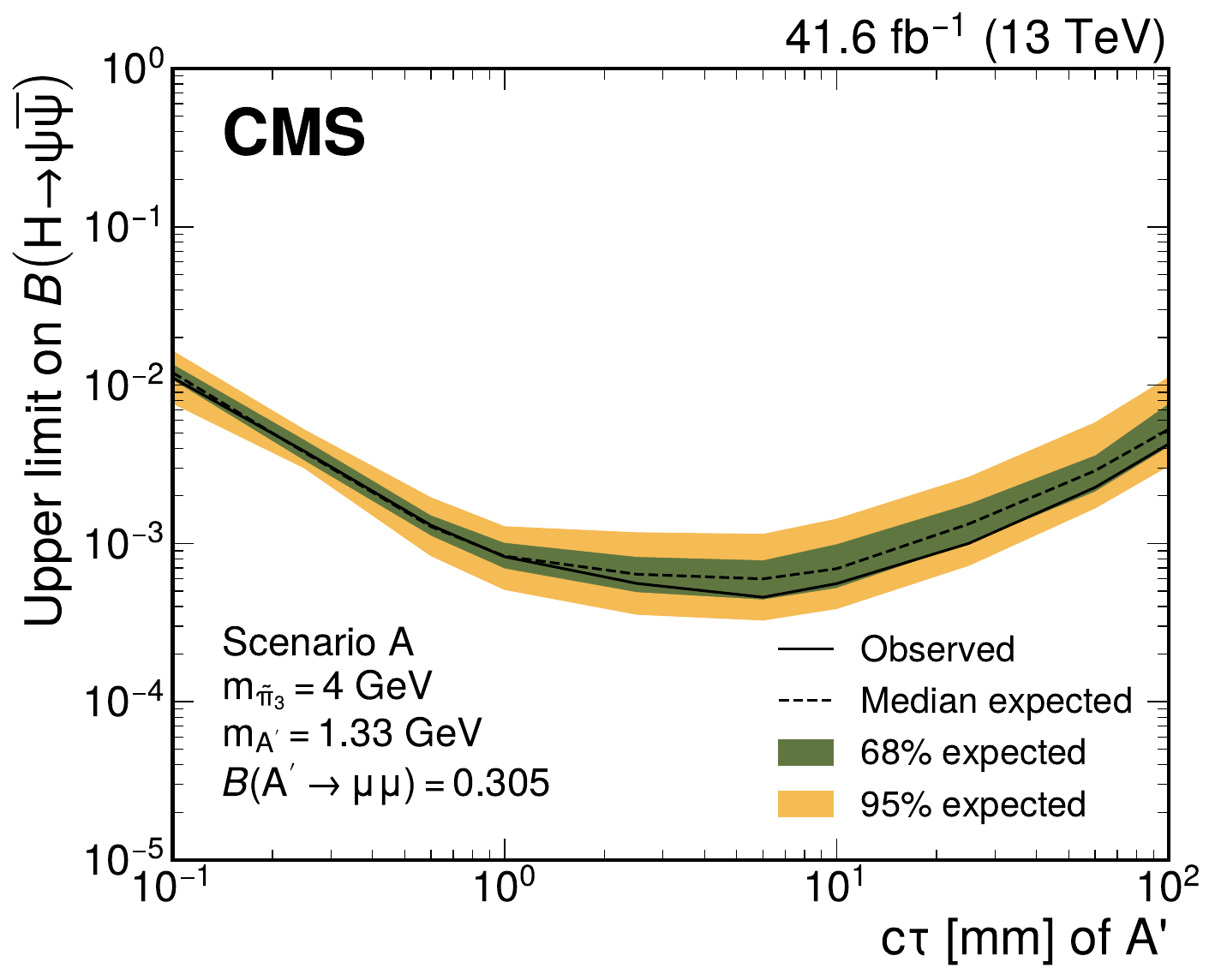}
\caption{Upper limits at 95\% \CL on the branching fraction $\mathcal{B}(\PH\to\dpart\overline{\dpart})$ as functions of the \dphoton $c\tau$ for representative $\Ppi_{3}$ and \dphoton
mass hypotheses, and branching fractions of the dark photon decaying into a muon pair in Scenario A. It is assumed that $m_{\Peta}=\TLambda=4m_{\Ppi_{2}}$ and ${\sin\theta=0.1}$, where $m_{\Peta}$ is the mass of the dark-sector pseudoscalar meson and $\theta$ is the mixing angle parametrising the isospin violation. The branching fraction $\mathcal{B}(\Ppi_{3}\to \dphoton\dphoton)$ is assumed to be 1.}
\label{scenarioA_per_mass}
\end{figure}

The limits for the Scenario A model as functions of the dark photon proper decay length are shown in Fig.~\ref{scenarioA_per_mass} for four representative mass hypotheses. The dark photon mass models chosen for Scenario A are in the low-mass regime (below 2\GeV). The limits obtained are in the range $10^{-2}$ to $10^{-4}$.

\begin{figure}[ht!]
\centering
\includegraphics[width=0.48\textwidth]{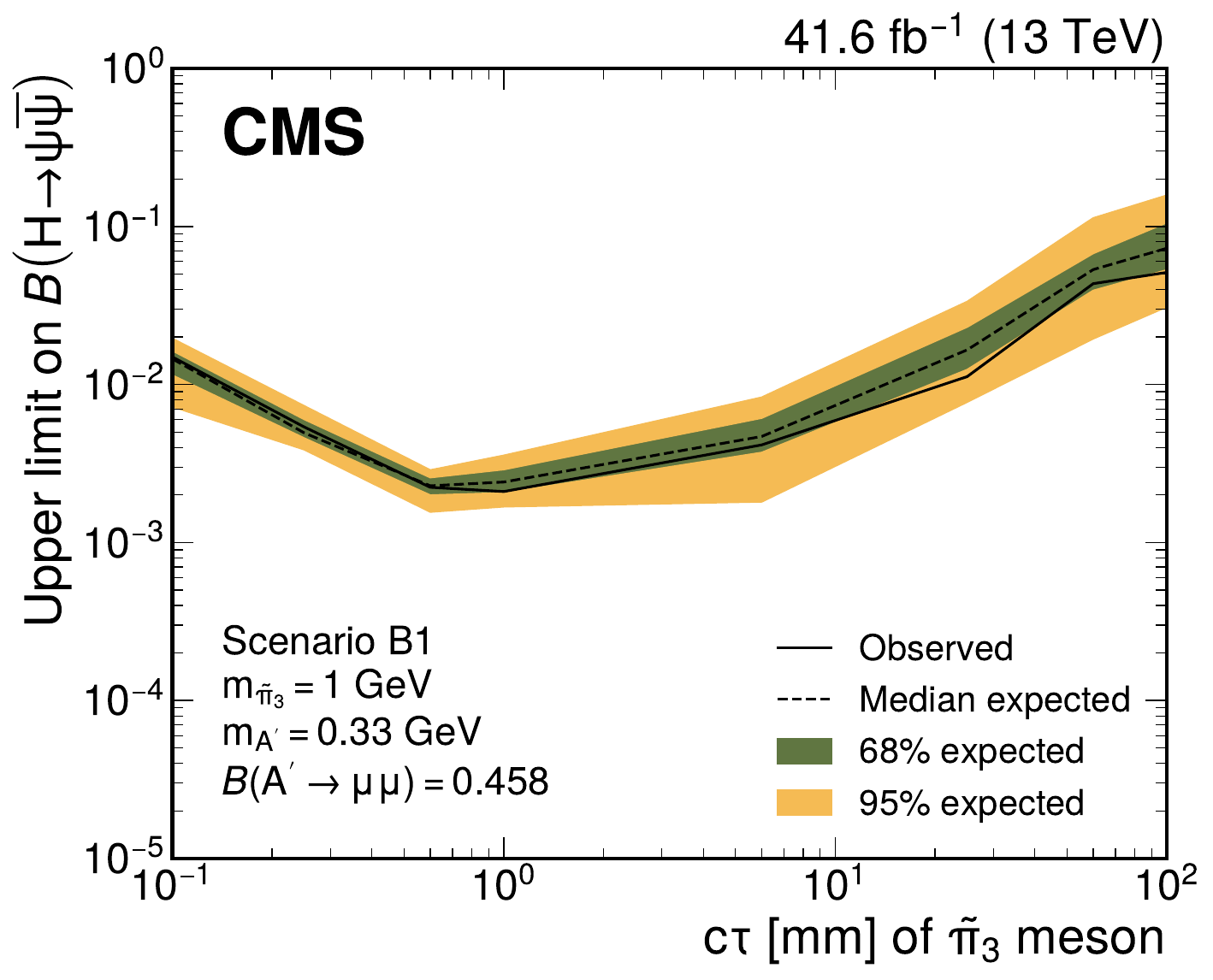}~~
\includegraphics[width=0.48\textwidth]{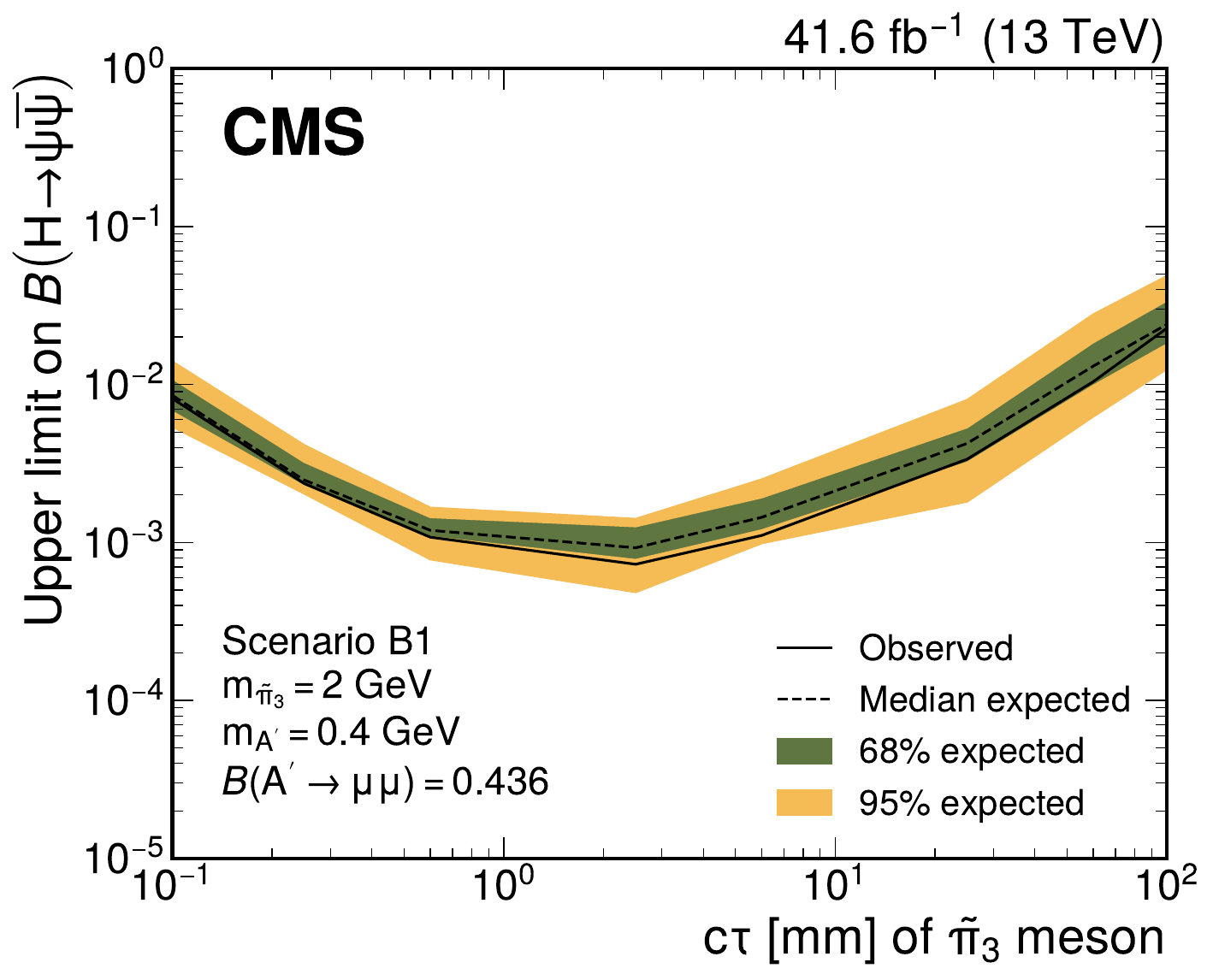}\\
\includegraphics[width=0.48\textwidth]{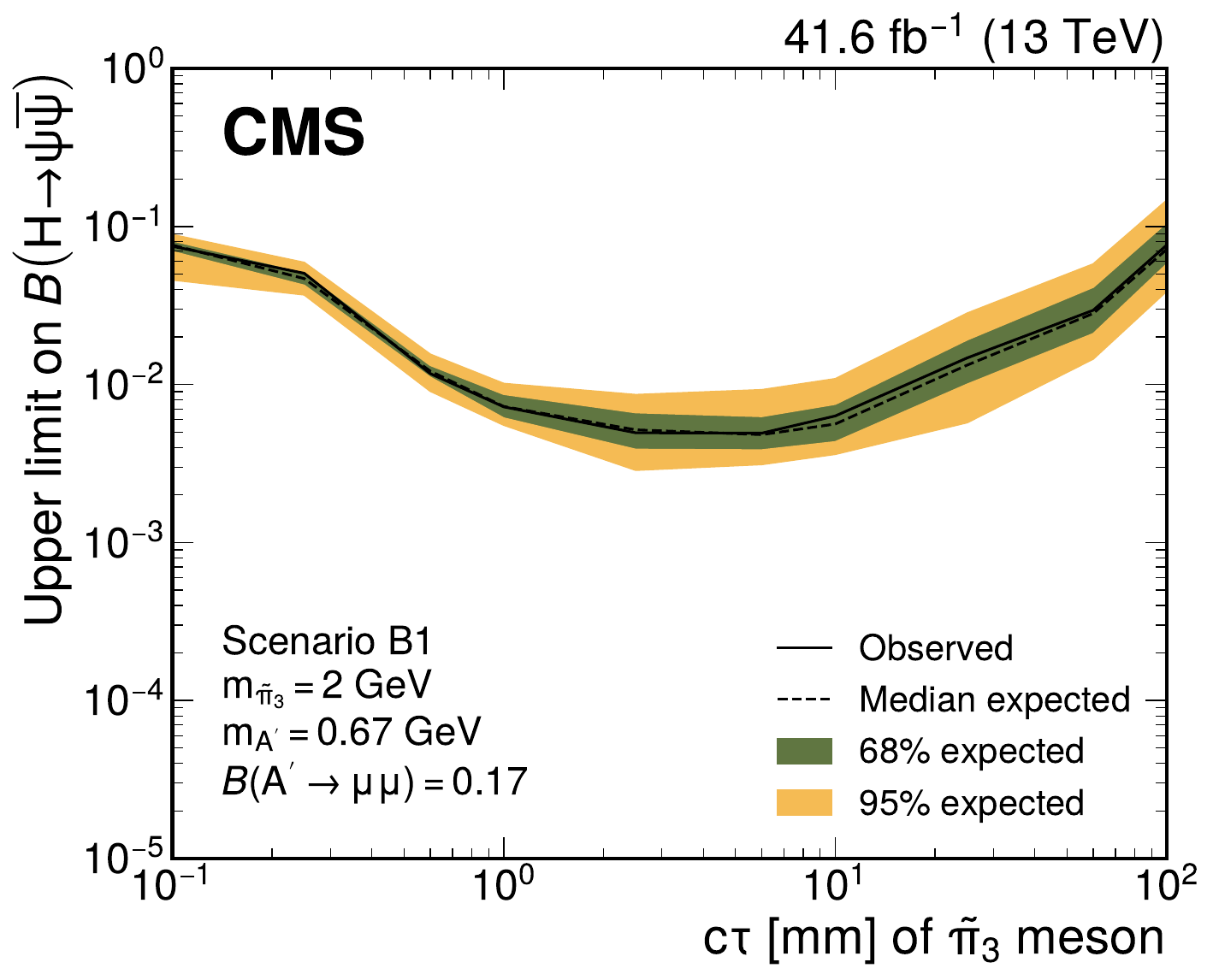}~~
\includegraphics[width=0.48\textwidth]{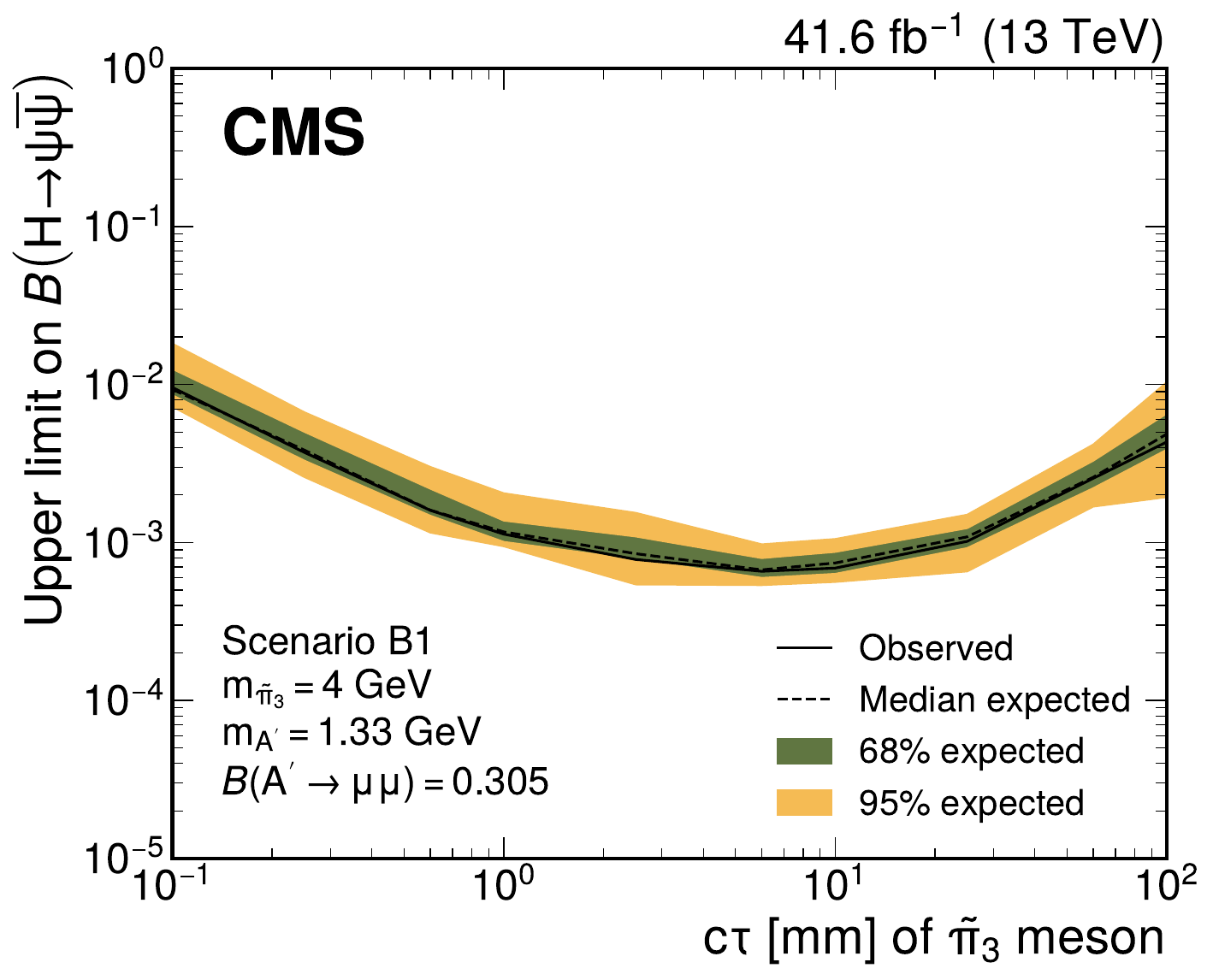}
\caption{Upper limits at 95\% \CL on the branching fraction $\mathcal{B}(\PH\to\dpart\overline{\dpart})$ as functions of the $\Ppi_{3}$ meson $c\tau$ for representative $\Ppi_{3}$ and \dphoton
mass hypotheses, and branching fractions of the dark photon decaying into a muon pair in Scenario B1. It is assumed that $m_{\Peta}=\TLambda=4m_{\Ppi_{2}}$ and ${\sin\theta=0.1}$, where $m_{\Peta}$ is the mass of the dark-sector pseudoscalar meson and $\theta$ is the mixing angle parametrising the isospin violation. The branching fraction $\mathcal{B}(\Ppi_{3}\to \dphoton\dphoton)$ is assumed to be 1.}
\label{scenarioB1_per_mass}
\end{figure}

The limits for the Scenario B1 model as functions of the dark meson $\Ppi_{3}$ proper decay length are shown in Fig.~\ref{scenarioB1_per_mass} for four representative mass hypotheses. The Scenario B1 model features a non-pointing dimuon vertex topology, and the benchmarks chosen are for low mass. The limits are found to be comparable to those for the Scenario A models, which shows sensitivity is achieved for non-pointing models as well as pointing models.

The constraints are more stringent in general for lower lifetime hypotheses owing to the higher SV reconstruction efficiency at smaller displacement. For very short lifetimes, the constraints are less stringent due to smaller signal efficiencies. The limits also tend to be more stringent at higher mass due to the lower background present. For similar dark meson masses, stronger sensitivity is achieved for the vector portal because there is a higher dark-meson multiplicity (as shown in Fig.~\ref{dark_meson_multiplicities}). 

The limits obtained for the vector portal model are complementary to those from a previous search performed by the CMS experiment in the muon detectors \cite{CMS:2024bvl} studying the same model, for which longer proper decay lengths are covered. This search probes new parameter space for proper decay lengths of the dark \Pomega meson below ${\sim}$100\mm and masses as low as 0.3\GeV. There is also an improvement in the limits of up to two orders of magnitude with respect to Ref.~\cite{CMS:2024bvl}, which gave the best LHC limits on the model, in the range of proper decay lengths between 100 and 500\mm. The first constraints to date are imposed on the Scenario A and Scenario B1 models.  

\begin{figure}[ht!]
\centering
\includegraphics[width=0.6\textwidth]{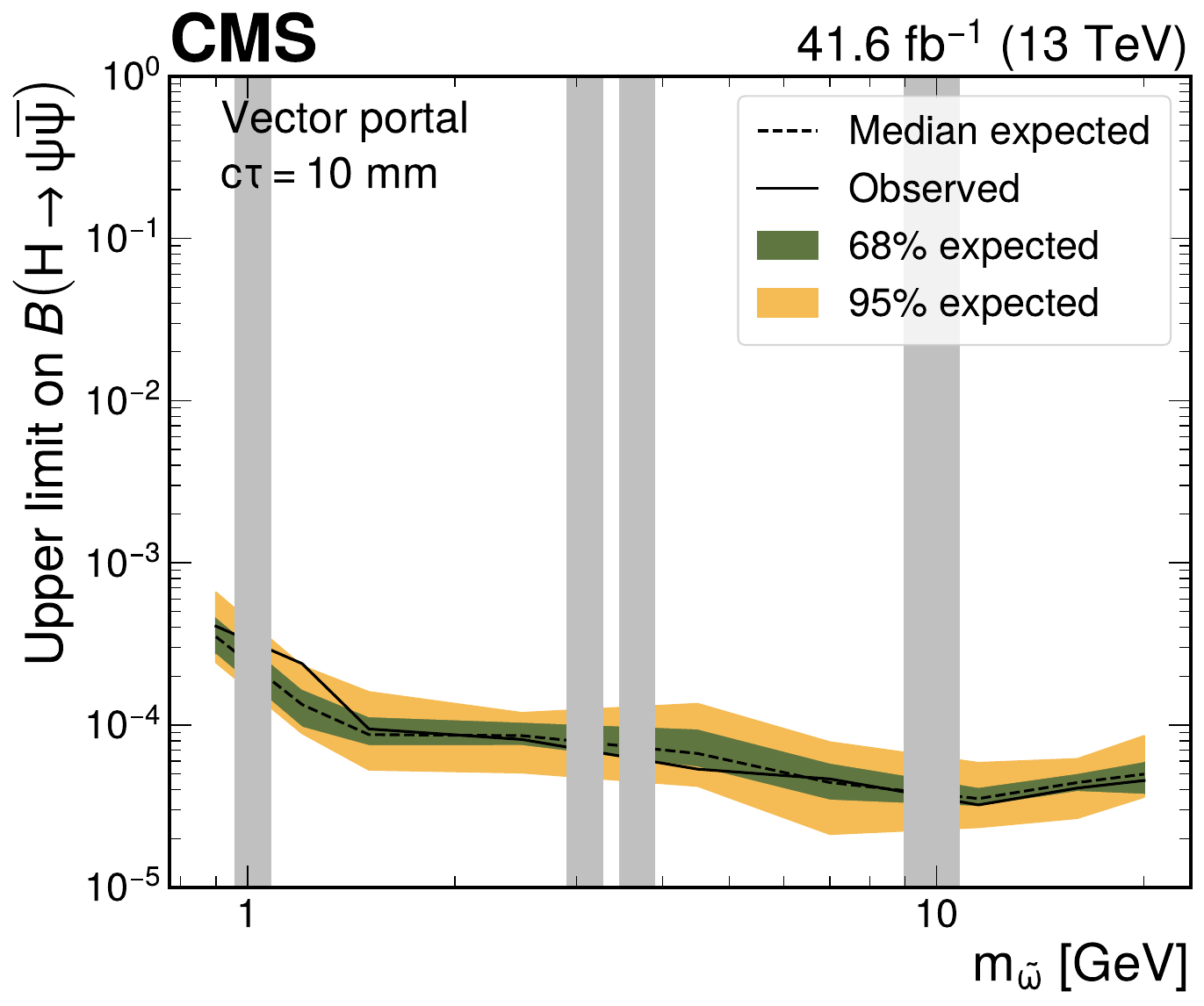}
\includegraphics[width=0.6\textwidth]{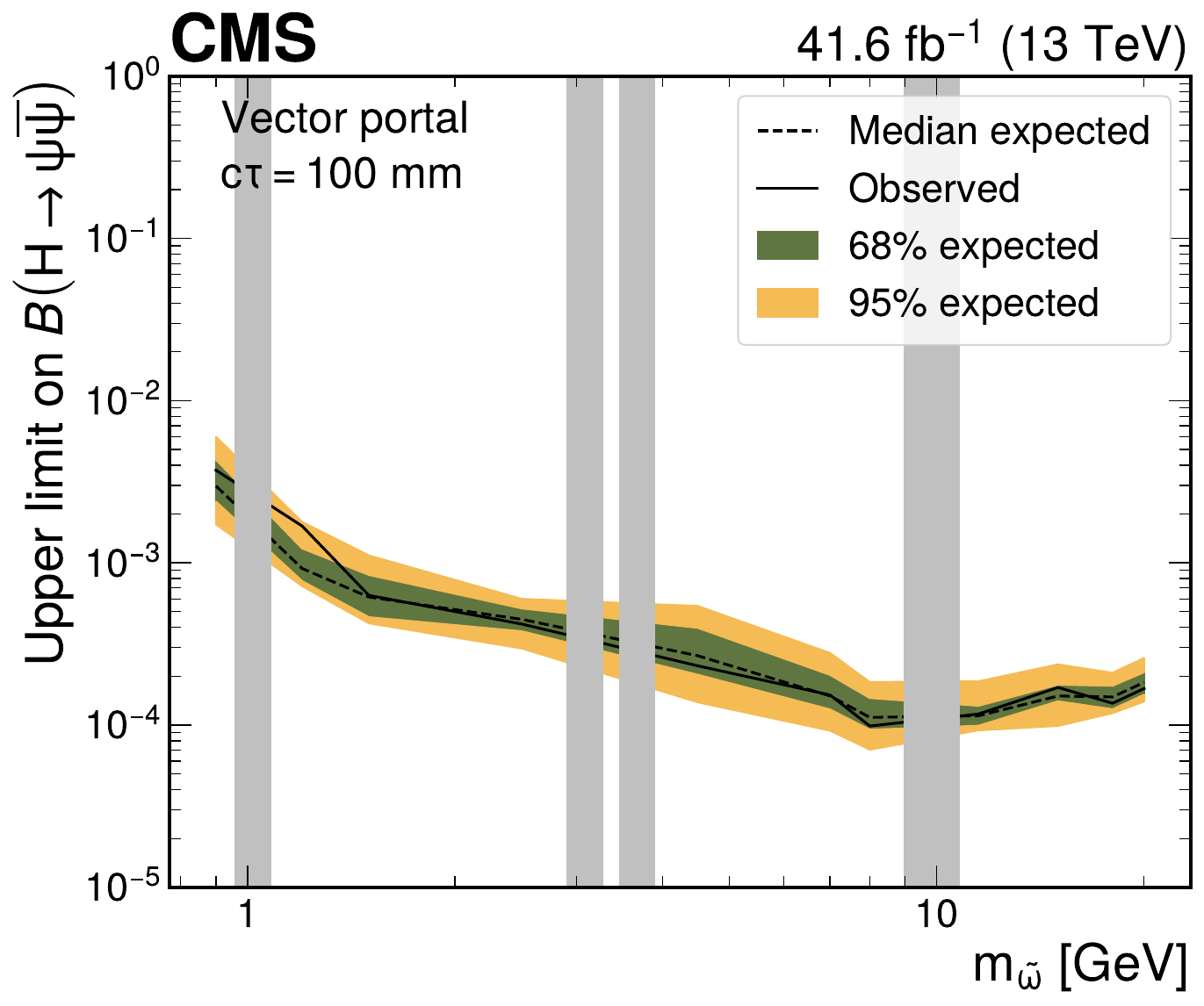}
\caption{Upper limits at 95\% \CL on the branching fraction $\mathcal{B}(\PH\to\dpart\overline{\dpart})$ as functions of the \Pomega meson mass for representative $c\tau$ hypotheses in the vector portal model. It is assumed that $m_{\Pomega}=\TLambda=m_{\Peta}$, where $m_{\Pomega}$, \TLambda, and $m_{\Peta}$ are parameters of the dark sector: the mass of the spin-one meson, confinement scale, and the mass of the spin-zero meson, respectively. The grey bands correspond to mass regions of known SM resonances, which are masked in the search.}
\label{vector_limits_per_lifetime}
\end{figure}

\begin{figure}[ht!]
\centering
\includegraphics[width=0.48\textwidth]{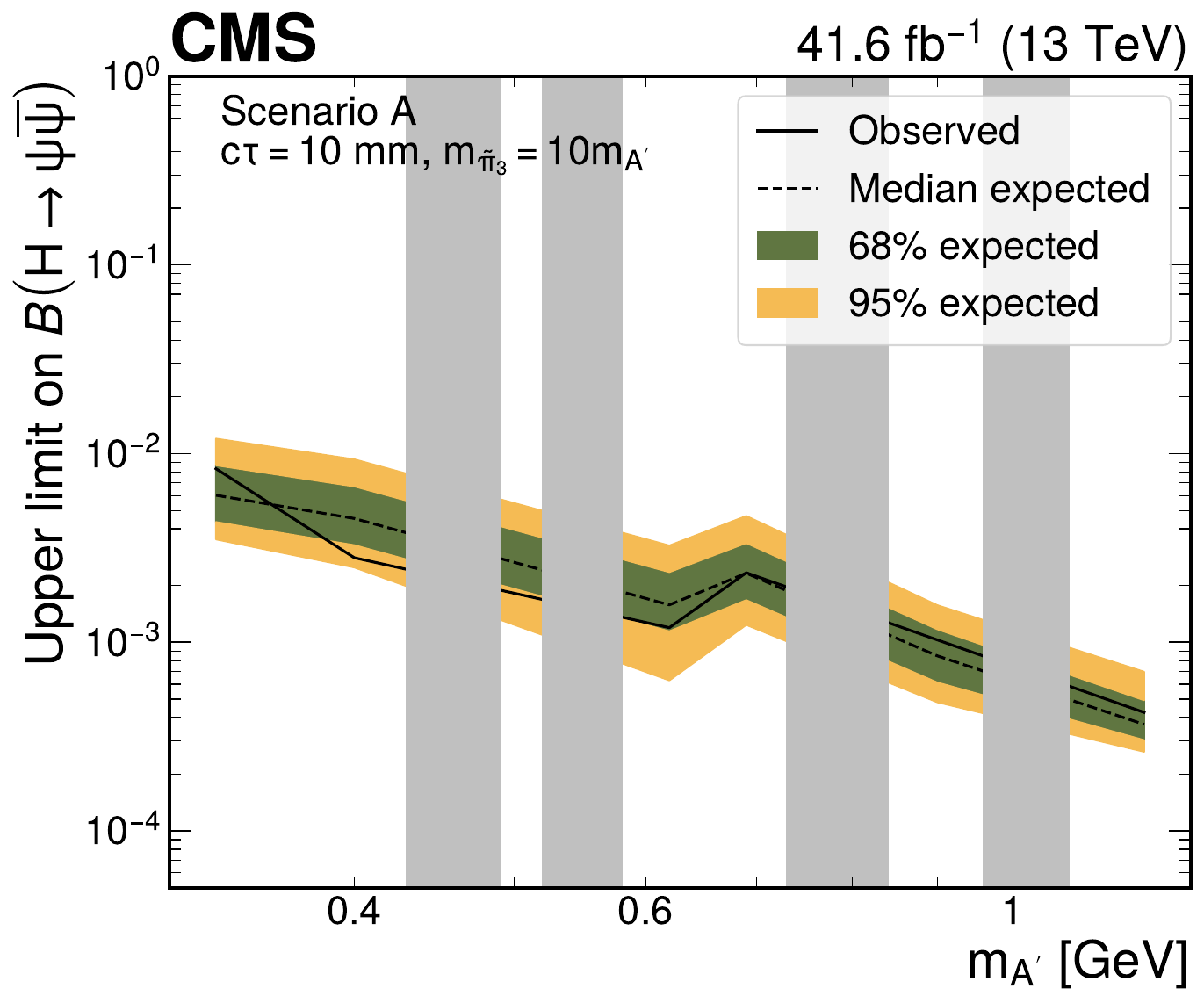}~~
\includegraphics[width=0.48\textwidth]{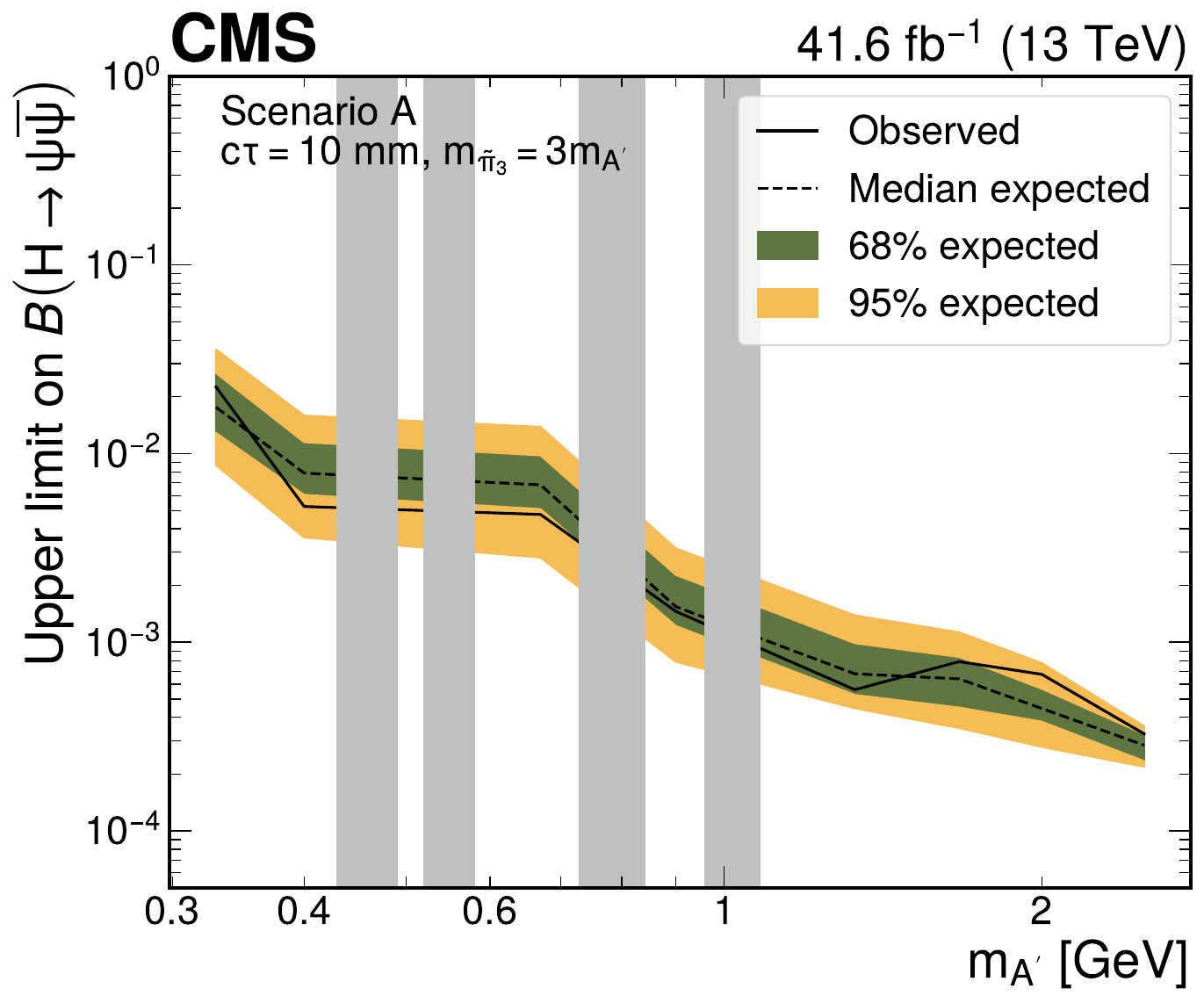}\\
\includegraphics[width=0.48\textwidth]{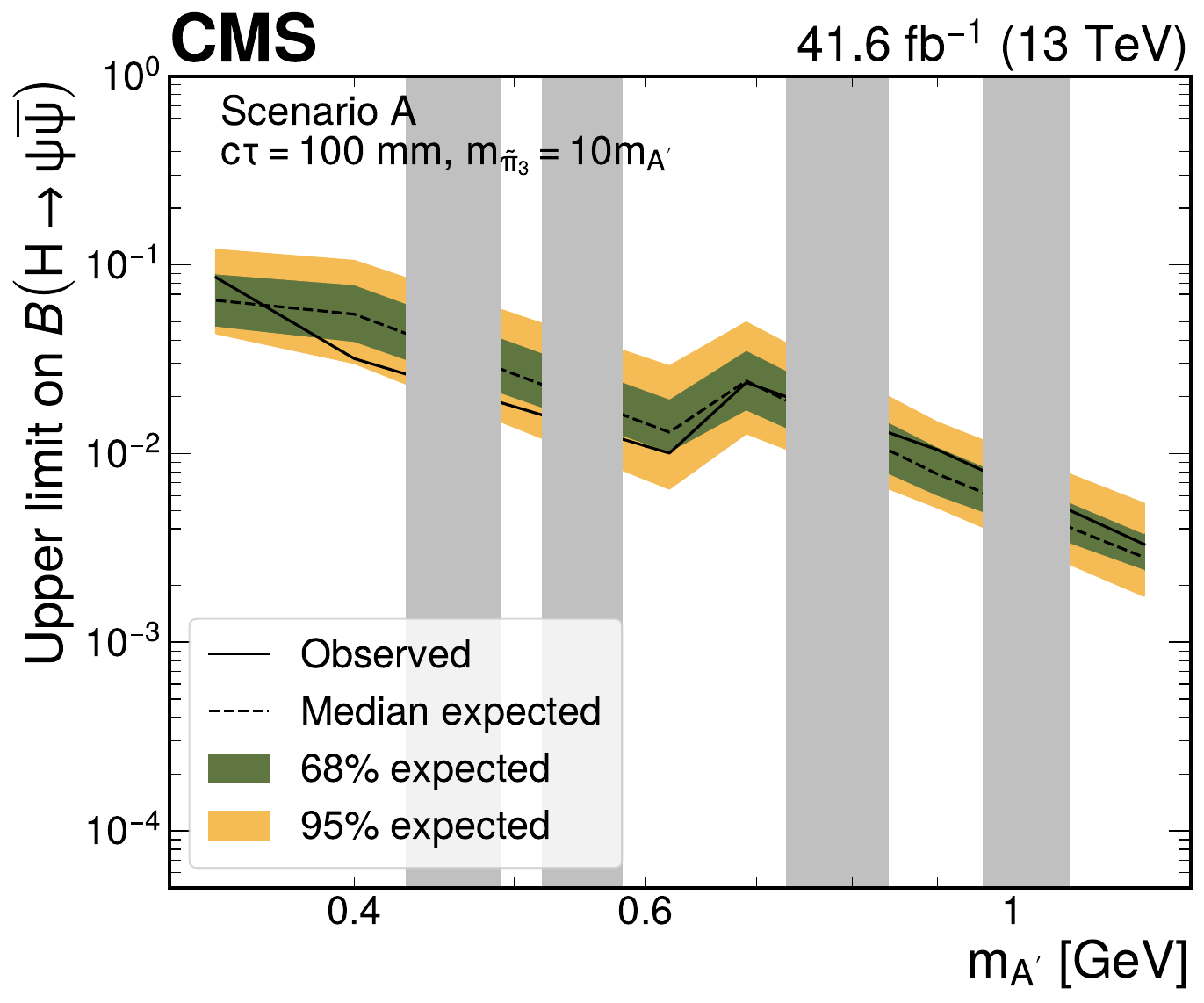}~~
\includegraphics[width=0.48\textwidth]{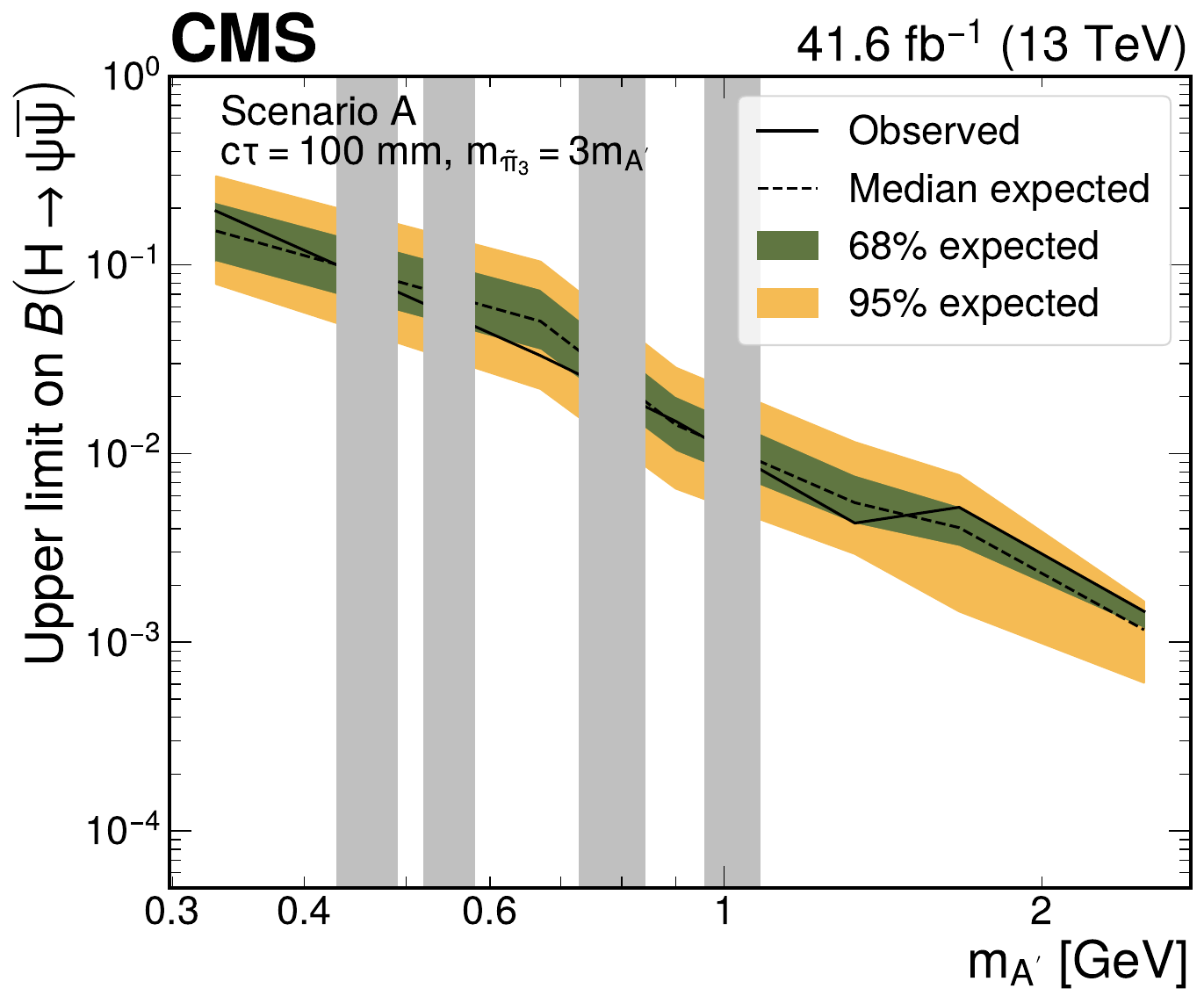}
\caption{Upper limits at 95\% \CL on the branching fraction $\mathcal{B}(\PH\to\dpart\overline{\dpart})$ as functions of the \dphoton mass in Scenario A, for representative \dphoton $c\tau$ hypotheses, and different ratios of the $\Ppi_{3}$ and \dphoton masses. The limits are shown for the cases where $m_{\Ppi_{3}}=10m_{\dphoton}$ (left column) and $m_{\Ppi_{3}}=3m_{\dphoton}$ (right column), respectively. It is assumed that $m_{\Peta}=\TLambda=4m_{\Ppi_{2}}$ and ${\sin\theta=0.1}$, where $m_{\Peta}$ is the mass of the dark-sector pseudoscalar meson and $\theta$ is the mixing angle parametrising the isospin violation. The branching fraction $\mathcal{B}(\Ppi_{3}\to \dphoton\dphoton)$ is assumed to be 1. The grey bands correspond to mass regions of known SM resonances, which are masked in the search.}
\label{scenarioA_limits_per_lifetime}
\end{figure}

\begin{figure}[ht!]
\centering
\includegraphics[width=0.48\textwidth]{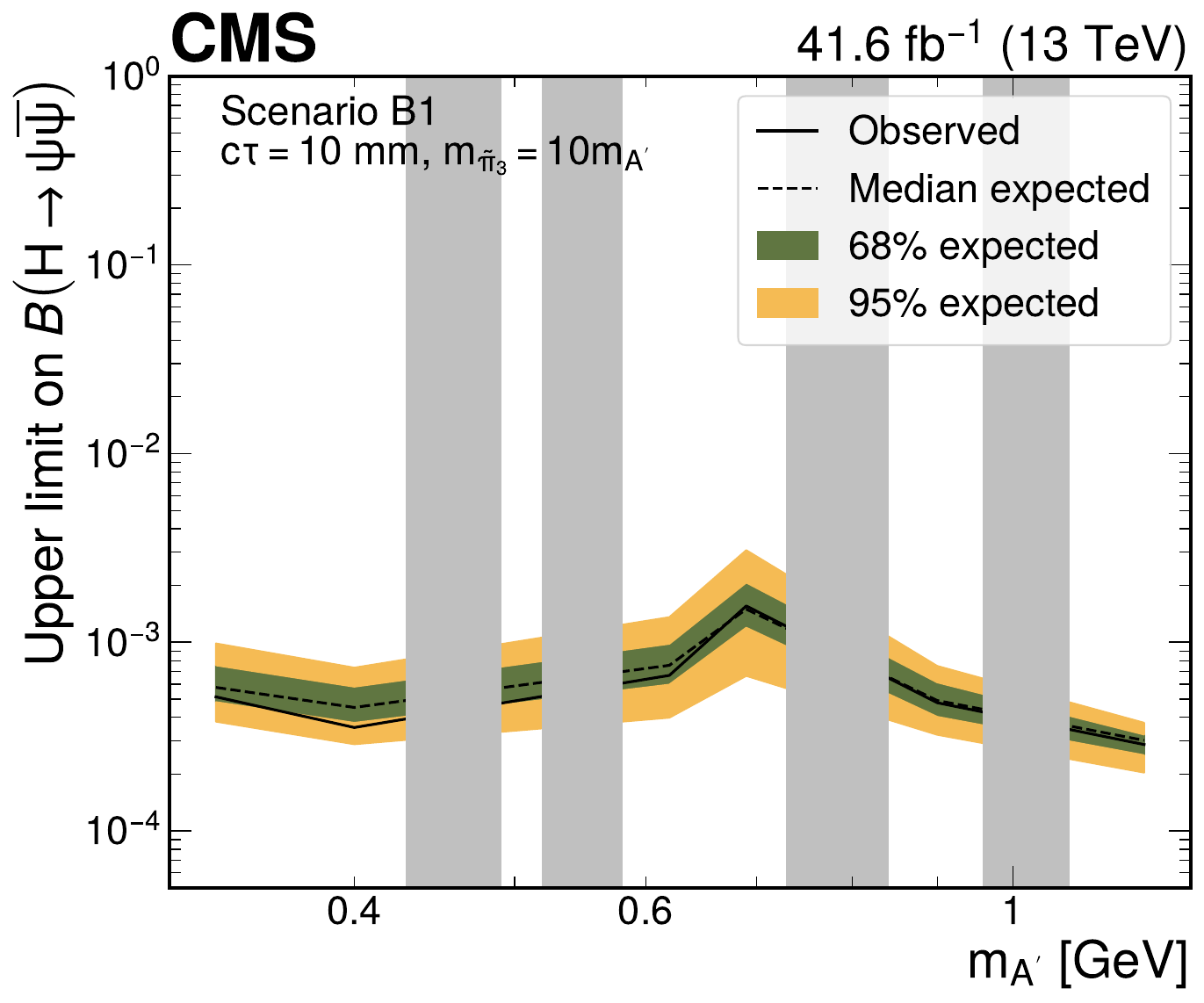}~~
\includegraphics[width=0.48\textwidth]{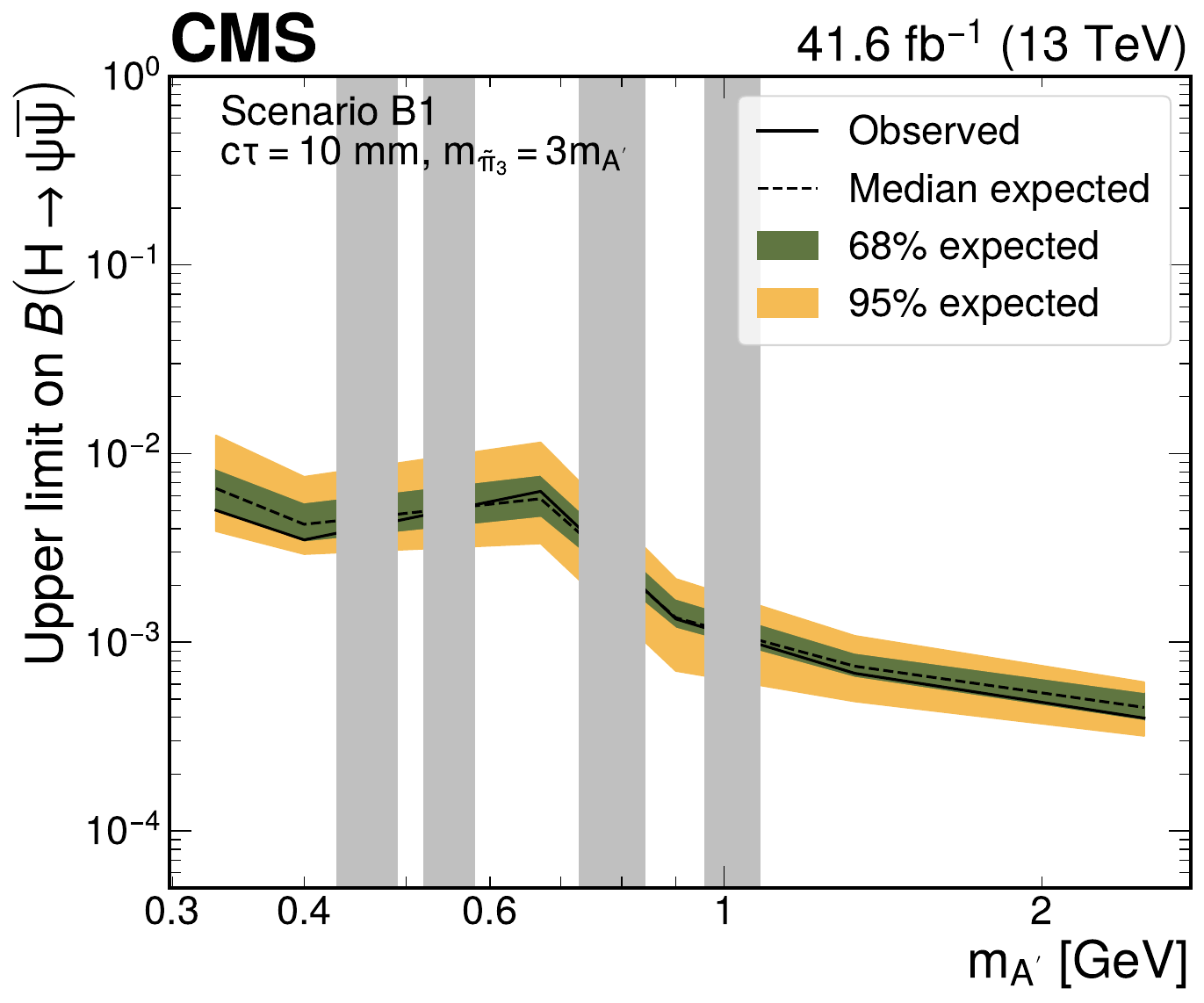}\\
\includegraphics[width=0.48\textwidth]{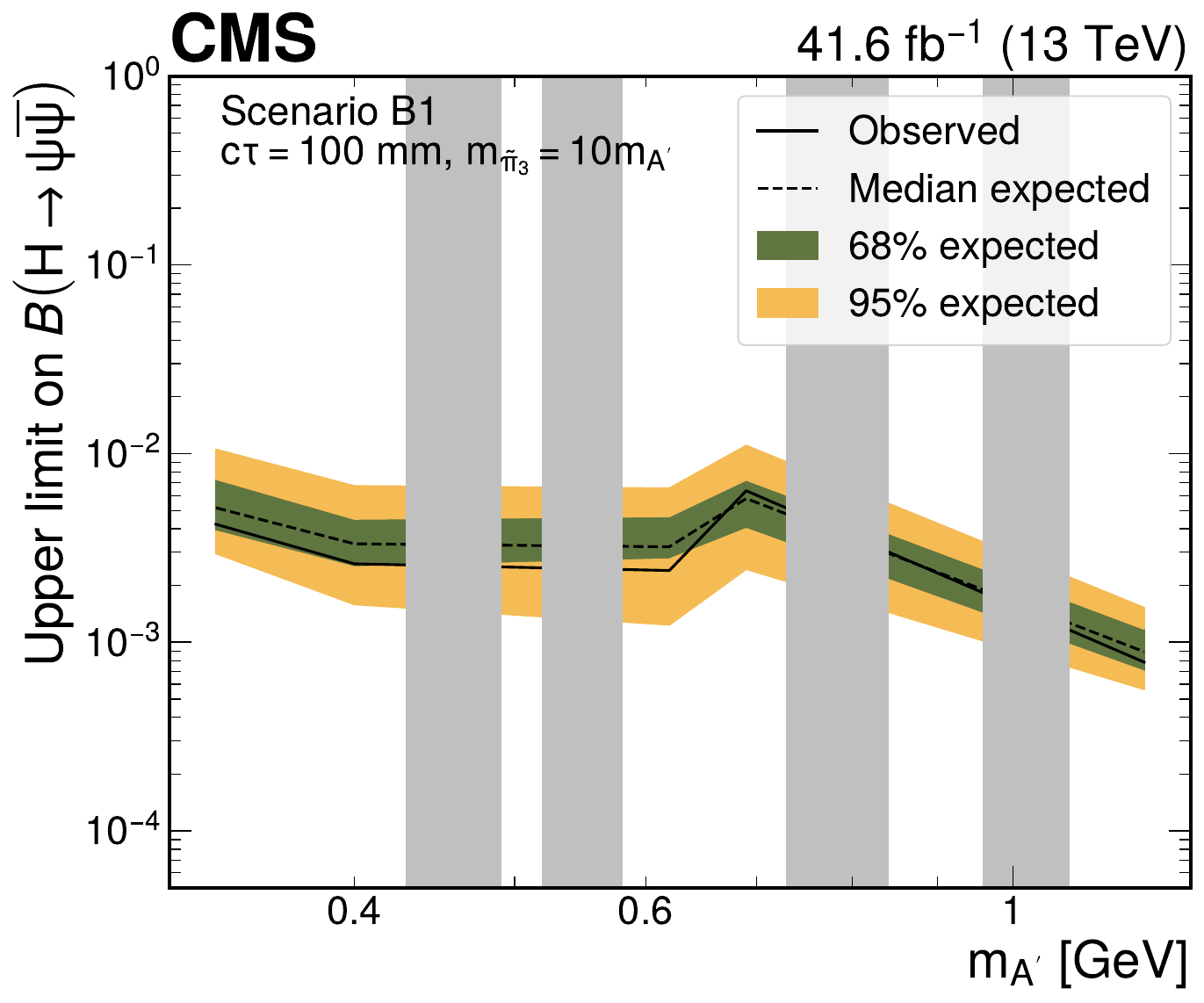}~~
\includegraphics[width=0.48\textwidth]{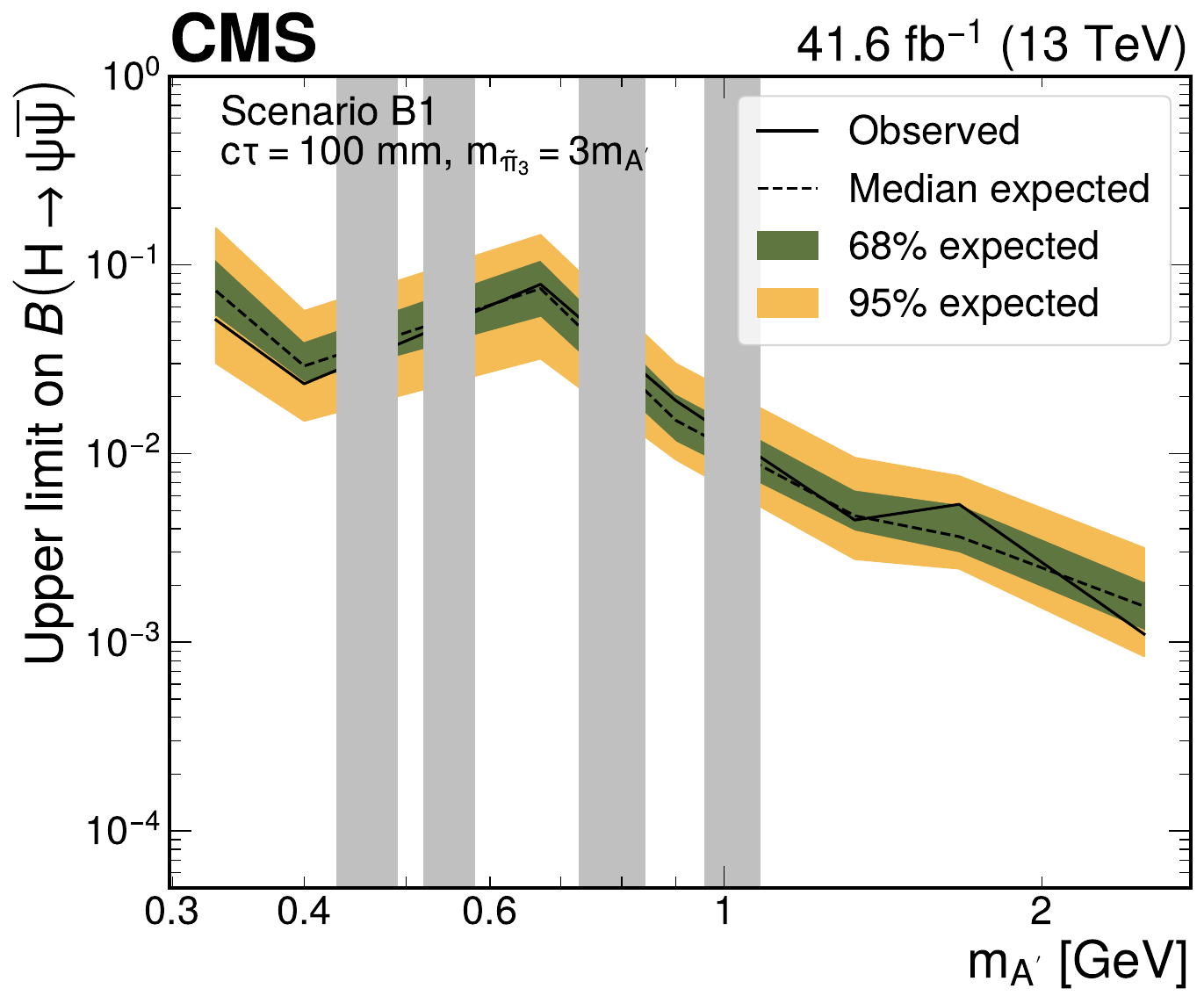}
\caption{Upper limits at 95\% \CL on the branching fraction $\mathcal{B}(\PH\to\dpart\overline{\dpart})$ as functions of the \dphoton mass in Scenario B1, for representative $\Ppi_{3}$ $c\tau$ hypotheses, and different ratios of the $\Ppi_{3}$ and \dphoton masses. The limits are shown for the cases where $m_{\Ppi_{3}}=10m_{\dphoton}$ (left column) and $m_{\Ppi_{3}}=3m_{\dphoton}$ (right column), respectively. It is assumed that $m_{\Peta}=\TLambda=4m_{\tilde{\pi}_{2}}$ and ${\sin\theta=0.1}$, where $m_{\Peta}$ is the mass of the dark-sector pseudoscalar meson and $\theta$ is the mixing angle parametrising the isospin violation. The branching fraction $\mathcal{B}(\Ppi_{3}\to \dphoton\dphoton)$ is assumed to be 1. The grey bands correspond to mass regions of known SM resonances, which are masked in the search.}
\label{scenarioB1_limits_per_lifetime}
\end{figure}

 Limits are also shown as functions of the dark vector meson mass in the vector portal model (Fig.~\ref{vector_limits_per_lifetime}), and the dark photon mass for Scenario A and Scenario B1 models in Fig.~\ref{scenarioA_limits_per_lifetime} and Fig.~\ref{scenarioB1_limits_per_lifetime}, respectively. The limits are shown for two representative proper decay length hypotheses. For Scenarios A and B1, two different cases are considered for the mass ratio between the dark pion $\Ppi_{3}$ and the dark photon \dphoton, which are $m_{\Ppi_{3}}=10m_{\dphoton}$ and $m_{\Ppi_{3}}=3m_{\dphoton}$ respectively. The limits are found to be stronger for the case with $m_{\Ppi_{3}}=10m_{\dphoton}$, because of the lower dark pion multiplicity, leading to a more energetic \dphoton. The final-state dimuon system is more Lorentz boosted and is more displaced as a result. The improvement in sensitivity from the higher displacement outweighs the effect of the lower multiplicity of SVs in this case, leading to an improvement in the sensitivity. The decrease in sensitivity at $m_{\dphoton}\sim0.7\GeV$ is due to a decrease in the branching fraction $\mathcal{B}(\dphoton\to\mu\mu)$, caused by competing decays of \dphoton to $\PGpp\PGpm$ or $\PGpp\PGpm\PGpz$.

\begin{figure}[ht!]
\centering
\includegraphics[width=0.8\textwidth]{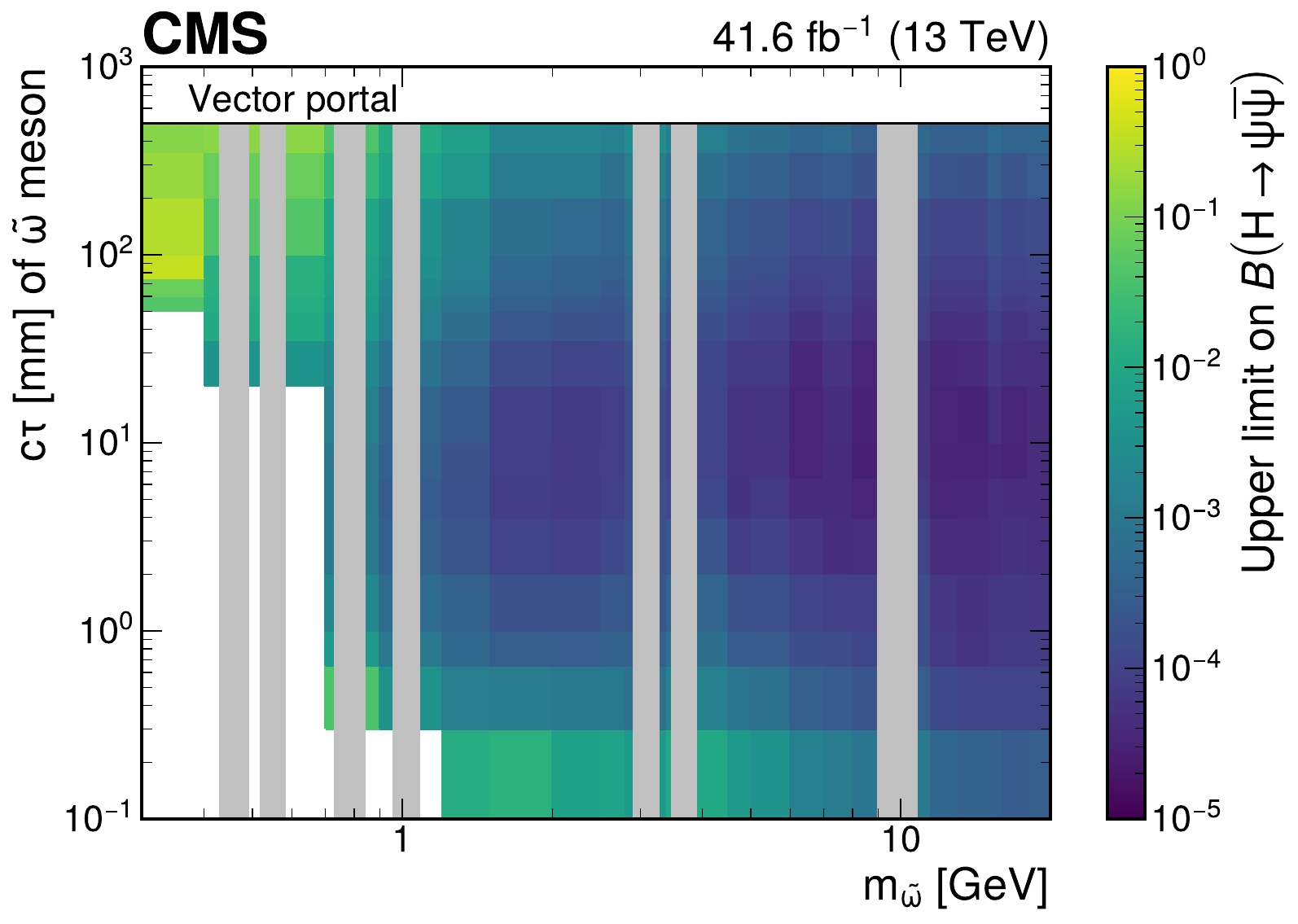}
\caption{Observed upper limits at 95\% \CL on the branching fraction $\mathcal{B}(\PH\to\dpart\overline{\dpart})$ as a function of the \Pomega meson mass and $c\tau$ in the vector portal model. The parameter space region that is omitted is kinematically forbidden in the model. It is assumed that $m_{\Pomega}=\TLambda=m_{\Peta}$, where $m_{\Pomega}$, \TLambda, and $m_{\Peta}$ are parameters of the dark sector: the mass of the spin-one meson, confinement scale, and the mass of the spin-zero meson, respectively. The grey bands correspond to mass regions of known SM resonances, which are masked in the search.}
\label{vector_2D_limits}
\end{figure}

\begin{figure}[ht!]
\centering
\includegraphics[width=0.48\textwidth]{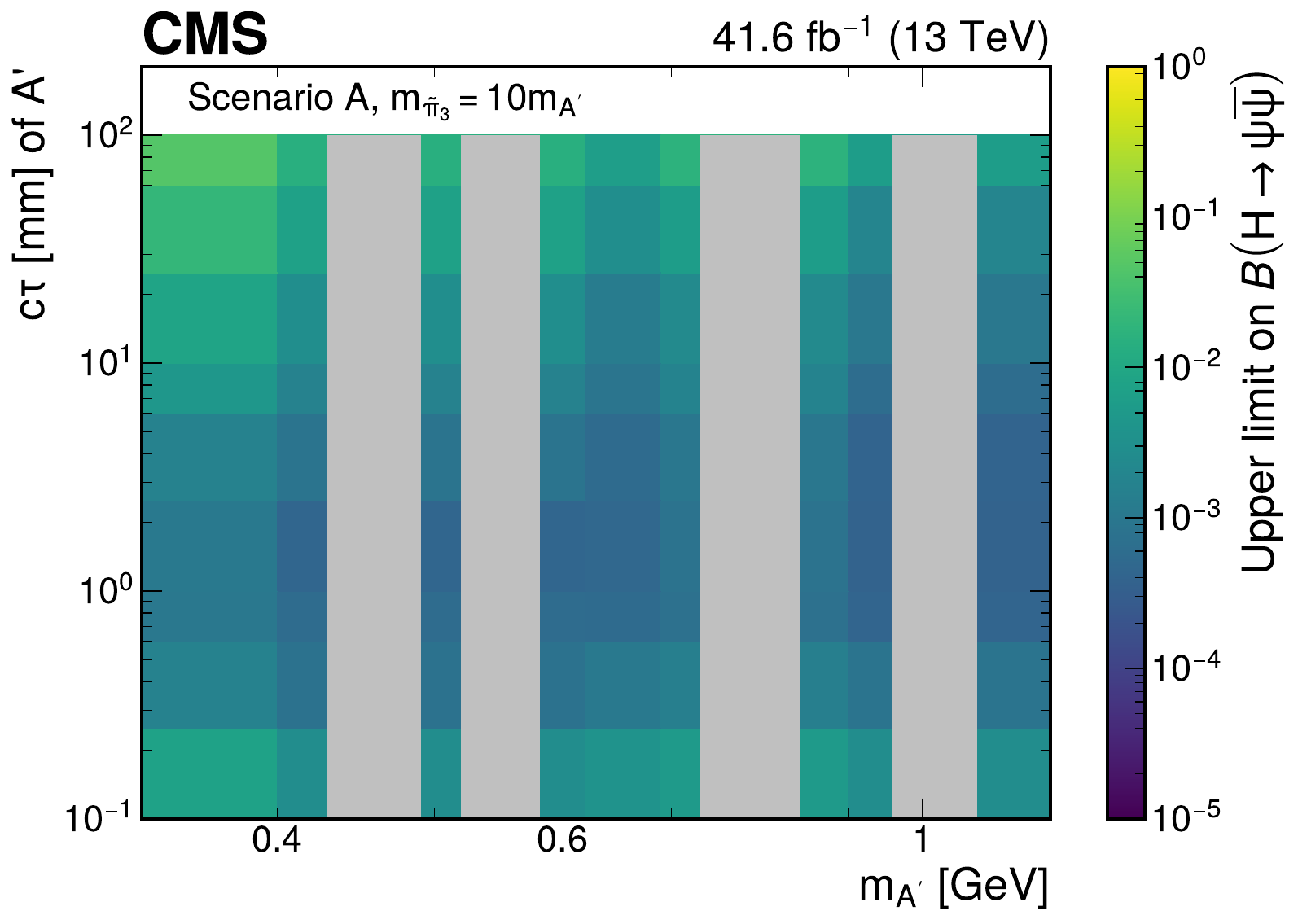}
\includegraphics[width=0.48\textwidth]{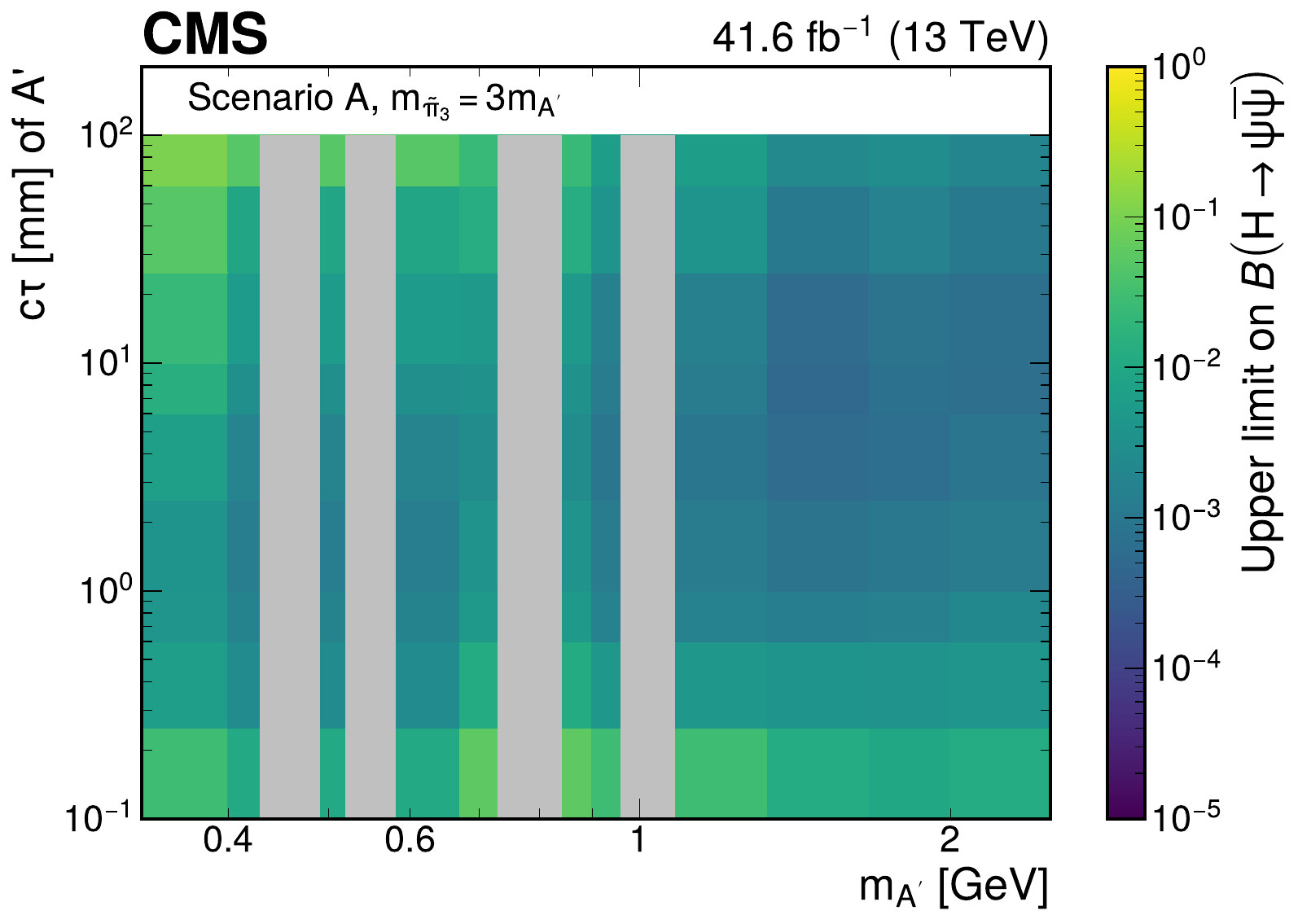}
\caption{Observed upper limits at 95\% \CL on the branching fraction $\mathcal{B}(\PH\to\dpart\overline{\dpart})$ as functions of the \dphoton mass and $c\tau$ for representative ratios of the $\Ppi_{3}$ and \dphoton masses in Scenario A. The limits are shown for the cases where $m_{\Ppi_{3}}=10m_{\dphoton}$ (left) and $m_{\Ppi_{3}}=3m_{\dphoton}$ (right), respectively. It is assumed that $m_{\Peta}=\TLambda=4m_{\Ppi_{2}}$ and ${\sin\theta=0.1}$, where $m_{\Peta}$ is the mass of the dark-sector pseudoscalar meson and $\theta$ is the mixing angle parametrising the isospin violation. The branching fraction $\mathcal{B}(\Ppi_{3}\to \dphoton\dphoton)$ is assumed to be 1. The grey bands correspond to mass regions of known SM resonances, which are masked in the search.}
\label{scenarioA_2D_limits}
\end{figure}

\begin{figure}[ht!]
\centering
\includegraphics[width=0.48\textwidth]{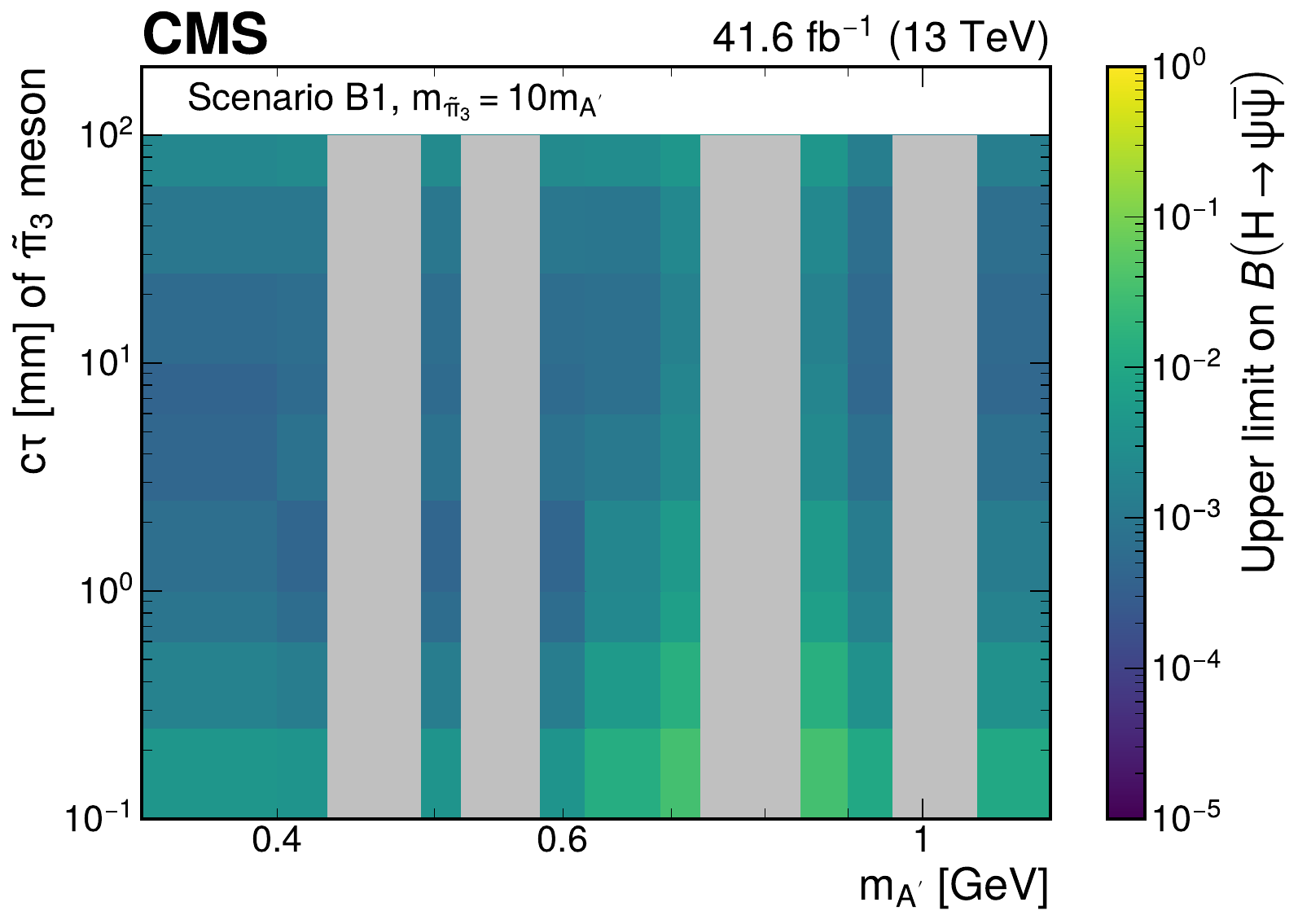}
\includegraphics[width=0.48\textwidth]{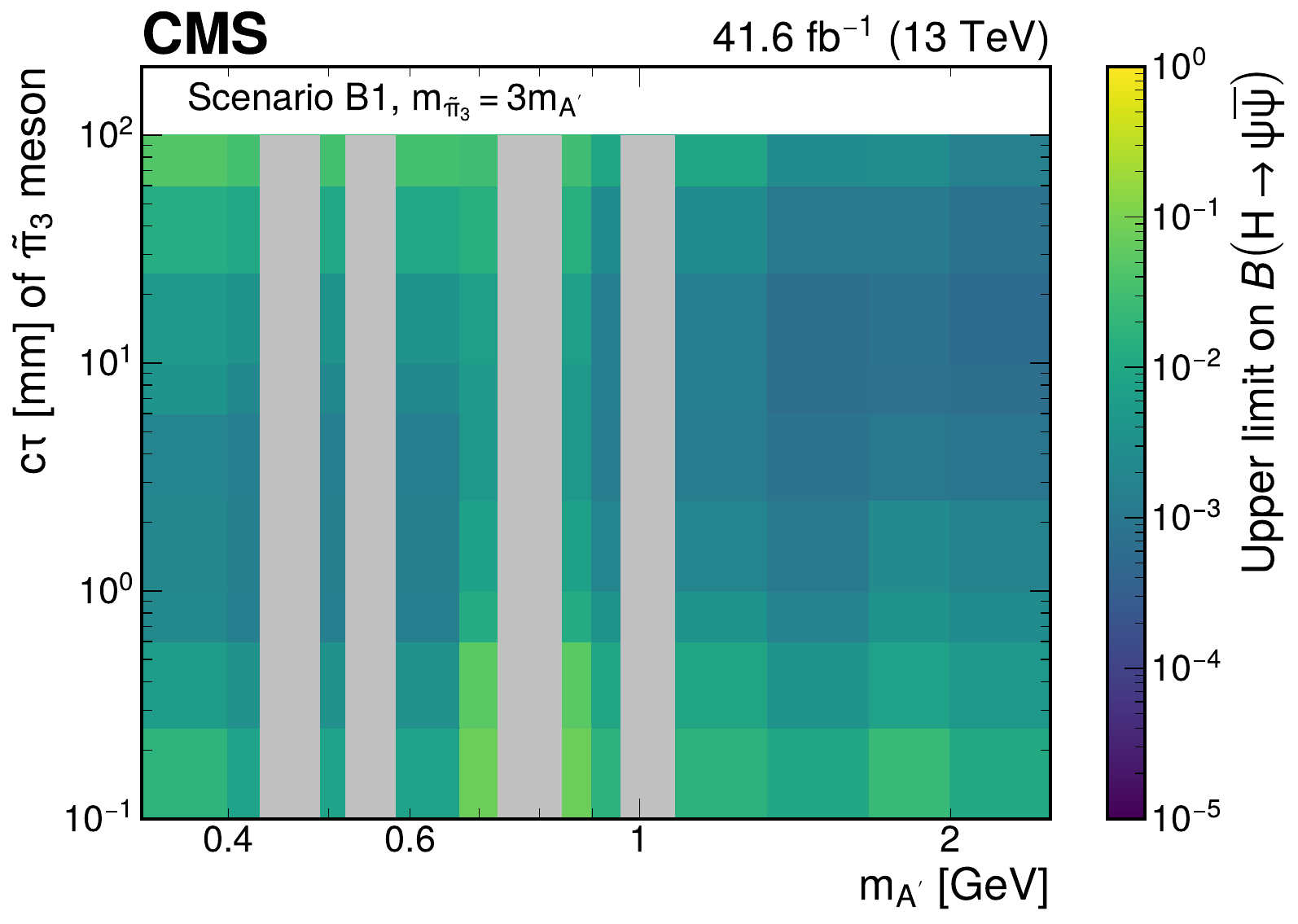}
\caption{Observed upper limits at 95\% \CL on the branching fraction $\mathcal{B}(\PH\to\dpart\overline{\dpart})$ as functions of the \dphoton mass and the $\Ppi_{3}$ $c\tau$ for representative ratios of the $\Ppi_{3}$ and \dphoton masses in Scenario B1. The limits are shown for the cases where $m_{\Ppi_{3}}=10m_{\dphoton}$ (left) and $m_{\Ppi_{3}}=3m_{\dphoton}$ (right), respectively. It is assumed that $m_{\Peta}=\TLambda=4m_{\Ppi_{2}}$ and ${\sin\theta=0.1}$, where $m_{\Peta}$ is the mass of the dark sector pseudoscalar meson and $\theta$ is the mixing angle parametrising the isospin violation. The branching fraction $\mathcal{B}(\Ppi_{3}\to \dphoton\dphoton)$ is assumed to be 1. The grey bands correspond to mass regions of known SM resonances, which are masked in the search.}
\label{scenarioB1_2D_limits}
\end{figure}

 Observed limits at 95\% \CL are shown in the two-dimensional parameter space of the dark vector meson mass and proper decay length for the vector portal in Fig.~\ref{vector_2D_limits}, the dark photon mass and proper decay length for Scenario A in Fig.~\ref{scenarioA_2D_limits}, and the dark photon mass and the dark pion ($\Ppi_{3}$) proper decay length for Scenario B1 in Fig.~\ref{scenarioB1_2D_limits}.

\clearpage

\section{Summary}
A search for dark showers has been performed with non-prompt muon pairs, using proton-proton collisions at the CERN LHC at a centre-of-mass energy of 13\TeV, collected by the CMS experiment in 2018, corresponding to an integrated luminosity of 41.6\fbinv. The data set is recorded using the data parking strategy, resulting in a sample of about $10^{10}$ recorded events, giving access to masses down to the sub-GeV scale. No significant excess beyond the standard model expectation is observed. Upper limits on the branching fraction of the Higgs boson decaying into dark partons are set as low as $10^{-4}$ at 95\% confidence level, providing the most stringent limits to date on the vector portal model with dark \Pomega mesons of mean proper decay length below 500\mm and masses between 0.3 and 20\GeV. For the first time, limits have been set for extended dark-shower models with two dark flavours (Scenario A and Scenario B1) that contain dark photons, probing their masses down to 0.33\GeV.

\begin{acknowledgments}
We congratulate our colleagues in the CERN accelerator departments for the excellent performance of the LHC and thank the technical and administrative staffs at CERN and at other CMS institutes for their contributions to the success of the CMS effort. In addition, we gratefully acknowledge the computing centres and personnel of the Worldwide LHC Computing Grid and other centres for delivering so effectively the computing infrastructure essential to our analyses. Finally, we acknowledge the enduring support for the construction and operation of the LHC, the CMS detector, and the supporting computing infrastructure provided by the following funding agencies: SC (Armenia), BMBWF and FWF (Austria); FNRS and FWO (Belgium); CNPq, CAPES, FAPERJ, FAPERGS, and FAPESP (Brazil); MES and BNSF (Bulgaria); CERN; CAS, MoST, and NSFC (China); MINCIENCIAS (Colombia); MSES and CSF (Croatia); RIF (Cyprus); SENESCYT (Ecuador); ERC PRG, TARISTU24-TK10 and MoER TK202 (Estonia); Academy of Finland, MEC, and HIP (Finland); CEA and CNRS/IN2P3 (France); SRNSF (Georgia); BMFTR, DFG, and HGF (Germany); GSRI (Greece); NKFIH (Hungary); DAE and DST (India); IPM (Iran); SFI (Ireland); INFN (Italy); MSIT and NRF (Republic of Korea); MES (Latvia); LMTLT (Lithuania); MOE and UM (Malaysia); BUAP, CINVESTAV, CONACYT, LNS, SEP, and UASLP-FAI (Mexico); MOS (Montenegro); MBIE (New Zealand); PAEC (Pakistan); MES, NSC, and NAWA (Poland); FCT (Portugal); MESTD (Serbia); MICIU/AEI and PCTI (Spain); MOSTR (Sri Lanka); Swiss Funding Agencies (Switzerland); MST (Taipei); MHESI (Thailand); TUBITAK and TENMAK (T\"{u}rkiye); NASU (Ukraine); STFC (United Kingdom); DOE and NSF (USA).

\hyphenation{Rachada-pisek} Individuals have received support from the Marie-Curie programme and the European Research Council and Horizon 2020 Grant, contract Nos.\ 675440, 724704, 752730, 758316, 765710, 824093, 101115353, 101002207, 101001205, and COST Action CA16108 (European Union); the Leventis Foundation; the Alfred P.\ Sloan Foundation; the Alexander von Humboldt Foundation; the Science Committee, project no. 22rl-037 (Armenia); the Fonds pour la Formation \`a la Recherche dans l'Industrie et dans l'Agriculture (FRIA) and Fonds voor Wetenschappelijk Onderzoek contract No. 1228724N (Belgium); the Beijing Municipal Science \& Technology Commission, No. Z191100007219010, the Fundamental Research Funds for the Central Universities, the Ministry of Science and Technology of China under Grant No. 2023YFA1605804, the Natural Science Foundation of China under Grant No. 12061141002, 12535004, and USTC Research Funds of the Double First-Class Initiative No.\ YD2030002017 (China); the Ministry of Education, Youth and Sports (MEYS) of the Czech Republic; the Shota Rustaveli National Science Foundation, grant FR-22-985 (Georgia); the Deutsche Forschungsgemeinschaft (DFG), among others, under Germany's Excellence Strategy -- EXC 2121 ``Quantum Universe" -- 390833306, and under project number 400140256 - GRK2497; the Hellenic Foundation for Research and Innovation (HFRI), Project Number 2288 (Greece); the Hungarian Academy of Sciences, the New National Excellence Program - \'UNKP, the NKFIH research grants K 131991, K 133046, K 138136, K 143460, K 143477, K 146913, K 146914, K 147048, 2020-2.2.1-ED-2021-00181, TKP2021-NKTA-64, and 2021-4.1.2-NEMZ\_KI-2024-00036 (Hungary); the Council of Science and Industrial Research, India; ICSC -- National Research Centre for High Performance Computing, Big Data and Quantum Computing, FAIR -- Future Artificial Intelligence Research, and CUP I53D23001070006 (Mission 4 Component 1), funded by the NextGenerationEU program (Italy); the Latvian Council of Science; the Ministry of Education and Science, project no. 2022/WK/14, and the National Science Center, contracts Opus 2021/41/B/ST2/01369, 2021/43/B/ST2/01552, 2023/49/B/ST2/03273, and the NAWA contract BPN/PPO/2021/1/00011 (Poland); the Funda\c{c}\~ao para a Ci\^encia e a Tecnologia, grant CEECIND/01334/2018 (Portugal); the National Priorities Research Program by Qatar National Research Fund;  MICIU/AEI/10.13039/501100011033, ERDF/EU, "European Union NextGenerationEU/PRTR", and Programa Severo Ochoa del Principado de Asturias (Spain); the Chulalongkorn Academic into Its 2nd Century Project Advancement Project, the National Science, Research and Innovation Fund program IND\_FF\_68\_369\_2300\_097, and the Program Management Unit for Human Resources \& Institutional Development, Research and Innovation, grant B39G680009 (Thailand); the Kavli Foundation; the Nvidia Corporation; the SuperMicro Corporation; the Welch Foundation, contract C-1845; and the Weston Havens Foundation (USA).
\end{acknowledgments}

\bibliography{auto_generated} 

\cleardoublepage \appendix\section{The CMS Collaboration \label{app:collab}}\begin{sloppypar}\hyphenpenalty=5000\widowpenalty=500\clubpenalty=5000
\cmsinstitute{Yerevan Physics Institute, Yerevan, Armenia}
{\tolerance=6000
A.~Hayrapetyan, V.~Makarenko\cmsorcid{0000-0002-8406-8605}, A.~Tumasyan\cmsAuthorMark{1}\cmsorcid{0009-0000-0684-6742}
\par}
\cmsinstitute{Institut f\"{u}r Hochenergiephysik, Vienna, Austria}
{\tolerance=6000
W.~Adam\cmsorcid{0000-0001-9099-4341}, J.W.~Andrejkovic, L.~Benato\cmsorcid{0000-0001-5135-7489}, T.~Bergauer\cmsorcid{0000-0002-5786-0293}, M.~Dragicevic\cmsorcid{0000-0003-1967-6783}, C.~Giordano, P.S.~Hussain\cmsorcid{0000-0002-4825-5278}, M.~Jeitler\cmsAuthorMark{2}\cmsorcid{0000-0002-5141-9560}, N.~Krammer\cmsorcid{0000-0002-0548-0985}, A.~Li\cmsorcid{0000-0002-4547-116X}, D.~Liko\cmsorcid{0000-0002-3380-473X}, M.~Matthewman, I.~Mikulec\cmsorcid{0000-0003-0385-2746}, J.~Schieck\cmsAuthorMark{2}\cmsorcid{0000-0002-1058-8093}, R.~Sch\"{o}fbeck\cmsAuthorMark{2}\cmsorcid{0000-0002-2332-8784}, D.~Schwarz\cmsorcid{0000-0002-3821-7331}, M.~Shooshtari\cmsorcid{0009-0004-8882-4887}, M.~Sonawane\cmsorcid{0000-0003-0510-7010}, W.~Waltenberger\cmsorcid{0000-0002-6215-7228}, C.-E.~Wulz\cmsAuthorMark{2}\cmsorcid{0000-0001-9226-5812}
% ...
\par}
\vskip\cmsinstskip
\dag:~Deceased\\
$^{1}$Also at Yerevan State University, Yerevan, Armenia\\
$^{2}$Also at TU Wien, Vienna, Austria\\
% ...
\end{sloppypar}
\end{document}